\documentclass[useAMS,usenatbib]{mn2e}

\usepackage{graphicx}
\usepackage{amsmath}
\usepackage{amssymb}

\usepackage{color}

\def\electron{e}
\newcommand{\figlab}[2]{\hspace{-2.5cm}\makebox[2.5cm][r]{\raisebox{#2}{\small #1\hspace{0.1cm}}}} 
\newcommand{\figlabl}[2]{\hspace{-3.5cm}\makebox[3.5cm][l]{\raisebox{#2}{\small #1\hspace{0.1cm}}}} 

\title[Time dependent gamma-ray production in the IC $e^\pm$ pair cascade]
{Time dependent gamma-ray production in the anisotropic IC $e^\pm$ pair cascade
initiated by electrons in active galaxies}

\author[J. Sitarek and W. Bednarek]{J. Sitarek$^{1,2}$
and W. Bednarek$^{2}$ \\
$^{1}$Max-Planck-Institut f\"ur Physik, D-80805 M\"unchen, Germany, jsitarek@mppmu.mpg.de\\
$^{2}$Department of Astrophysics, University of \L\'od\'z, PL-90236 \L \'od\'z, Poland, bednar@astro.phys.uni.lodz.pl}

\begin{document}

\date{Accepted . Received ; in original form }

\pagerange{\pageref{firstpage}--\pageref{lastpage}} \pubyear{2010}

\maketitle

\label{firstpage}

\begin{abstract}
New high energy emission features have been recently discovered by the Cherenkov telescopes from active galaxies e.g., a few minutes variability time scale of TeV emission from Mrk 501 and PKS 2155-304, sub-TeV $\gamma$-ray emission from GeV peaked blazar 3C 279, and TeV emission from two nearby active galaxies, M87 and Cen~A, which jets are inclined at a relatively large angle to the line of sight. 
These results have put a new light on the high energy processes occurring in central parts of active galaxies stimulating more detailed studies of $\gamma$-ray emission models.
Here we report the results of a detailed analysis concerning the most general version of the model for the $\gamma$-ray production by leptons injected in the jet which interact with the thermal radiation from an accretion disk (the so called {\it external inverse Compton model}). 
We investigate the $\gamma$-ray spectra produced in an anisotropic Inverse Compton (IC) $e^\pm$ pair cascade in the whole volume above the accretion disk.
The cascade $\gamma$-ray spectra are obtained for different locations of the observer in respect to the direction of the jet. 
We also study the time evolution of this $\gamma$-ray emission caused by the propagation of the relativistic leptons along the jet and the delays resulting from different places of the origin of $\gamma$-rays above the accretion disk. 
We discuss the main features of such a cascade model assuming constant injection rate of electrons along the jet.
We are investigating two models for maximum energies of injected electrons: with a constant value independent on the distance along the jet or limited by the synchrotron energy losses considered locally in the jet. 
The model is discussed in the context of blazars observed at small and large inclination angles taking as an example the parameters of the two famous sources Cen~A and 3C 279.

\end{abstract}

\section{Introduction}

Recent observations of active galaxies in the TeV $\gamma$-rays confirmed earlier reports on an extremely variable nature of this emission. 
It can change on various time scales starting from a few minutes  (e.g. \cite{gai96,al07,ah07}) up to months/years \citep{hs09}.
This variable emission can correspond to the time scales of, either the periods of enhanced energy generation in the disk-jet system, or acceleration time of particles, or the passing time of relativistic particles through the medium which creates the target for efficient $\gamma$-ray production.
Moreover, the time scales can be modified due to relativistic effects caused by the movement of the emission region towards the observer and, for very distant sources, cosmological propagation.
The precise location of the emission site within the active nuclei remains unclear.
Theoretical arguments suggest that the emission place is rather distant from the black hole.
$\gamma$-ray radiation produced closer cannot escape to the observer due to a very strong, soft radiation field originating from the inner part of the accretion disk.
On the other hand, this strong radiation field can create a target on which $\gamma$-rays could be produced.
Recent observations of the nearby active galaxy, M87, whose jet is observed at a relatively large angle, suggest that the TeV $\gamma$-ray emission region can likely appear within a few tens of Schwarzschild radius from the black hole \citep{ah06,acc08,al08a}. 
It is likely that the emission region is close to the accretion disk inside the inner jet as suggested by the recently discovered connection between the TeV $\gamma$-ray flare and the subsequent appearance of the radio blobs \citep{acc09}. 
Also Cen~A, another nearby active galaxy inclined at large angle, has been recently detected as a TeV $\gamma$-ray source \citep{ah09}.
This emission is as well interpreted as originated very close to the central black hole \citep{ra09}.
 
Recently the LAT Telescope on board of the {\it Fermi} observatory is operating with an order of magnitude better sensitivity than the EGRET telescope.
So far, after 11 months of operations it observed 709 AGNs in the GeV range \citep{ab10}.
Among them 300 are BL Lac objects and 296 are flat-spectrum radio quasars.
On top of this, 8 objects have been associated as misaligned blazars. 
Therefore, investigation of the anisotropic high energy radiation processes in the  active galaxies is very timely. 

It is widely believed that $\gamma$-rays are produced in active galaxies by particles (leptons or hadrons) accelerated in the jet launched perpendicularly to the plane of the accretion disk. The acceleration mechanism is not well known. 
Usually two idealized cases of angular distribution of particles in the jet are discussed, either isotropic in the jet frame or in the recti-linear acceleration regions generally directed along the jet.  
The first case corresponds to the acceleration of particles in a shock wave. 
Due to a mild relativistic motion of the emission region this scenario is sometimes referred to as a {\it slow jet model}.
The latter case may correspond to the acceleration in magnetic field reconnection regions.
This scenario is sometimes called as {\it a fast jet model} because of  an ordered, super-relativistic motion of particles. 

$\gamma$-rays can be produced in different radiation mechanisms. 
A few basic scenarios have been proposed soon after discovery of the GeV $\gamma$-ray emission from blazars by the EGRET telescope on the CGRO.
Among the famous leptonic models are: 
synchrotron self-Compton (SSC) model \citep{mgc92}, 
external Compton (EC) model with scattering of the disk radiation \citep{dsm92}, 
or radiation around the disk \citep{sbr94}. 
Also hadronic models have been discussed, e.g. proton initiated cascade (PIC) model \citep{mb92}. 
Starting from that discovery period, several modifications of these first more general scenarios have been discussed in detail with the applications to specific objects. A more realistic three dimensional cascade model occurring in the strong  
radiation of the accretion disk are starting to be considered only recently \citep{sb10, rb10}.
In this paper we are specially interested in the formation of the $\gamma$-ray spectrum in terms of the external Compton accretion disk model (see for details of this model \citet{ds93}).

It has been shown that for typical parameters of blazars detected in GeV-TeV $\gamma$-ray energies, the optical depths for $\gamma$-rays injected relatively close to the disk at an arbitrary place can be clearly above unity, see e.g. recent calculations for 3C 279 and 3C 273 \citep{sb08} or for Cen~A \citep{sb10}. 
Therefore, the cascade processes can be important in these sources provided that injection of primary particles occurs relatively close to the central engine as suggested by the recent observations of TeV $\gamma$-ray emission from some objects (e.g. Cen~A, M87).

We have already investigated the $\gamma$-ray production in blazars in a scenario of IC $e^\pm$ pair cascade initiated by primary $\gamma$-rays produced e.g., in the Synchrotron self-Compton model. 
These $\gamma$-rays develop IC $e^\pm$ pair cascade in the thermal radiation from the accretion disk in the whole volume above the disk~\citep{sb10}. 
We have calculated the $\gamma$-ray spectra produced in such a model for arbitrary location of the observer in respect to the direction of the jet.
Furthermore the influence of a magnetic field above the accretion disk on the development of the IC $e^\pm$ pair cascade has been also investigated.

In the present paper we assume that the cascade process in the accretion disk is initiated by leptons (electrons and positrons) accelerated in the jet as postulated by the external IC model \citep{ds93}. 
These leptons produce the first generation of $\gamma$-rays by scattering an anisotropic disk radiation.
Therefore the angular distribution and spectrum of those primary $\gamma$-rays is dependent on the disk/jet system properties. They significantly differ from the spectra obtained in terms of the previous calculations~\citep{sb10} in which primary $\gamma$-rays were injected isotropically in the blob frame. 
These first generation $\gamma$-rays initiate anisotropic IC $e^\pm$ pair cascade in the radiation field of the accretion disk. In a framework of this model we investigate spectral, angular and also time properties of the $\gamma$-ray emission.
Therefore, our present calculations follow significantly more complicated and complete model than offered by the previously considered scenario with the isotropic injection of primary $\gamma$-rays produced in the SSC model. 

\section{A model for gamma-ray production in IC $\electron^\pm$ pair cascade}\label{sec:model}

We adopt a classical optically thick and geometrically thin accretion disk model around super massive black hole as a dominant source of radiation in the central region of the active galactic nuclei (see \cite{ss73}). 
The emission of the accretion disk is treated as a black body with a power law temperature dependence on the distance,~$r$, from the black hole, $T=T_{\rm in}(r/r_{\rm in})^{-3/4}$, where $T_{\rm in}$ is the temperature at the inner disk radius $r_{\rm in}$. 
We realize that the radiation field around the accretion disk can be much more complicated than produced in such a simple disk model. 
For example, the inner region of the accretion disk can become significantly hotter and geometrically thick as expected in some modern models for the advection dominated disks (e.g. \citet{ny94}). 
Also the disk may not be geometrically thin, e.g. slim disks \citep{abr88}. 

This limitation of our calculations to such a simple disk scenario is motivated by the complexity of the overall cascade model which we intend to study in this paper. 
Other versions of the model, with modified accretion disk radiation field and/or the additional components of the radiation field produced e.g. isotropically around the disk, can be studied in the future along the prescription described in this paper. 

We assume that this disk thermal radiation create the only target for relativistic primary electrons. 
The blob is assumed to be point-like, i.e. its dimensions are much smaller than the characteristic distance scale (e.g. the injection height). 
The case of an extended injection blob with a complicated geometry can be also studied by a simply integration of our results for the point-like injection over the whole injection region. 
Unfortunately, such process will require to introduce additional free parameters, which describe the structure of the injection blob. 
We leave such more complicated studies (but probably also more realistic) for the future paper.
Primary electrons are confined in a compact blob moving along the jet (perpendicular to the disk plane) with a constant velocity $v$. 
Directions of injected electrons are isotropic in the blob frame.
Note however, that for $v\cong c$, due to the relativistic beaming, distribution of electrons in the disk frame is highly anisotropic, with more electrons moving in the direction of the jet. In the disk reference frame the energy of the electron depends on its instantaneous direction.

Primary electrons with specific spectrum interact with the disk radiation in the inverse Compton (IC) scattering process producing high energy $\gamma$-rays. 
These $\gamma$-rays propagate, in principle, at an arbitrary angle to the disk axis (following the instantaneous direction of the parent electron at the moment of the IC scattering). 
If the acceleration of electrons occurs relatively close to the accretion disk, then the optical depths for produced primary $\gamma$-rays can have substantial values at specific directions. 

Recently, we studied the optical depths for $\gamma$-rays propagating at an arbitrary directions in respect to the accretion disk for a few well known sources such as 3C 273, 3C 279, Cen~A~\citep{sb08,sb10}. 
It became clear that for the parameters characteristic for these objects a part of produced $\gamma$-rays can be absorbed in the disk radiation creating $e^\pm$ pairs. Thus, the first generation of $e^\pm$ pairs is created outside the jet in the whole volume above the accretion disk. 
These secondary $e^\pm$ pairs propagate in the radiation field and the magnetic field above the accretion disk producing synchrotron radiation and the next generation of $\gamma$-ray photons. As a result, IC $e^\pm$ pair cascade develop in the volume above the accretion disk in which the additional role is taken by synchrotron process of secondary $e^\pm$ pairs. The details of such 
a model, in which the cascade is initiated by primary $\gamma$-rays produced in the jet (e.g. in the Synchrotron self-Compton scenario)  has been recently considered by \citet{sb10}. It was shown that synchrotron energy losses of secondary cascade $e^\pm$ pairs can be important only close to the disk surface where the dipole magnetic field is relatively strong. Farther out of the disk, the magnetic energy density drops faster than the disk radiation energy density. Therefore, we neglect the influence
of the magnetic field on the IC $e^\pm$ pair cascade process in the present studies.

In this paper we consider another version of that scenario.  
The cascade process is initiated not by primary $\gamma$-rays but by primary electrons accelerated in the blob moving relativistically along the jet. 
We assume that electrons are injected with a constant rate into the blob which moves along the jet. Electrons are trapped within the blob without any substantial escape. They cool in the accretion disk radiation field producing $\gamma$-rays which escape the blob into the whole volume above the disk. 
Other scenarios could be in principle also considered, e.g. continuous injection of electrons only at the base of the jet for certain period of time, or their injection 
into the extended blob.
Unfortunately, the possible variety of the basic assumptions on the injection model of electrons can not be limited at present stage of our knowledge by the theory of the acceleration process of particles in the AGNs. 
Therefore, we limit our calculations in the present paper to the simple possible scenario, i.e. constant injection rate of electrons along the jet.
We develop a numerical code which follow the IC $e^\pm$ pair cascade in the 3D volume above the accretion disk. The method of the numerical simulation of the cascade scenario is described in Appendix~\ref{dodatek}. This code is also able to follow the time structure of the high energy radiation which escapes towards the observer. All cascade particles ($\gamma$-rays, $e^\pm$ pairs) are tracked in this code up to the point when they escape far away from the disk (where the radiation field is already very weak) or their energies fall below assumed threshold equal to 10~GeV. 

As noted above, electrons are isotropic in the blob frame. 
They are injected with a power law spectrum which spectral index is fixed to a constant value independent on the distance from the base of the jet.
 As an example, we apply the differential spectral index equal to $-2$ for the spectrum extending from the minimum electron energy equal to 10 GeV, up to a given maximum energy. 
These electrons cool on the IC process which efficiency depends on the distance from the accretion disk. 
So then, their equilibrium spectrum in the blob also depends on the distance from the base of the jet. 
We assume that electrons are trapped within the blob during its propagation through the accretion disk radiation field. 
We neglect the possible escape of electrons from the blob at this stage of calculations. 
It would require a detailed knowledge of the blob parameters (e.g. its dimensions, the magnetic field strength within the blob) to define the precise mechanism of escape of the electrons. 
In the future more realistic model, this effect should be taken into account.
We expect that the energy dependent escape should steepen the equilibrium spectra of electrons within the blob which will reflect in obtained $\gamma$-ray spectra. 
On the other hand, escaping electrons will propagate in a complicated way around the accretion disk (depending on the magnetic field structure) producing additional $\gamma$-ray photons. 
We are planning to study this effect in a future paper.
 
The equilibrium spectrum of electrons for different locations within the jet is determined  numerically. 
Two models for the maximum energy of injected electrons are considered. 
In the first model, the maximum energy of electrons in the blob is fixed on 10 TeV.
We apply such a high value in order to have possibility of investigation of cascade process also in the Klein-Nishina regime. 
In the second, more realistic model, the maximum energy is obtained by balancing the energy gains of electrons from the acceleration mechanism and their energy losses on the synchrotron process occurring in the magnetic field of the jet. We note that the magnetic field inside the jet is expected to be significantly stronger than in the remaining volume above the accretion disk due to its strong confinement by the jet. 
The energy gains of electrons from the acceleration mechanism in the blob can be parametrized by: 
\begin{eqnarray}
\left({\frac{dE}{dt}}\right)_{\rm acc} = \xi  c E/R_{\rm L}\approx 
10^{13}\xi B~~~{\rm eV~s^{-1}},
\label{eq4}
\end{eqnarray}
\noindent
where $\xi$ is the acceleration coefficient, $R_{\rm L}$ is the Larmor radius of electrons $E$ is the energy of electron, and $B$ is the magnetic field strength (in units of Gauss) at the acceleration region. The synchrotron energy loss rate is
\begin{eqnarray}
\left({\frac{dE}{dt}}\right)_{\rm syn} = {\frac{4}{3}}c\sigma_{\rm T}U_{\rm B}\gamma_{\rm e}^2\approx 8.6\times 10^{-4}B^2\gamma_{\rm e}^2~~~{\rm eV~s^{-1}},
\label{eq5}
\end{eqnarray}
\noindent
where $U_{\rm B}$ is the energy density of magnetic field, $\sigma_{\rm T}$ is the Thomson cross section, $c$ is the velocity of light, and $\gamma_{\rm e}$ is the Lorentz factor of electrons. By comparing Eqs.~\ref{eq4} and~\ref{eq5}, we get the limit on the maximum energy of accelerated electrons due to the synchrotron losses:
\begin{eqnarray}
E_{\rm syn}^{\rm max}\approx 5(\xi_{-2}/B)^{1/2}~~~{\rm TeV,} 
\label{eq6}
\end{eqnarray}
\noindent
where $\xi = 0.01\xi_{-2}$. 
In a similar way we can obtain the limit on the maximum electron energy by comparing their energy gains with energy losses on IC process.
The Klein-Nishina cross section drop at higher energies (KN effect).
Therefore, unless the energy density of disk radiation do not dominate completely over the energy density of the magnetic field in the jet, this limitation is less restrictive.

We estimate the magnetic field in the jet by assuming some level of the equipartition between the magnetic field and radiation at the disk inner radius,
i.e  $U_{\rm B} = \eta U_{\rm rad}^{\rm disk}$, 
where $\eta$ is a parameter which describes the level of equipartition. Based on this equality, we simply relate the magnetic field at $r_{\rm in}$ to the inner disk temperature, $T_{\rm in}$ by,
$B_{\rm in}\approx 40\sqrt{\eta} T^2_4$ G, where $T_{\rm in} = 10^4T_4$ K. 
If the blob is filling the cross section of a conical jet, we estimate the value of the magnetic field along the jet axis,
\begin{eqnarray}
B(h) = B_{\rm in}r_{\rm in}^2/(r_{\rm in}+h\sin\alpha)^2.
\end{eqnarray}
The opening angle of the jet  $\alpha$ is assumed of the order of $\sim 5^{\rm o}$.
Those magnetic fields are strong enough to confine primary electrons inside the blob.
For such magnetic field model in the jet, we calculate the maximum energies of electrons
at different locations along the jet (see Eq.~\ref{eq6}). In the cascade calculations we assume $\xi = 0.01$.
The first model for electron injection (model I) corresponds to the situation of a weak magnetic field in the jet (in respect to radiation field, i.e. $\eta << 1$). Another possibility is the acceleration of particles in the rectilinear reconnection regions along the magnetic field lines without significant synchrotron energy losses. 
The second model (model II) corresponds to strong magnetic field in the jet.
For the calculations performed in the framework of the second model we assume $\eta = 1$.

\section{Gamma-ray spectra expected at different inclination angles}\label{sec_3}

The details of calculations of $\gamma$-ray spectra produced in such an anisotropic IC $e^\pm$ pair cascade are described in Appendix~\ref{dodatek}. 
The injection spectrum of primary electrons, its cut-off energy at specific distances from the base of the jet, and other assumptions of the considered model are described in Sect.~2. 
The calculations are shown for the case of electron injection spectra normalized to 1 erg in the blob reference frame. 
The spectra of cascade $\gamma$-rays escaping to the observer will certainly depend on the observation angle (measured in respect to the jet axis) and the velocity of the blob. 
Therefore, we consider separately the cases of active galaxies with slower blobs but with the jets inclined  at relatively large angles (e.g. such as Cen~A, M87) and with faster blobs but inclined at relatively small angles to the line of sight. 
We note that the $\gamma$-ray cascading processes should occur efficiently, since the $\gamma$-spheres for these sources can extend to large distances from the center of the accretion disk (see Figs.~4 and 5 in \citet{sb08} for 3C 279 and Fig.~1 in \citet{sb10} for Cen~A). 

\subsection{Blazars at large inclination angles (Cen~A)}

We apply the parameters expected for the misaligned blazar Cen~A, as an example of an AGN seen at large inclination angle. Cen~A was observed in GeV $\gamma$-rays by the EGRET instrument on board of the Compton GRO. The measured spectral index of $-2.58$ \citep{h99} is rather soft.
The emission was weak, with no significant flares \citep{sr99}. 
Observations with the LAT instrument on board of the {\it Fermi} satellite show similar behavior with a spectral index of $-2.71$ \citep{ab10}. Cen~A is the second radio galaxy (after M~87) observed in the VHE gamma-rays. With the integral flux of $0.8\%$ of Crab Nebula flux, it is one of the weakest VHE gamma-ray sources observed so far \citep{ah09}. The spectral index of VHE gamma-rays ($-2.7$) is consistent with the values obtained at the GeV energies.

It is supposed that the central engine of Cen~A harbors a super massive black hole with the mass estimated on $M_{\rm Cen A} = (5.5\pm 3.0)\times 10^7 M_\odot$~\citep{ca08}. A clear jet is observed in this radio galaxy. 
It propagates at the inclination angle estimated on $\alpha_{jet}\approx 15^\circ-80^\circ$~\citep{ho06}.
We apply the inner disk temperature equal to $T_{\rm in} = 3\times 10^4$~K, which is of the order of that one observed directly in other AGNs (e.g. 3C 273). The total disk luminosity with these parameters is estimated on 
$L_{\rm D} = 4\pi \sigma_{\rm SB}r_{\rm in}^2T_{\rm in}^4\approx 3.6\times 10^{41}$ erg s$^{-1}$,  where $r_{\rm in} = 4.5\times 10^5M_{\rm Cen A}/M_\odot$ cm = $2.5\times 10^{13}$cm.
Therefore, the disk luminosity is comparable to the measured nuclear X-ray luminosity probably originating within the inner jet \citep{ev04}. 
Note that the disk is observed at a relatively large angle, which effectively reduces the observed disk luminosity. This accretion disk emission can be additionally obscured by the circumnuclear disk. Therefore, it may not be directly observed. The Cen~A jet velocity is estimated on $\beta\approx 0.5c$ \citep{ti98, ha03}.

In the case of blazars inclined at large angles ($\alpha > 10^\circ$), we consider blobs moving with mild velocities along the jet (i.e the Lorentz factor of the blob, $\gamma_{\rm b}$, is 
not much larger than 1). 
As noted above, electrons in the blob frame are isotropic having a power law spectrum. 
As an example, we consider the differential spectral index equal to $-2$ 
above the minimum electron energy equal to 10 GeV and the maximum energy described by the model I or 
model II (see Sect.~2). Injected electrons cool down on the IC process in the accretion disk radiation field 
when moving with the blob along the jet axis. The first generation of $\gamma$-rays initiate IC $e^\pm$ pair cascades in the whole volume above the disk. 

\begin{figure}
\centering
\includegraphics[scale=0.38, trim=  0 33 1 0, clip]{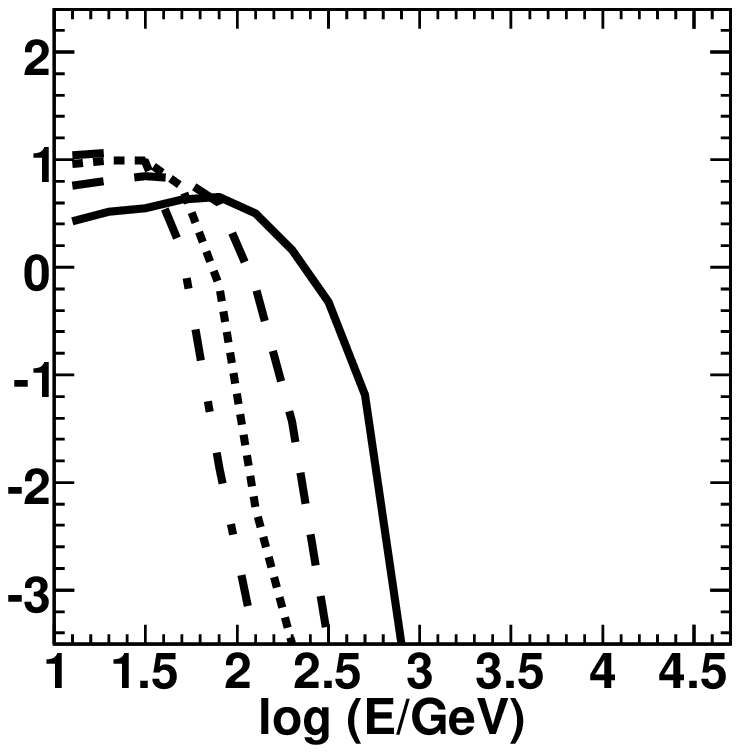}\!\!\figlab{$v=0, H_0=3r_{in}$}{2cm}
\includegraphics[scale=0.38, trim= 31 33 1 0, clip]{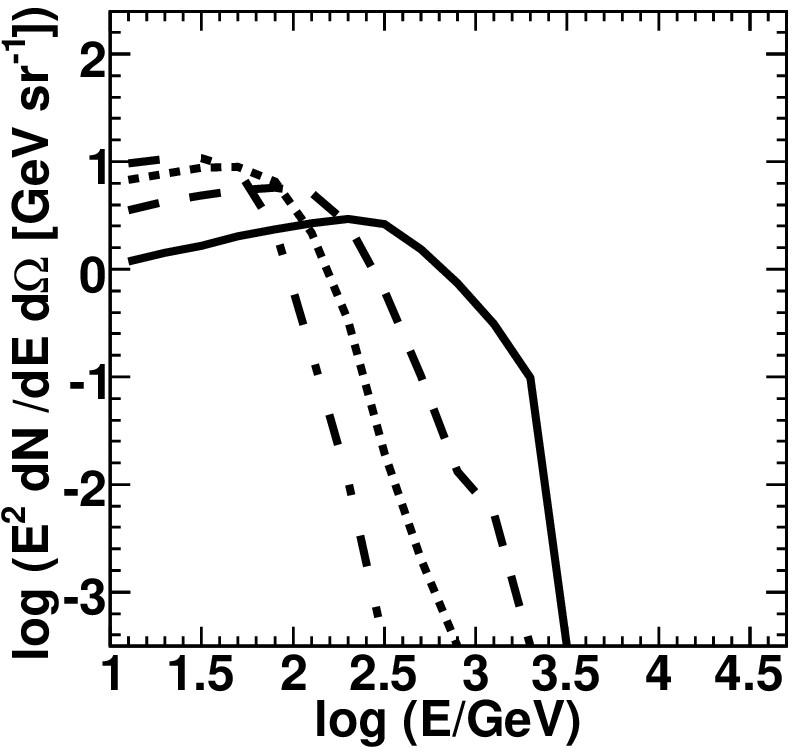}\!\!\figlab{$H_0=10r_{in}$}{2cm}   
\includegraphics[scale=0.38, trim= 31 33 0 0, clip]{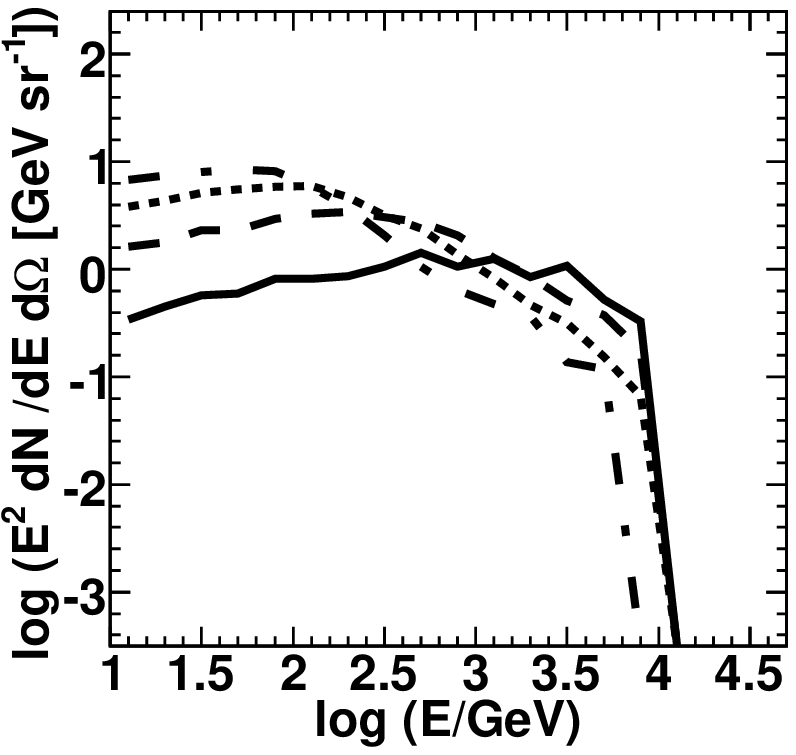}\figlab{$H_0=100r_{in}$}{2cm}\\   
						   
\includegraphics[scale=0.38, trim=  0 32 1 0, clip]{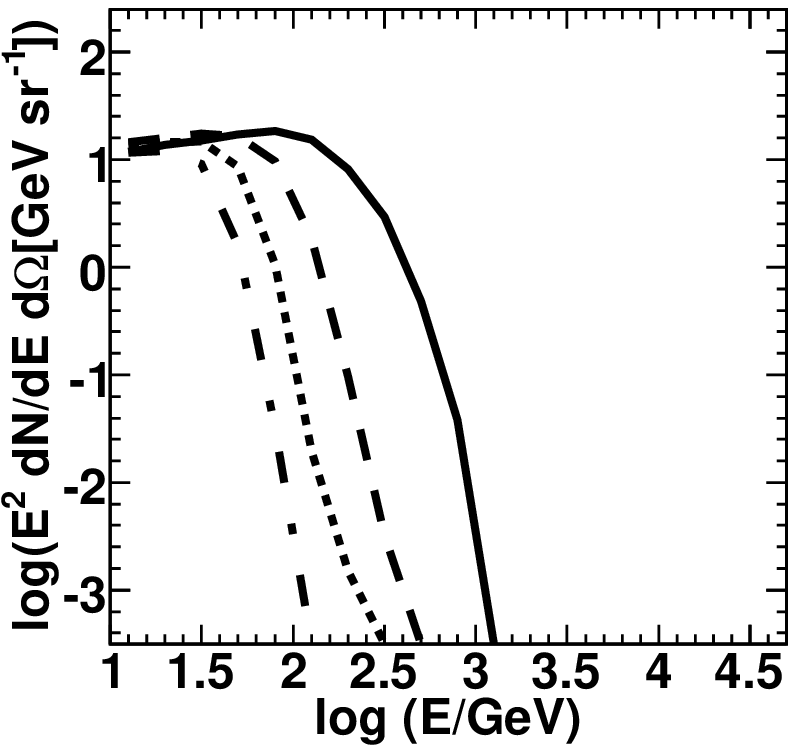}\!\!\figlab{$v=0.5c$}{2cm}
\includegraphics[scale=0.38, trim= 31 32 1 0, clip]{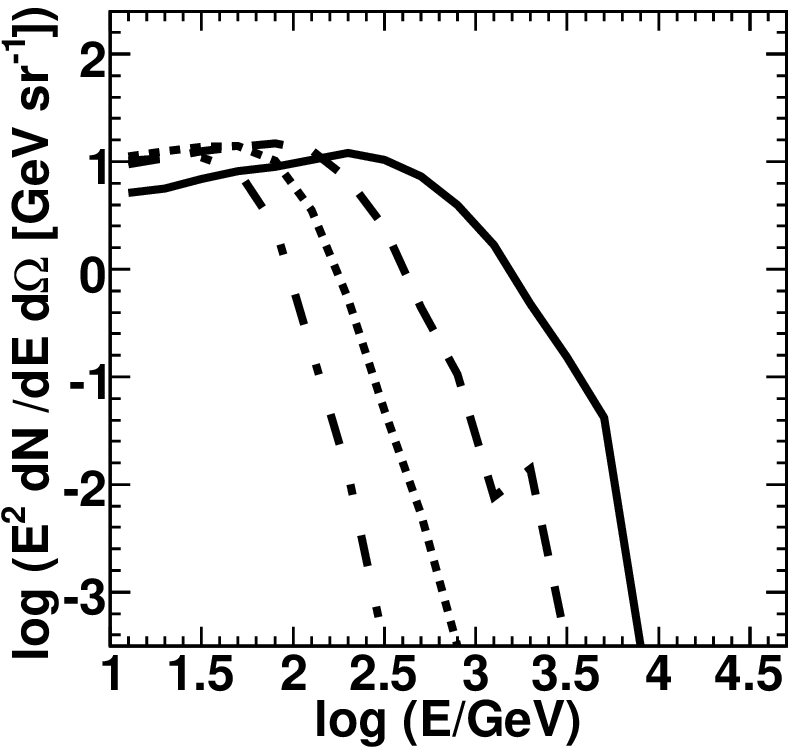}\!\!
\includegraphics[scale=0.38, trim= 31 32 0 0, clip]{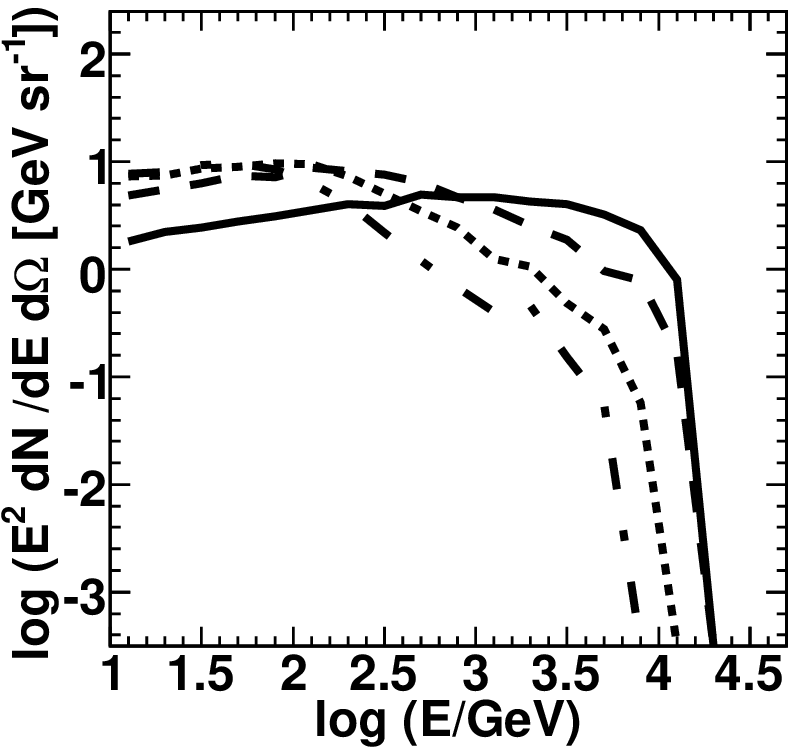}\\
						   
\includegraphics[scale=0.38, trim=  0  0 1 0, clip]{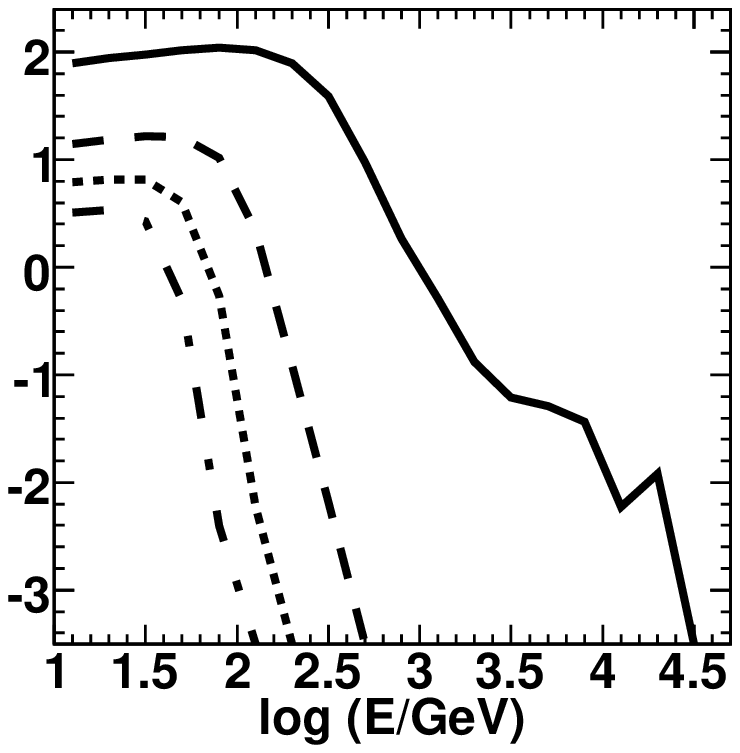}\!\!\figlab{$v=0.9c$}{2.5cm}
\includegraphics[scale=0.38, trim= 31  0 1 0, clip]{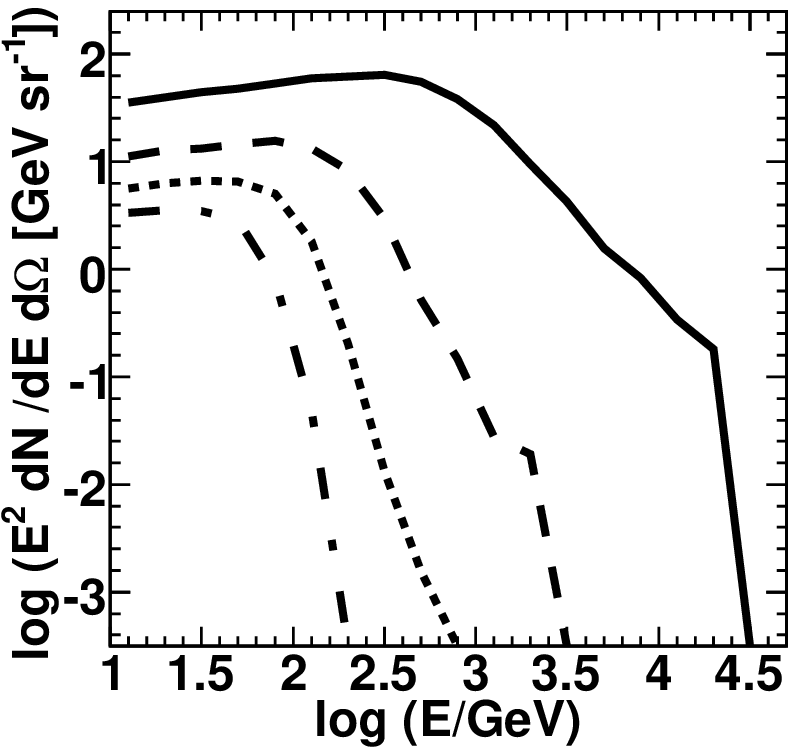}\!\!
\includegraphics[scale=0.38, trim= 31  0 0 0, clip]{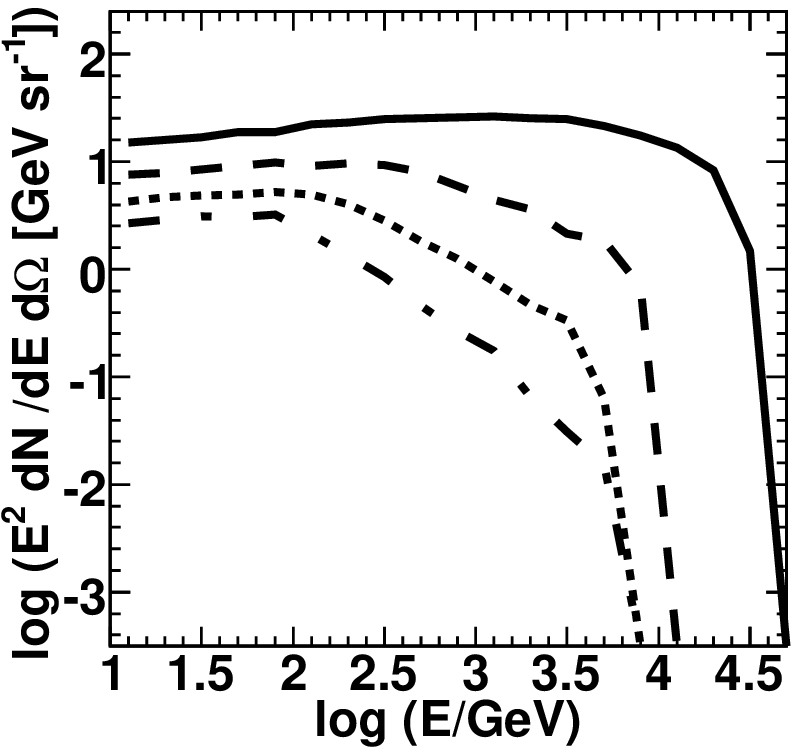}\\

\caption{Gamma-ray spectra from the IC $e^\pm$ pair cascade initiated by electrons injected 
with differential spectral index $-2$ between 10 GeV and 10 TeV (in the blob frame). Electrons are injected isotropically in the blob starting at a distance from its base equal to: 3, 10, 100 
$r_{in}$ (figures from left to right) with a velocity 0, $0.5c$, and $0.9c$. They cool gradually on the IC process. The $\gamma$-ray spectra are shown at the range of the observation angles $\xi<40^\circ$ (solid), 
$40^\circ<\xi<60^\circ$ (dashed), $60^\circ<\xi<75^\circ$ (dotted), and 
$75^\circ<\xi<90^\circ$ (dot-dashed). The calculations have been performed for the disk parameters as expected in the case of Cen~A (see text).}
\label{fig1}
\end{figure}

We consider specific cases of the blobs propagating along the jet with the velocity $v$, starting at different height, $H_0$, above the accretion disk. 
At first, we consider a case of electrons originally injected in the blob at fixed distance $H_0=3, 10, 100r_{in}$ from its base. The blob is static or propagates along the jet with the velocity $v=0.5, 0.9c$.
Electrons cool on the IC process when propagating inside the blob.
In Fig.~\ref{fig1} we show the $\gamma$-ray spectra escaping at different range of observation angles from such IC $e^\pm$ pair cascade. Let us note the main features of these cascade $\gamma$-ray spectra.
When electrons are injected close to the disk (see the case $H_0=3r_{\rm in}$), the cut-offs in the $\gamma$-ray spectra appear at energy for which the optical depth for $\gamma$-rays is comparable to unity. This cut-off clearly depends on the observation angle $\xi$. It shifts to lower energies with increasing angle. This is due to a more efficient absorption effects, since $\gamma$-rays can interact at larger angles with more energetic soft photons coming from the bright, central part of the accretion disk. 
In the case of electrons injected farther from the black hole (e.g. $H_0\gtrsim 100$), the $\gamma$-ray spectra extend clearly through the TeV energy range. The $\gamma$-ray spectra produced at larger inclination angles and the intermediate distances from the black hole show clear break.
It corresponds to the transition between the Klein-Nishina and the Thomson regimes of IC scattering. 
In contrast, there is no break present for $\gamma$-ray spectra in the case of a small $\xi$ and large $H_0$.
In this case electrons are not able to interact efficiently with the higher energy photons coming from the inner part of the accretion disk. The final cut-offs in the $\gamma$-ray spectra are related to the maximum energies of primary electrons in the disk frame.

For blobs moving with small velocities, the GeV $\gamma$-ray flux emitted at larger angles can even dominate over the $\gamma$-ray flux emitted at lower angles. 
This is caused by more efficient cascading process for photons which propagate at large inclination angles to the disk axis. For faster jets, the effects of relativistic beaming start to become clearly visible. 
In this case, directions of electrons in the disk reference frame are more preferably pointed along the jet. 
Therefore, also the $\gamma$-rays are produced more preferably along the jet direction. 
Due to this beaming effect, the $\gamma$-ray flux observed at higher inclination angles is much lower than that one observed along the jet. Moreover, energies of $\gamma$-ray photons moving along the jet (observed in the disk reference frame) are relativistically boosted. 
As a result, the cut-offs of cascade $\gamma$-ray spectra appear at higher energies.

\begin{figure}
\centering
\includegraphics[scale=0.53, trim= 0 12 0 0, clip]{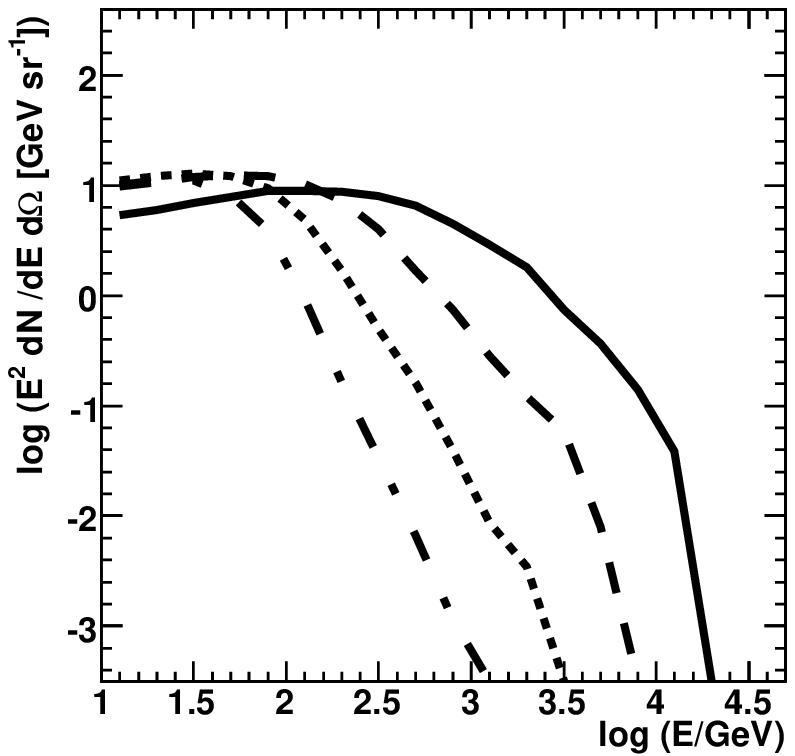}\figlab{$v=0.5c, H_0=1-30r_{in}$}{3.4cm}
\includegraphics[scale=0.53, trim= 20 12 0 0, clip]{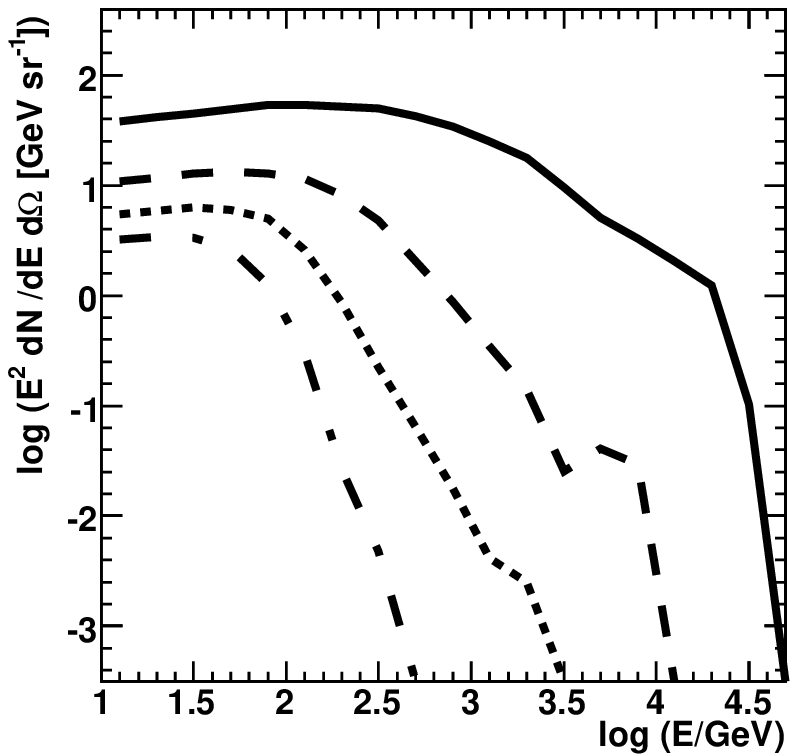}\figlab{$v=0.9c, H_0=1-30r_{in}$}{3.4cm}\\
\includegraphics[scale=0.53, trim= 0 12 0 0, clip]{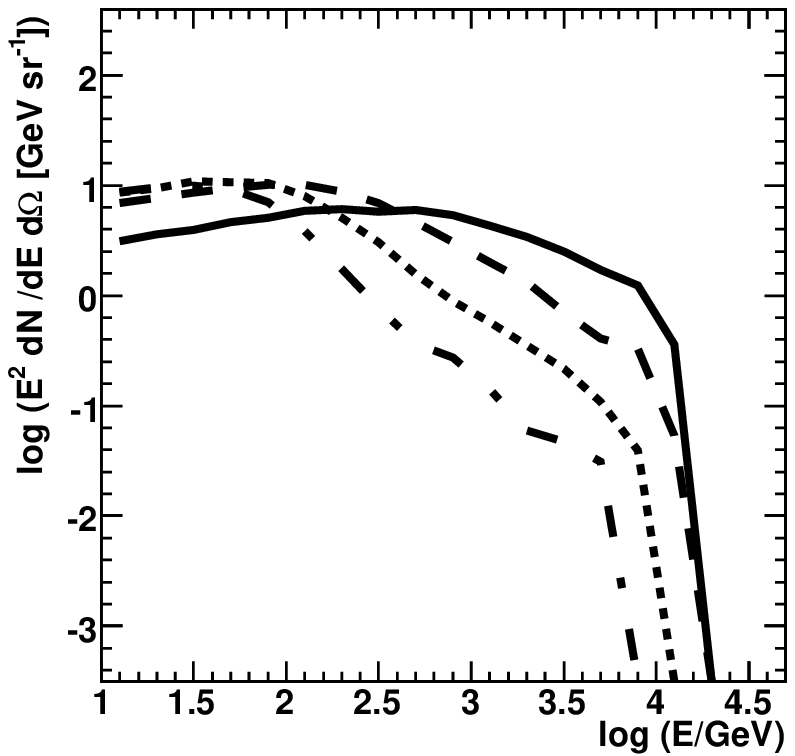}\figlab{$v=0.5c, H_0=1-100r_{in}$}{3.4cm}
\includegraphics[scale=0.53, trim= 20 12 0 0, clip]{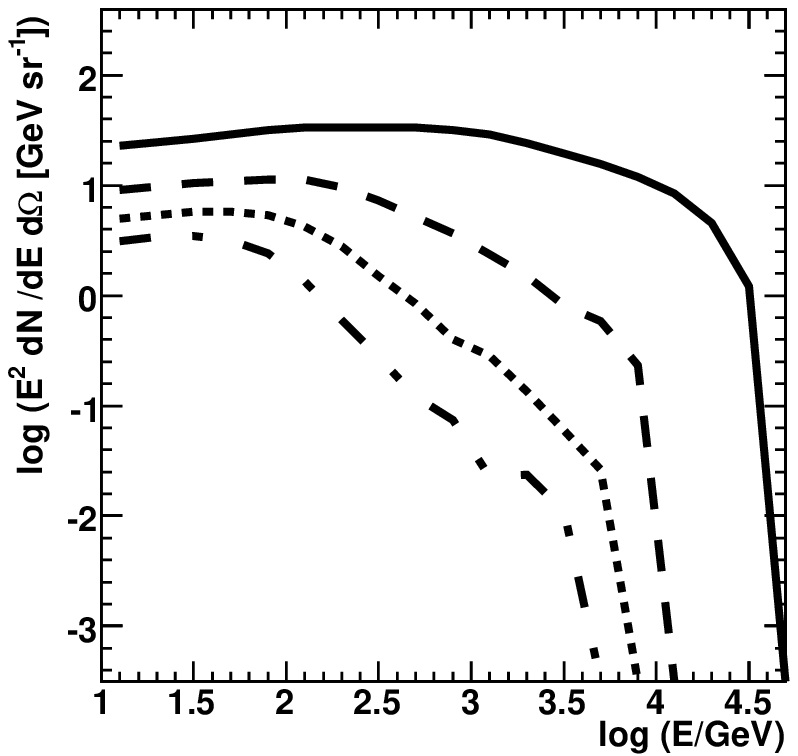}\figlab{$v=0.9c, H_0=1-100r_{in}$}{3.4cm}\\
\includegraphics[scale=0.53, trim= 0 0 0 0, clip]{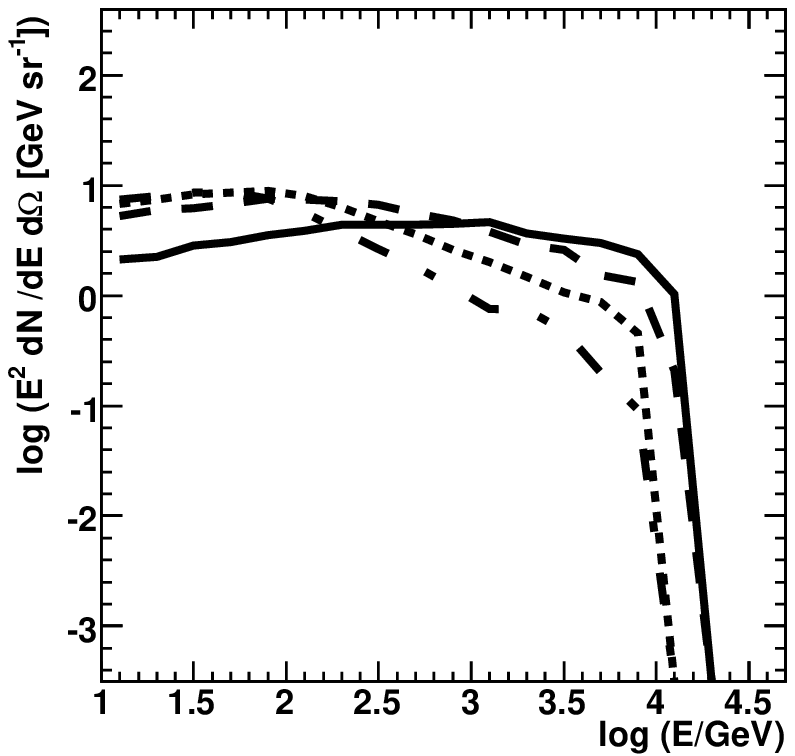}\figlab{$v=0.5c, H_0=1-300r_{in}$}{3.6cm}
\includegraphics[scale=0.53, trim= 20 0 0 0, clip]{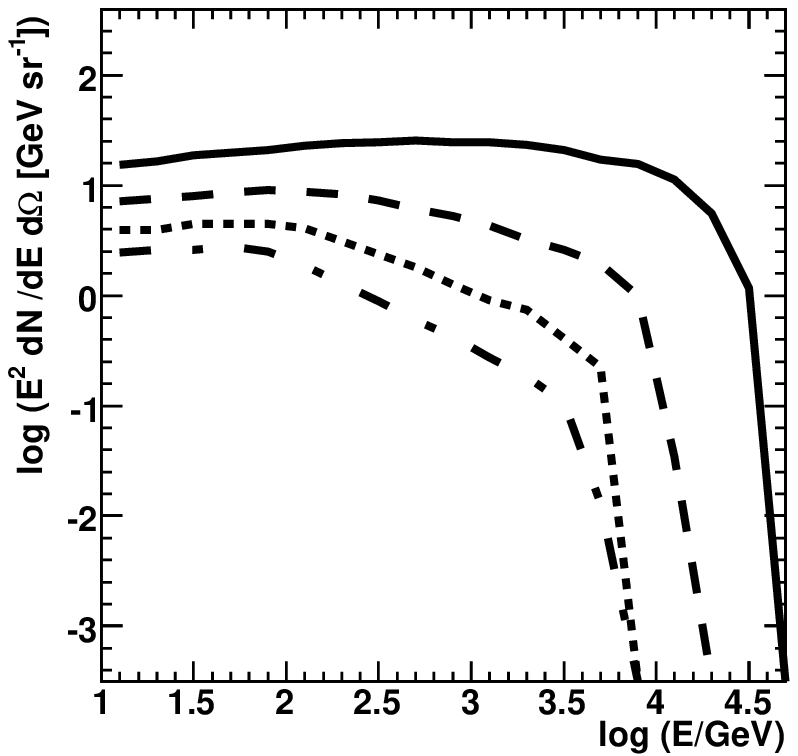}\figlab{$v=0.9c, H_0=1-300r_{in}$}{3.6cm}
\caption{
As in Fig.~\ref{fig1} but for the extended injection region of relativistic electrons along the jet. 
It is assumed that the injection process occurs between 1-30$r_{in}$ (top figures), 1-100$r_{in}$ (middle) and 1-300$r_{in}$ (bottom). 
The blob propagates with the velocity: 0.5c (left figures) and  0.9c (right figures). 
The $\gamma$-ray spectra are shown within the range of observation angles:
$\xi<40^\circ$ (solid), 
$40^\circ<\xi<60^\circ$ (dashed), 
$60^\circ<\xi<75^\circ$ (dotted), 
$75^\circ<\xi<90^\circ$ (dot-dashed).}
\label{fig2_model1}
\end{figure}

In Fig.~\ref{fig2_model1}, we show the $\gamma$-ray spectra expected at different observation angles assuming that primary electrons are injected into the blob when it moves through a range of distances above the accretion disk between $H_1$ and $H_2$. 
Their cooling process is followed up to distances at which they obtain energies below 10 GeV.
The $\gamma$-ray spectra are shown for $v=0.5$, and $0.9c$, $H_1=1r_{in}$ and $H_2=30, 100, 300r_{in}$. 
As expected from the analysis of the spectra calculated for fixed injection place (see 
Fig.~\ref{fig1}), the shape of the $\gamma$-ray spectra
strongly depends on the range of injection distances of electrons. 
The $\gamma$-ray spectra extend to higher energies for the injection of primary electrons farther from the accretion disk. They are also flatter.
Still, the GeV $\gamma$-ray fluxes produced at large inclination angles dominate for slow jets. 
However, the $\gamma$-ray spectra at the TeV energies become steeper. They also have a lower energy cut-off for large inclination angles. Slower jets produce lower $\gamma$-ray fluxes with steeper spectra at TeV energies due to a weaker relativistic beaming and a more significant KN effect. 
We conclude that the location of the break in the $\gamma$-ray spectrum and its spectral index above $\sim 100$ GeV is clearly related to the inclination angle of the jet. 
It may serve as a diagnostic tool for the inclination of the accretion disk with respect to the observer. On the other hand, the information on the inclination angle can allow to estimate a possible break in the expected $\gamma$-ray spectrum from a specific source.

\begin{figure}
\centering

\includegraphics[scale=0.53, trim= 0 12 0 0, clip] {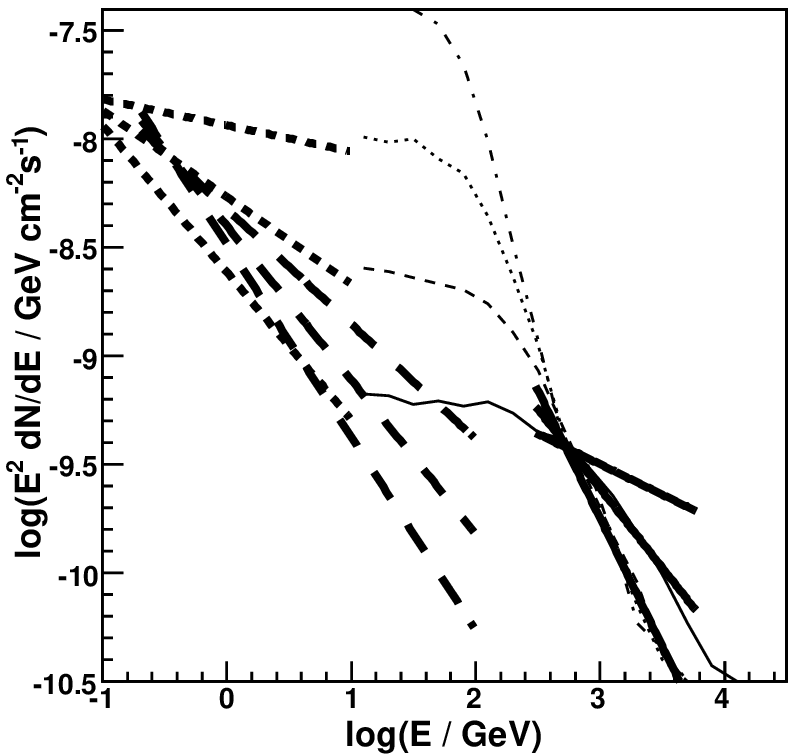}\figlabl{$v=0.5c$}{0.75cm}\figlabl{$H_0=1-100r_{in}$}{0.35cm}\figlab{$\frac{dN_e}{dE_e}\!\!\sim\!\!E_e^{-2.5}$}{3.4cm}
\includegraphics[scale=0.53, trim= 16 12 0 0, clip]{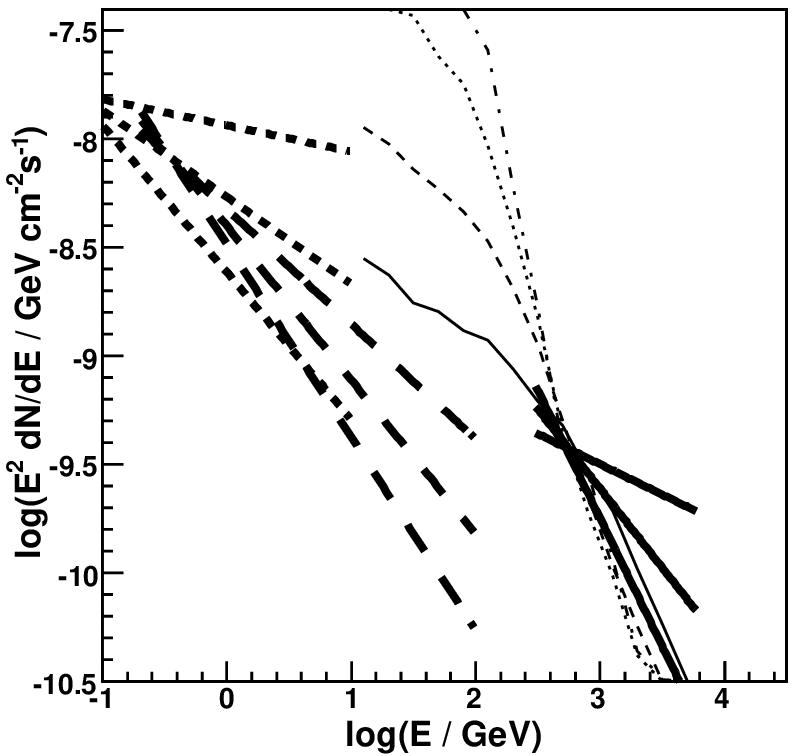}\figlabl{$v=0.5c$}{0.75cm}\figlabl{$H_0=1-100r_{in}$}{0.35cm}\figlab{$\frac{dN_e}{dE_e}\!\!\sim\!\!E_e^{-3}$}{3.4cm}\\
\includegraphics[scale=0.53, trim= 0 0 0 0, clip]  {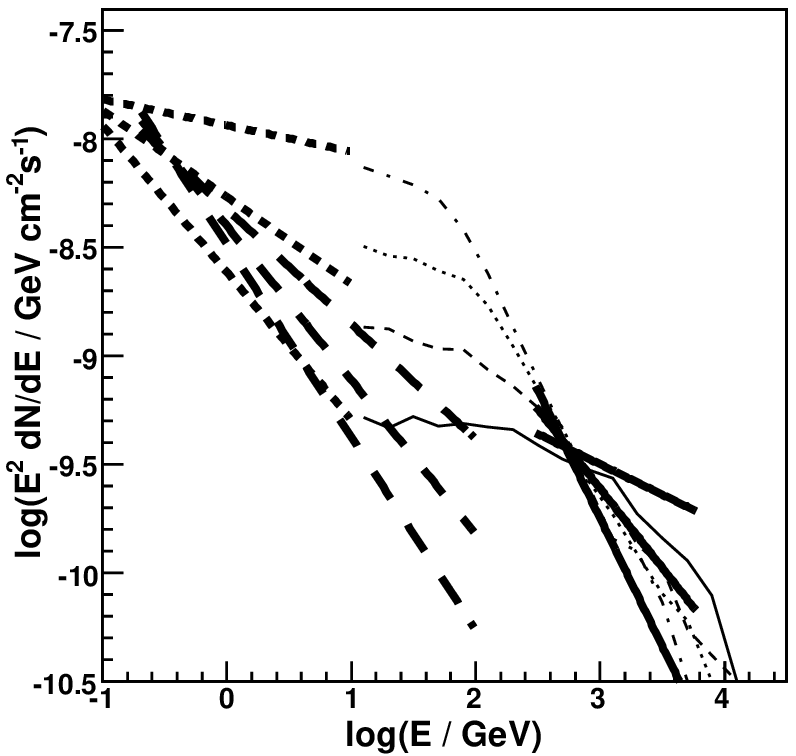}\figlabl{$v=0.5c$}{0.95cm}\figlabl{$H_0=1-300r_{in}$}{0.55cm}\figlab{$\frac{dN_e}{dE_e}\!\!\sim\!\!E_e^{-2.5}$}{3.6cm}
\includegraphics[scale=0.53, trim= 16 0 0 0, clip] {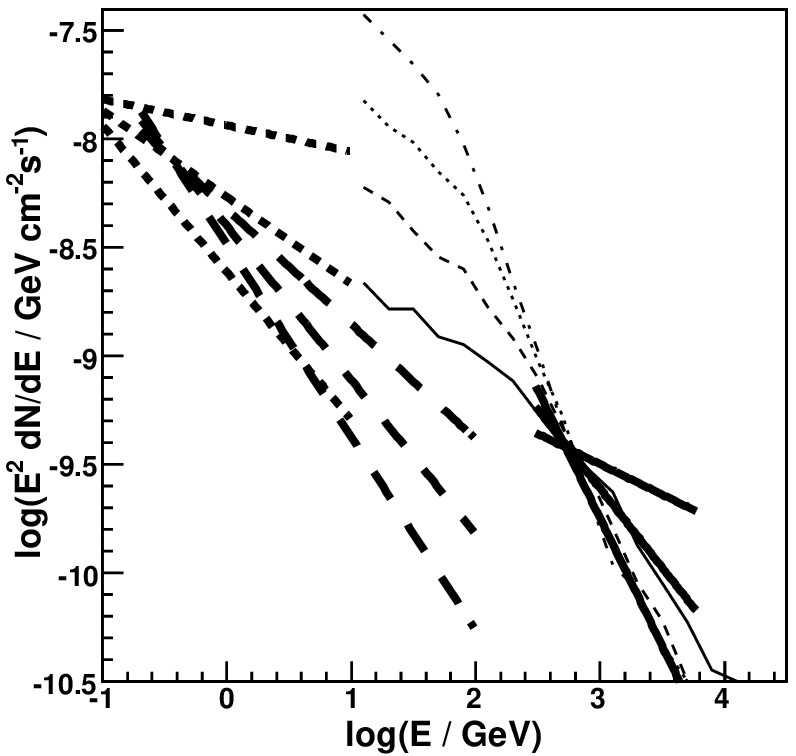}\figlabl{$v=0.5c$}{0.95cm}\figlabl{$H_0=1-300r_{in}$}{0.55cm}\figlab{$\frac{dN_e}{dE_e}\!\!\sim\!\!E_e^{-3}$}{3.6cm}
\caption{					    					      
Comparison of cascade gamma-ray spectra (thin curves) with the observations of Cen~A. The IC e$^{\pm}$ pair  cascade is initiated by electrons injected with a power law with an index $-2.5$ (left figures) and $-3$ (right) inside the blob propagating with a velocity: $0.5c$.
The injection process occurs between 1-100$r_{in}$ (top figures), and 1-300$r_{in}$ (bottom). 
The $\gamma$-ray spectra are shown within the range of observation angles:
$\xi<40^\circ$ (thin solid), 
$40^\circ<\xi<60^\circ$ (thin dashed), 
$60^\circ<\xi<75^\circ$ (thin dotted), 
$75^\circ<\xi<90^\circ$ (thin dot-dashed).
Thick curves (with error bow-ties) show the observations of Cen~A:
EGRET \citep[dotted, ][]{sr99}, {\it Fermi} \citep[dashed, ][]{ab10}, and H.E.S.S.~\citep[solid, ][]{ah09}
The cascade spectra are normalized to the total energy output in the 0.5 -- 5TeV range as measured by the H.E.S.S. telescopes.
}
\label{fig3_cena_comp}
\end{figure}

In Fig.~\ref{fig3_cena_comp} we compare the calculated cascade $\gamma$-ray spectra with the observations of Cen~A. The velocity of the blob is fixed on $v=0.5c$ (as in \citet{ha03}). 
Electron are injected with the spectral index $-2.5$, and $-3$. Reasonable agreement with the spectral shapes observed by the EGRET and the H.E.S.S. is obtained for low observation angles ($< 40^\circ$) and the spectral index of injected electrons equal to $-3$.
The $\gamma$-ray spectra produced in our model for the larger inclination angles of the accretion disk-jet
system are too steep to be able to explain these measurements. Note however, that the results reported by the EGRET, {\it Fermi} and H.E.S.S. are not simultaneous. So then, this conclusion should not be considered as a final.

\begin{figure}
\centering
\includegraphics[scale=0.53, trim= 0 12 0 0, clip]{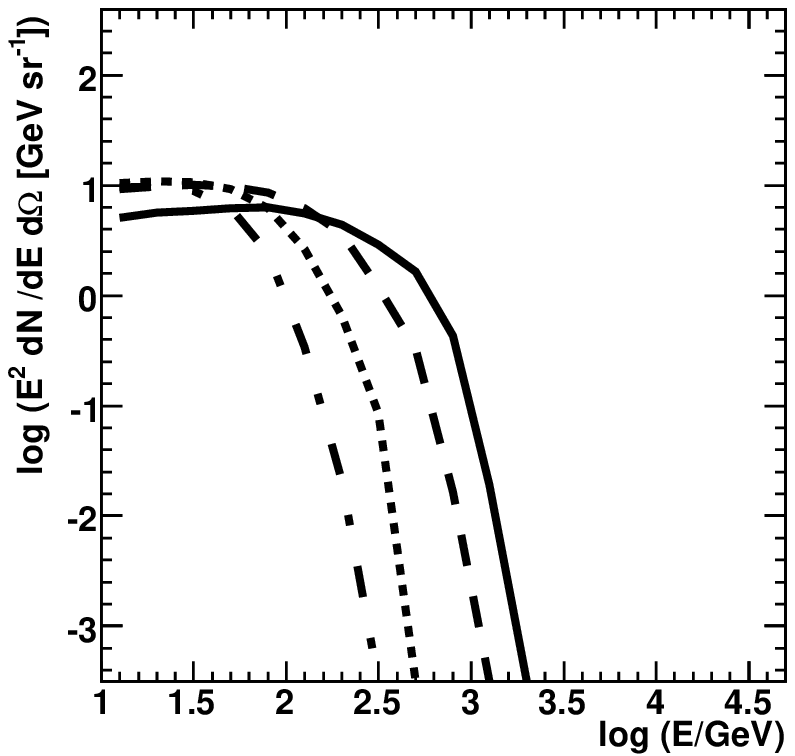}\figlab{$v=0.5c, H_0=1-30r_{in}$}{3.4cm}
\includegraphics[scale=0.53, trim= 20 12 0 0, clip]{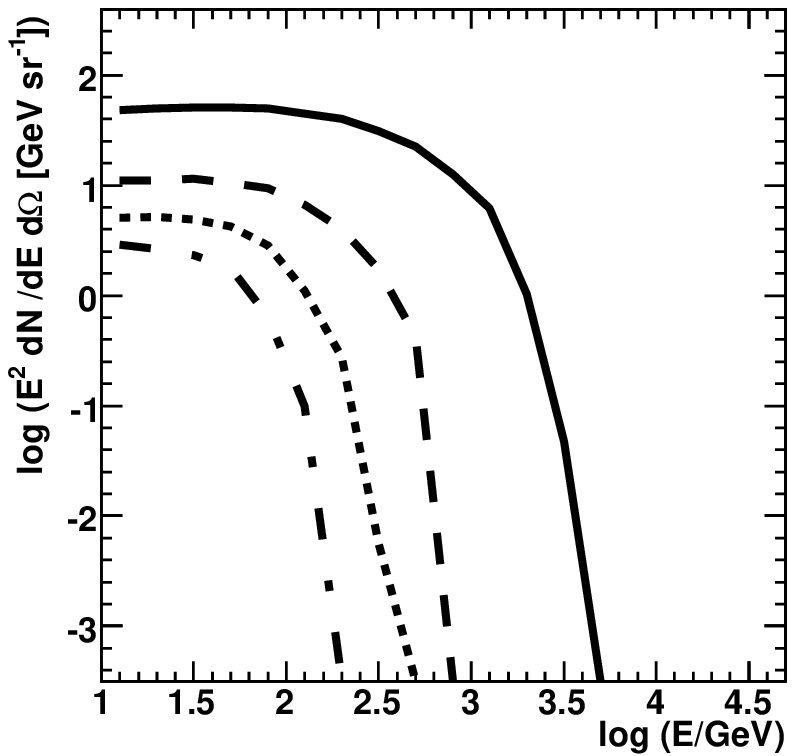}\figlab{$v=0.9c, H_0=1-30r_{in}$}{3.4cm}\\
\includegraphics[scale=0.53, trim= 0 12 0 0, clip]{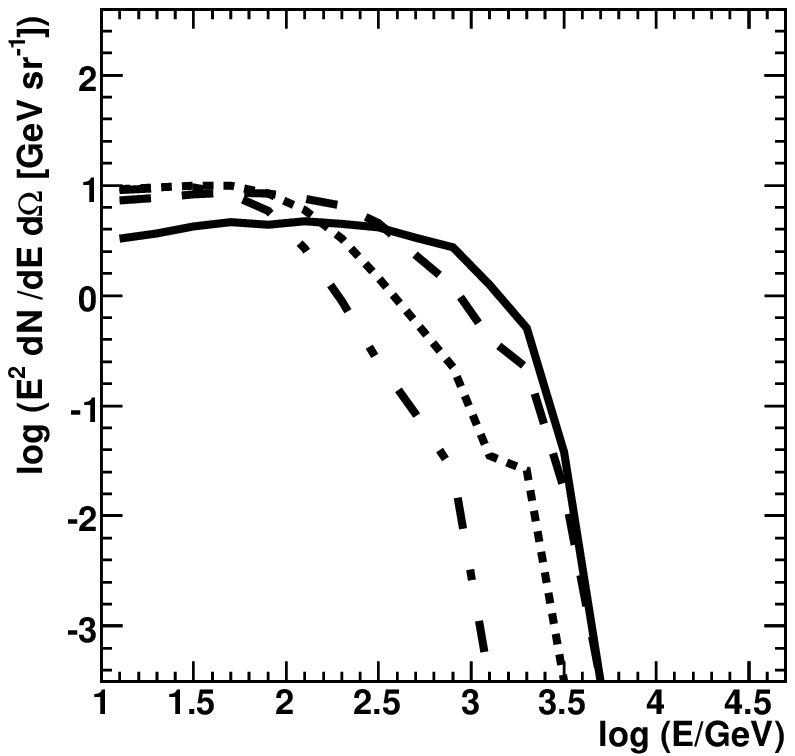}\figlab{$v=0.5c, H_0=1-100r_{in}$}{3.4cm}
\includegraphics[scale=0.53, trim= 20 12 0 0, clip]{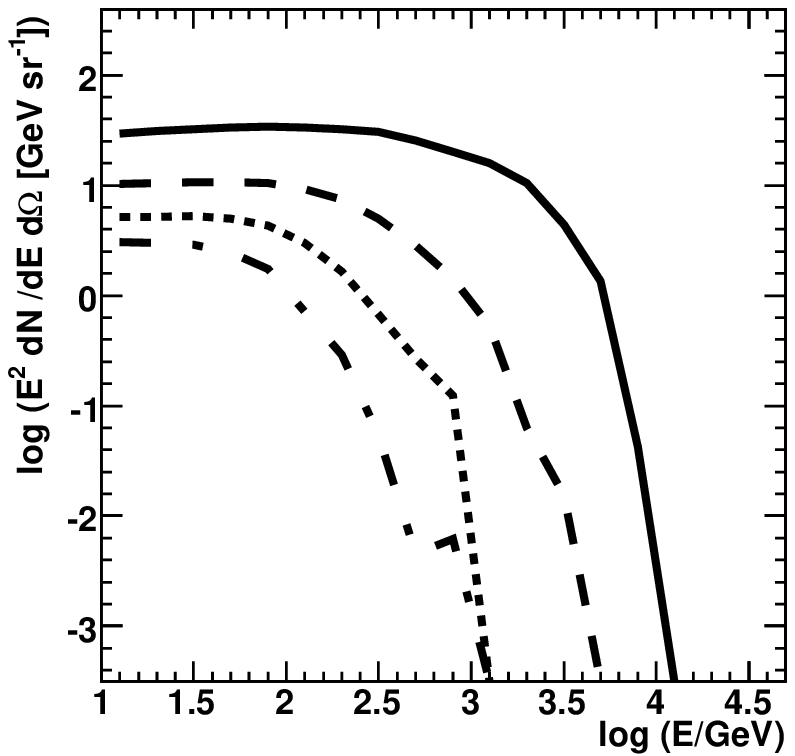}\figlab{$v=0.9c, H_0=1-100r_{in}$}{3.4cm}\\
\includegraphics[scale=0.53, trim= 0 0 0 0, clip]{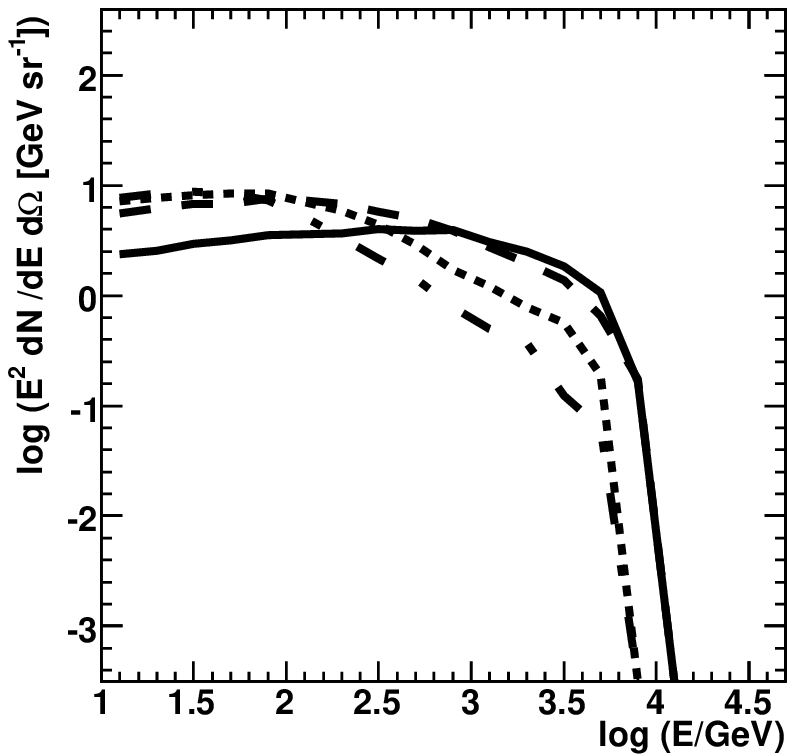}\figlab{$v=0.5c, H_0=1-300r_{in}$}{3.6cm}
\includegraphics[scale=0.53, trim= 20 0 0 0, clip]{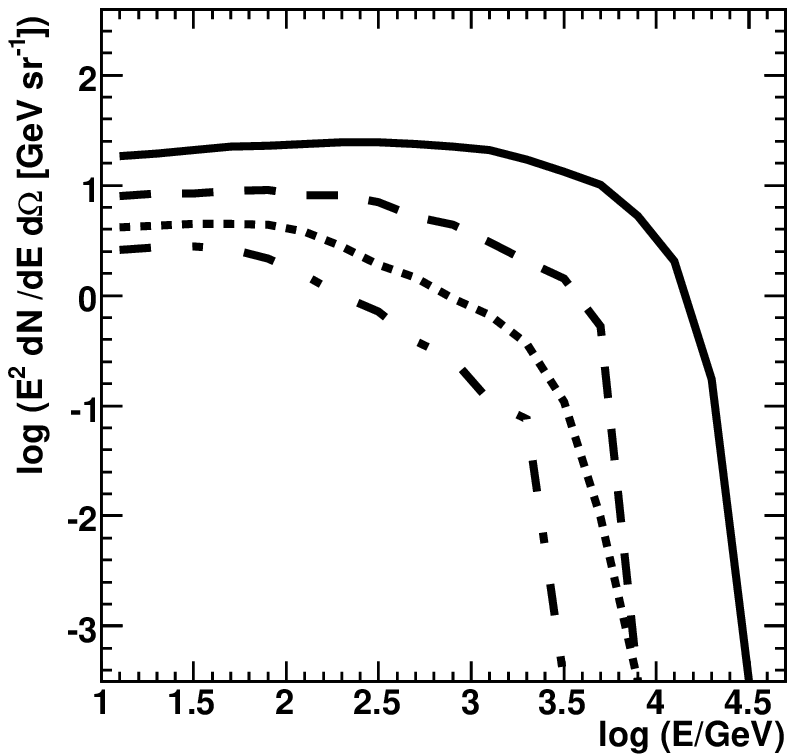}\figlab{$v=0.9c, H_0=1-300r_{in}$}{3.6cm}
\caption{
As in Fig.~\ref{fig2_model1} but for the model II of the maximum energy of electrons injected into the blob. 
These maximum energies depend on the distance along the jet according to Eq.~\ref{eq6}. 
The magnetic field inside the jet is defined by $\eta=1$.
}
\label{fig4_model2hi}
\end{figure}

The $\gamma$-ray spectra obtained for electrons with maximum energies limited by synchrotron energy losses (model II) are shown in Fig.~\ref{fig4_model2hi}. 
In order to consider more extreme example of model II, we apply the synchrotron energy losses calculated for the value of $\eta = 1$. For this value of $\eta$, the cut-off in the primary electron spectrum is at  lower energy than in the case of  model I. It results in slightly steeper $\gamma$-ray spectra with the cut-offs at lower energies. The effect becomes more pronounced in the case of a blob propagating closer to the black hole, due to stronger magnetic fields and therefore lower maximum energies of accelerated electrons.

\subsection{Blazars at small inclination angles (3C 279)}

As an example of a blazar observed at a small inclination angle we take the parameters of the famous blazar 3C 279. With a redshift of $z=0.538$, it  is a rather distant source.
It is one of the earliest discovered blazar in the $\gamma$-rays \citep{h92}.
A strong $\gamma$-ray flare with a very flat spectrum (differential spectral index $-1.89$) has been detected from 3C~279 already in June 1991.  
The flux increased during $\sim 1$ week period and declined during the following 2 days \citep{k93}.
Even stronger $\gamma$-ray flare has been observed by the EGRET in 1996 with a similar time structure and spectral index \citep{w98}. 
3C~279 has been positively detected by the EGRET telescope in all observation periods \citep{h01a}, showing a variety of flux and spectral index stages. 
The shortest variability time scales observed in these flares were below $\sim 1$ day \citep{w98, h01b}. 

More recent observations performed with the LAT telescope (high energy gamma-ray instrument on board of the {\it Fermi}) show a different behavior.
In December 2008 a longer flare, with a $\sim 15$ days period of constantly rising flux followed by a slower decay lasting for $\sim 30$ days, was reported during period of a high activity (the LAT monitored source list, \citet{cc08}). 
A similar, long flare has been observed by {\it Fermi}-LAT in July/August 2009 \citep{i09}.

3C~279 has been also recently detected by the MAGIC telescope \citep{al08b}, being at present the most distant sub-TeV $\gamma$-ray source.  Due to a severe absorption in the extragalactic background light (EBL) radiation, the observed spectrum in sub-TeV range is very soft, with a spectral index of $-4.11\pm0.68$.

For this source we apply the disk luminosity $L_{\rm disk}=2\times 10^{45}\mathrm{\rm erg\, s^{-1}}$ and the temperature $T_{\rm in}=2\times 10^4$ K at the disk inner radius \citep{pian99}.  The inner radius of the accretion disk in 3C~279 is estimated on $r_{\rm in}=4.2 \times 10^{15} \mathrm{\rm cm}$.

\subsubsection{Spectra at the source}

\begin{figure}
\centering
\includegraphics[scale=0.38, trim=  0 33 1 0, clip]{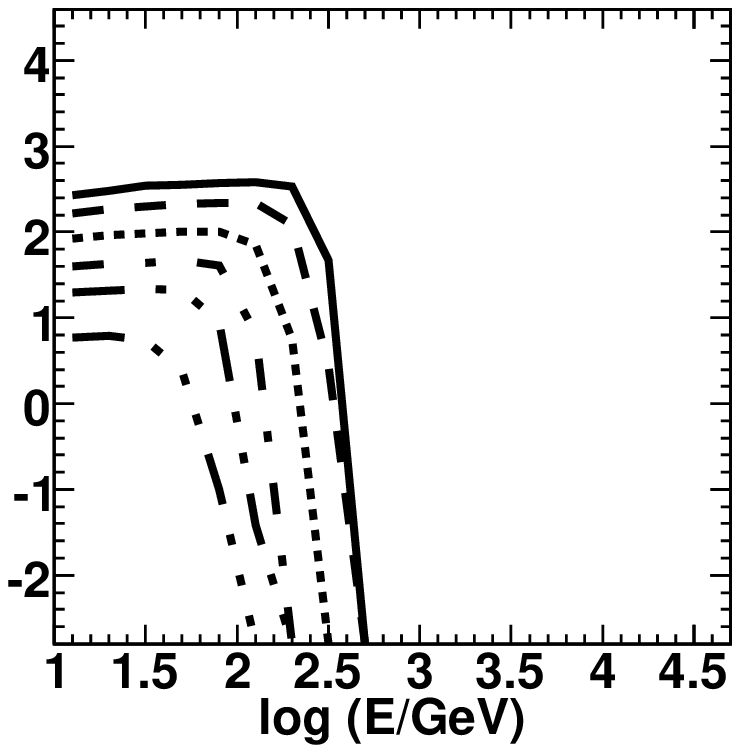}\!\!\figlab{$H_0=3r_{in}$}{2cm}\figlab{$v=0.9c$}{1.6cm}
\includegraphics[scale=0.38, trim= 31 33 1 0, clip]{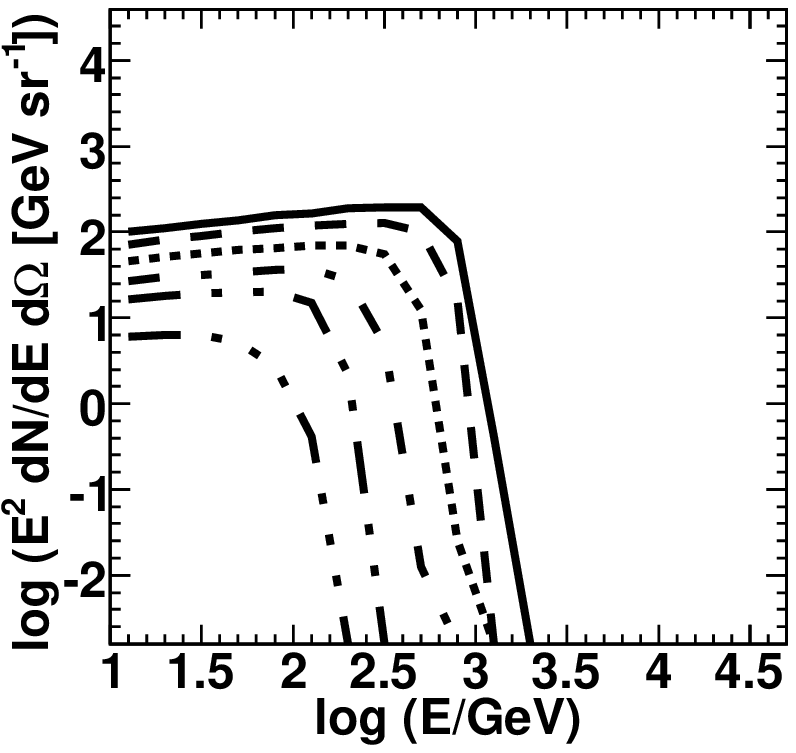}\!\!\figlab{$H_0=10r_{in}$}{2cm}
\includegraphics[scale=0.38, trim= 31 33 0 0, clip]{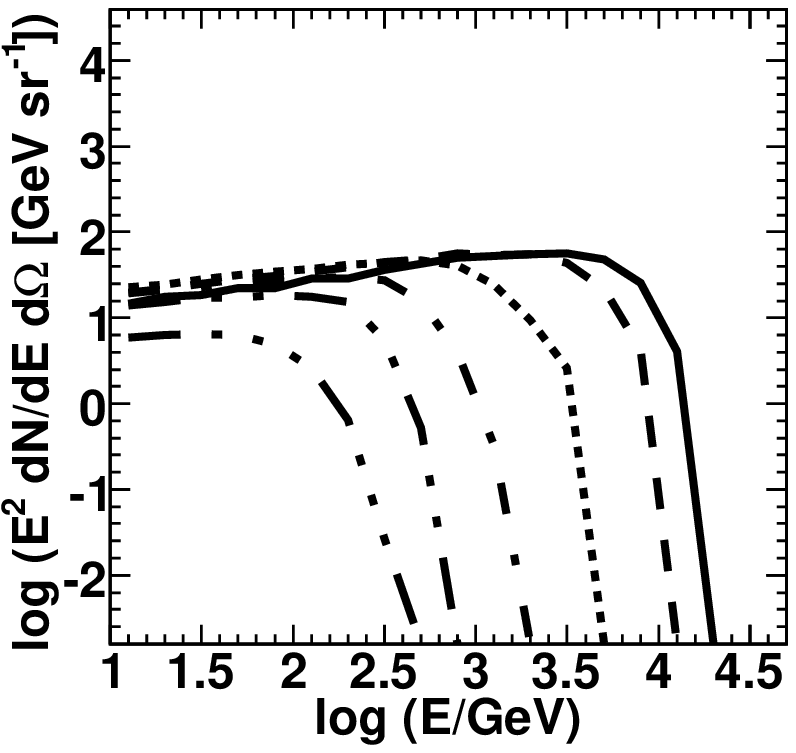}\figlab{$H_0=100r_{in}$}{2cm}\\   
						   
\includegraphics[scale=0.38, trim=  0 32 1 0, clip]{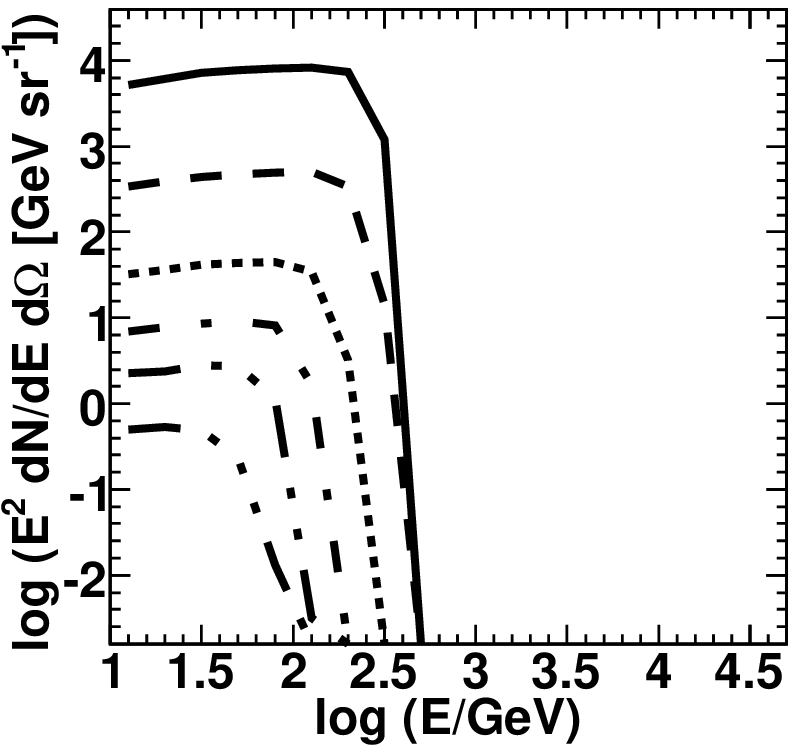}\!\!\figlab{$v=0.99c$}{2cm}
\includegraphics[scale=0.38, trim= 31 32 1 0, clip]{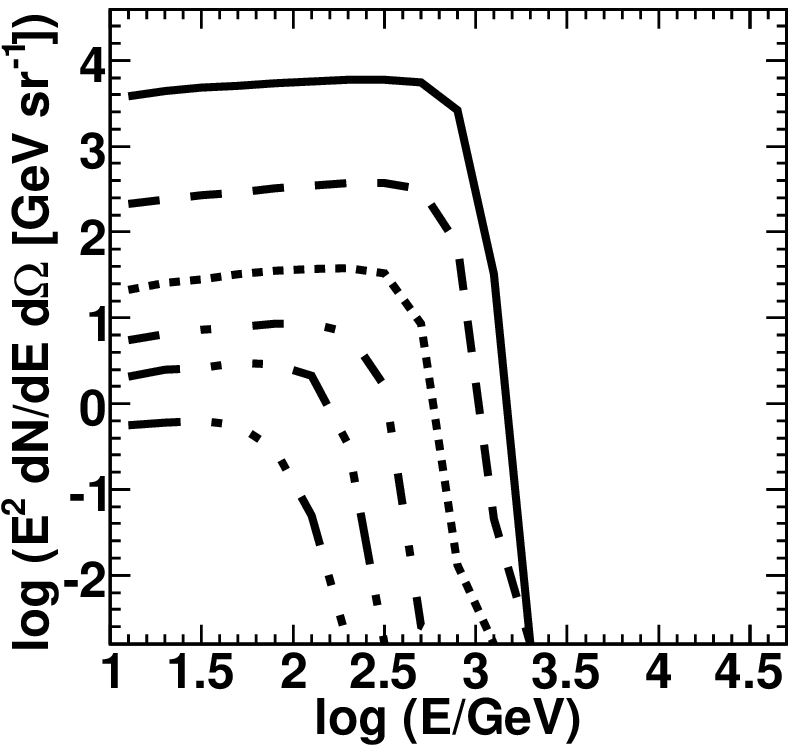}\!\!
\includegraphics[scale=0.38, trim= 31 32 0 0, clip]{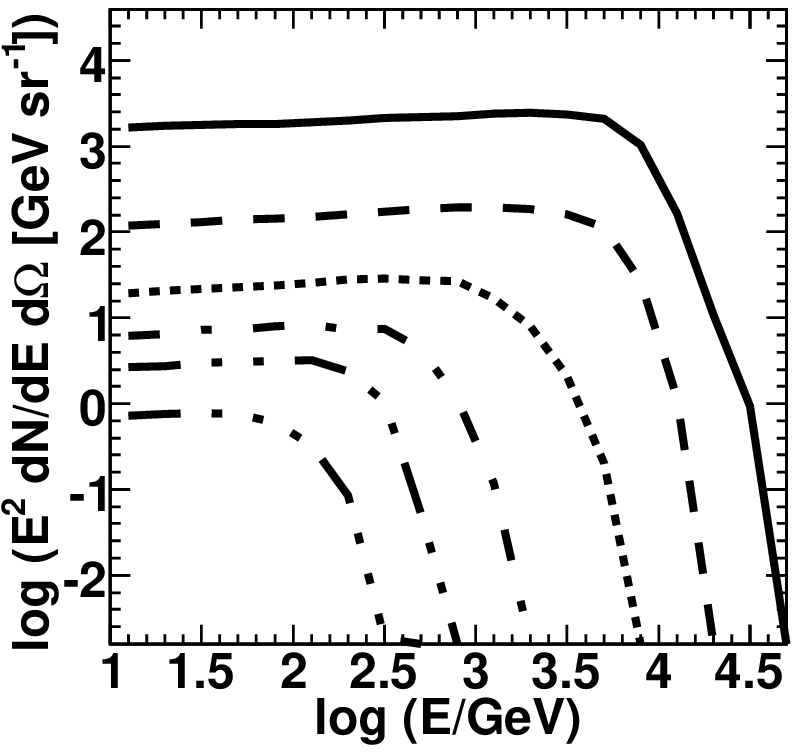} \\
						   
\includegraphics[scale=0.38, trim=  0  0 1 0, clip]{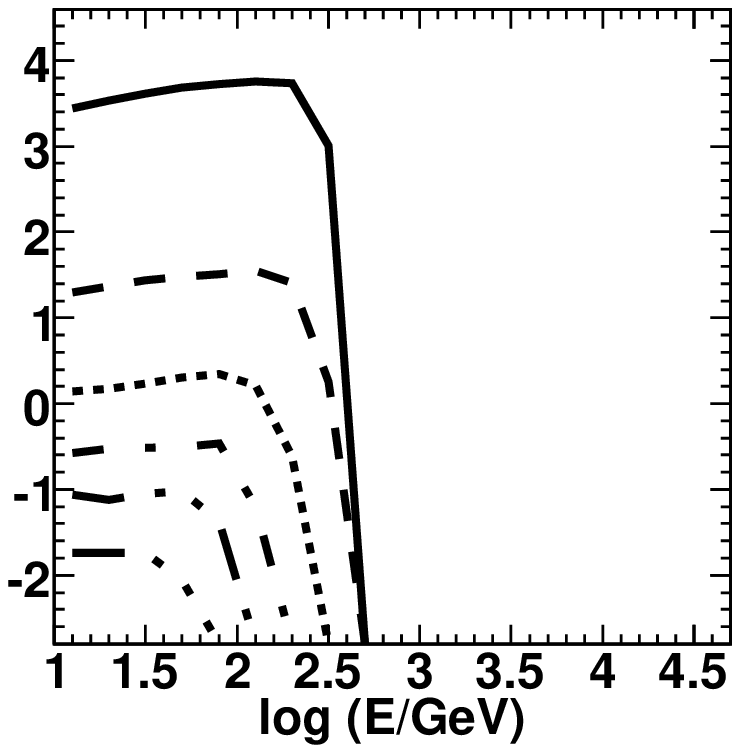}\!\!\figlab{$v=0.998c$}{2.5cm}
\includegraphics[scale=0.38, trim= 31  0 1 0, clip]{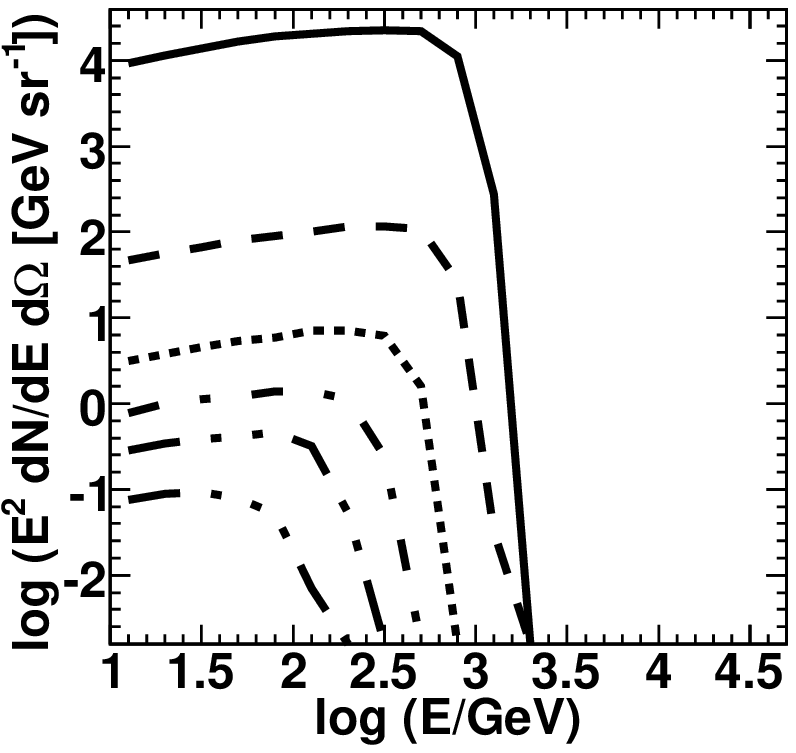}\!\!
\includegraphics[scale=0.38, trim= 31  0 0 0, clip]{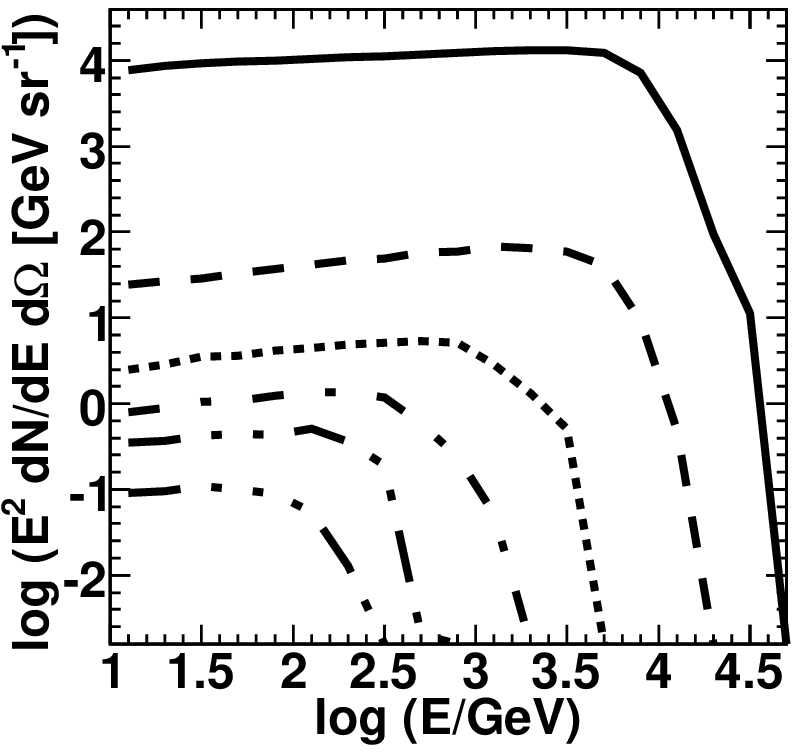} 
\caption{Gamma-ray spectra from IC $e^\pm$ pair cascade initiated by electrons injected isotropically in a blob with spectral index $-2$ between 10 GeV and 10 TeV (in the blob frame) for disk parameters as expected for 3C~279.
A blob propagates along the jet starting from a distance 3, 10, and 100$r_{\rm in}$ from its base (panels from left to right), with a velocity 0.9c (top panels), 0.99c (middle), and 0.998c (bottom).  
The spectra are shown for the range of inclination angles $\xi<10^\circ$ (solid), 
$10^\circ<\xi<20^\circ$ (dashed), 
$20^\circ<\xi<30^\circ$ (dotted), $30^\circ<\xi<40^\circ$ (dot-dashed),
$40^\circ<\xi<50^\circ$ (dot-dot-dashed), and $\xi>50^\circ$ (dot-dot-dot-dashed).}
\label{spectra_3c279_point}
\end{figure}

As in the case of Cen A, let's at first consider a simple scenario of the injection source of electrons at fixed distance from the base of the jet.
We assume the injection spectrum of electrons has a differential spectral index $-2$ between 10~GeV and 10~TeV. In Fig.~\ref{spectra_3c279_point} we present the angle dependent $\gamma$-ray spectra produced by electrons injected at a distance from the accretion disk $H_0=3$, 10, 100 $r_{in}$.
The calculations have been performed for various values of the blob velocities $v=0.9c$ (corresponding to the Lorentz factor $\gamma=2.3$), $0.99c$ ($\gamma=7.1$), and $0.998c$ ($\gamma=15.8$). These Lorentz factors are more characteristic for distant blazars observed at small inclination angles.
Those sources can be detected due to the strong relativistic beaming, which significantly enhances the observed flux. The shape of the cascade $\gamma$-ray spectra remains rather stable, with a cut-off energy determined by the 
injection height and the observation angle. We note, that the change of the spectral index due to the KN effect is not so clearly observed as in the case of Cen A. It is only visible in the $\gamma$-ray spectrum produced by electrons injected at the distance $100r_{\rm in}$ from the disk. This much smaller KN effect is due to the lower energies of the disk photons seen by electrons in the reference frame of the blob moving with significantly larger Lorentz factors than in the case of Cen~A.

\begin{figure}
\includegraphics[scale=0.53, trim= 0 12 0 0, clip]   {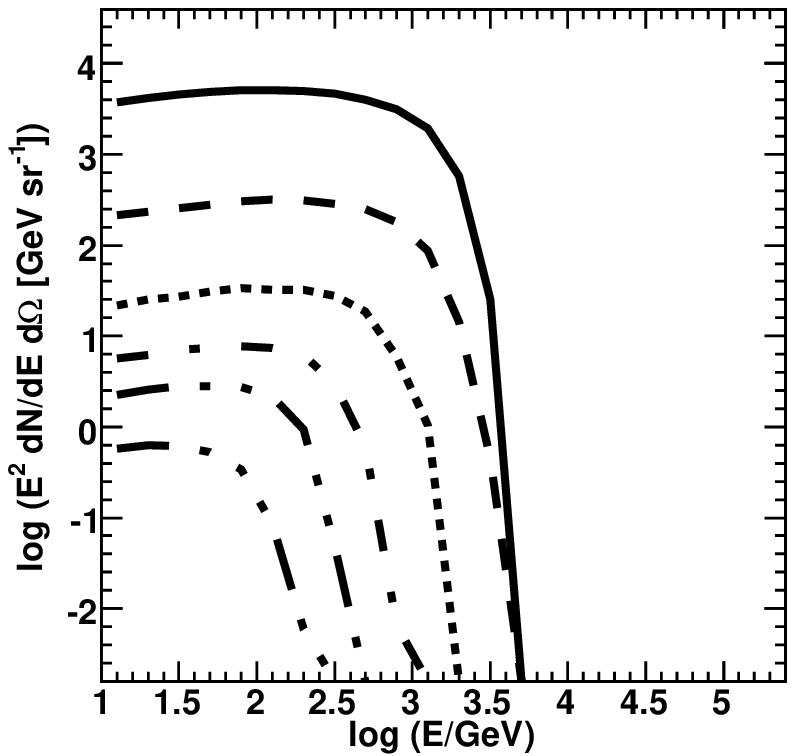}\figlab{$H_0=1-30r_{in}$}{3.4cm}\figlab{$v=0.99c$}{3.1cm}
\includegraphics[scale=0.53, trim= 20 12 0 0, clip] {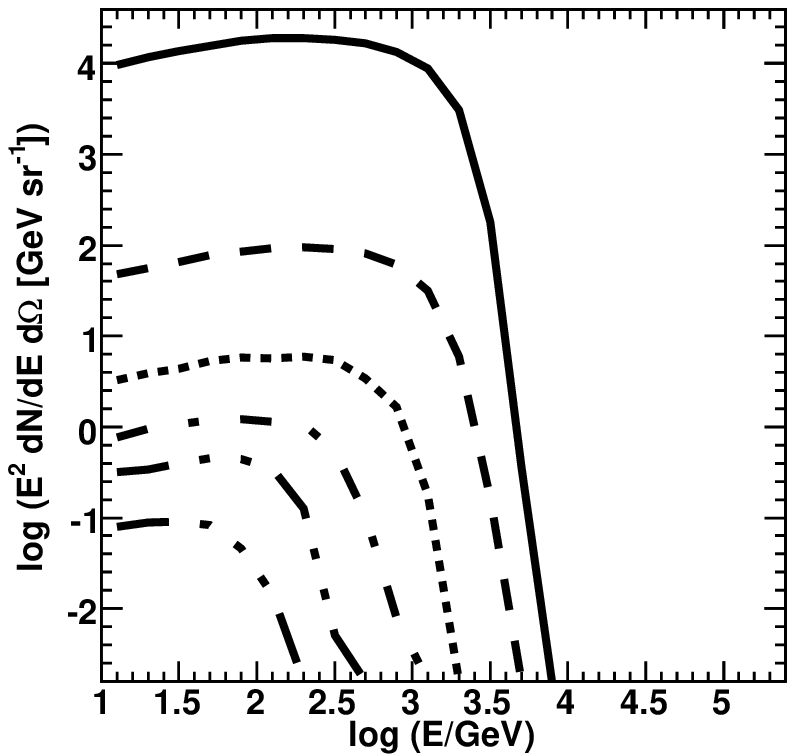}\figlab{$1-30r_{in}$}{3.4cm}\figlab{$0.998c$}{3.1cm}\\ 
\includegraphics[scale=0.53, trim= 0 12 0 0, clip]  {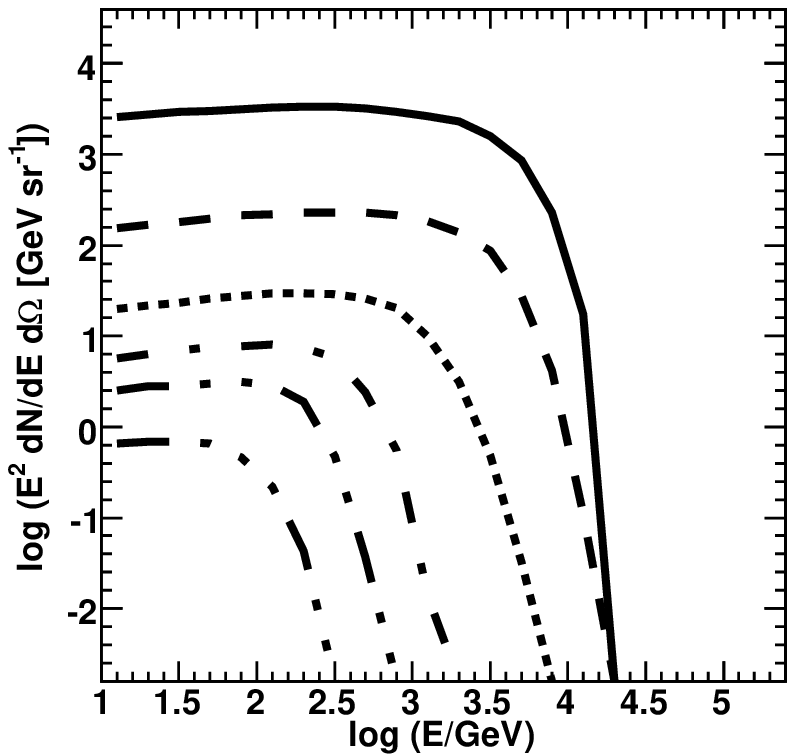}\figlab{$1-100r_{in}$}{3.4cm}\figlab{$0.99c$}{3.1cm}
\includegraphics[scale=0.53, trim= 20 12 0 0, clip]{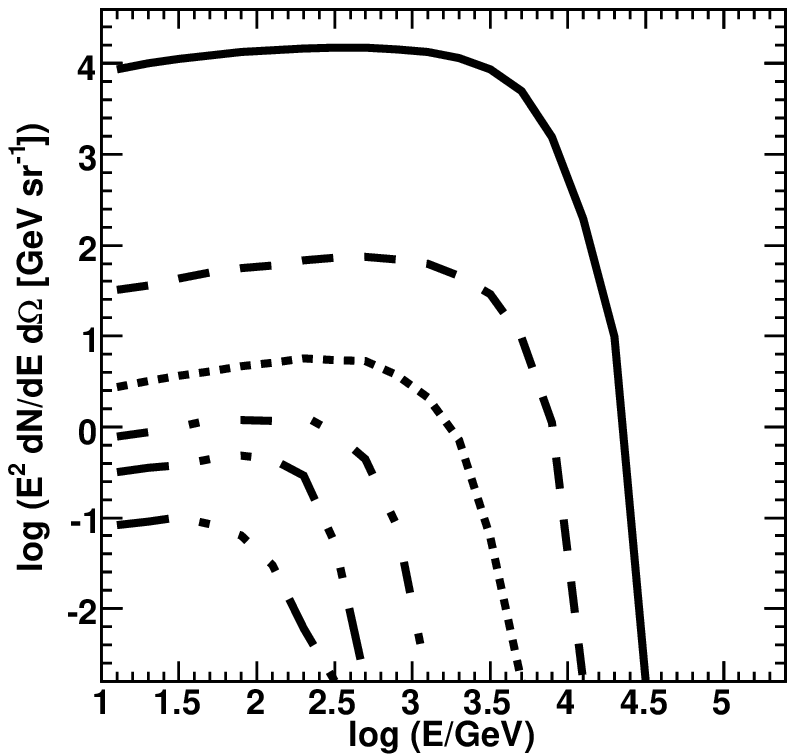}\figlab{$1-100r_{in}$}{3.4cm}\figlab{$0.998c$}{3.1cm}\\
\includegraphics[scale=0.53, trim= 0 0 0 0, clip]   {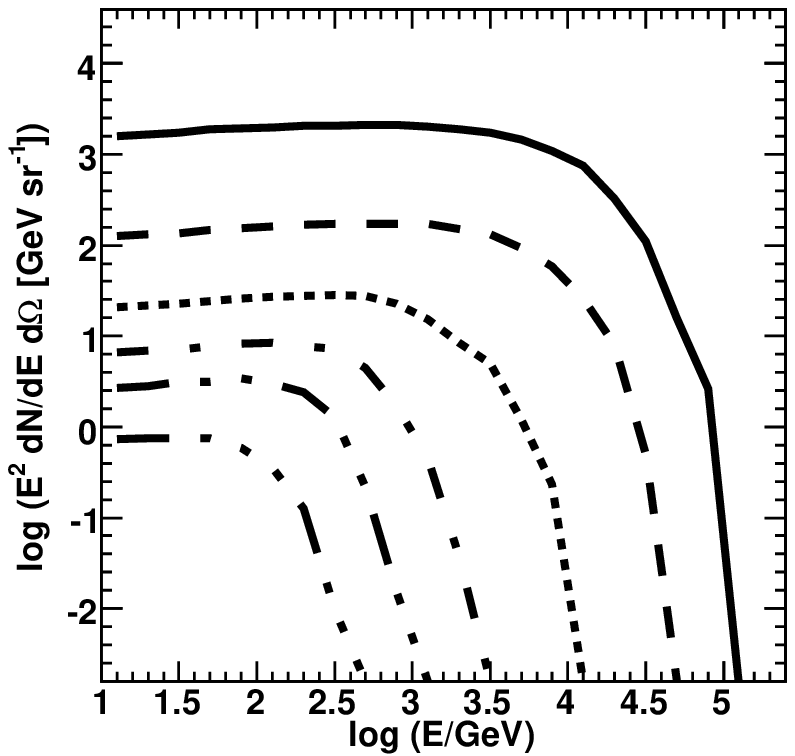}\figlab{$1-300r_{in}$}{3.75cm}\figlab{$0.99c$}{3.45cm}	 
\includegraphics[scale=0.53, trim= 20 0 0 0, clip] {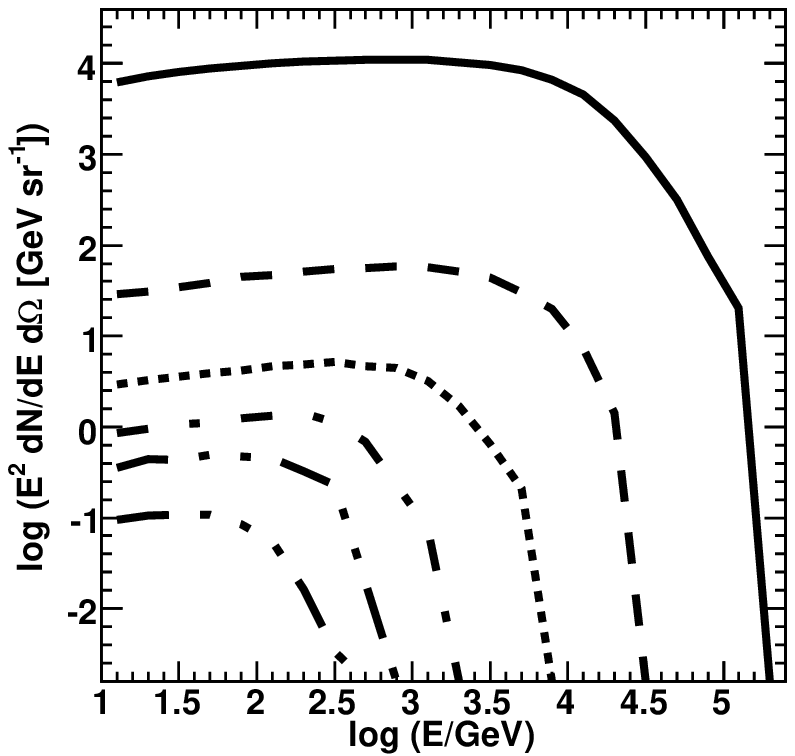}\figlab{$1-300r_{in}$}{3.75cm}\figlab{$0.998c$}{3.45cm}
\caption{
As in Fig.~\ref{spectra_3c279_point} but for the extended injection region of relativistic electrons along the jet. The injection of electrons occurs between 1-30$r_{in}$ (top figures), 
1-100$r_{in}$ (middle), and 1-300$r_{in}$ (bottom).
The blob propagates with a velocity 0.99c ($\gamma=7.1$, left figures) and  
0.998c ($\gamma=15.8$, right figures).}
\label{spectra_3c279_cont}
\end{figure}

We also calculate the $\gamma$-ray spectra in the case of a continuous injection of electrons into the blob moving through a range of distances within the jet.
The $\gamma$-ray spectra escaping in this case at different range of angles are shown in Fig.~\ref{spectra_3c279_cont}. The $\gamma$-ray spectra extend to lower  energies for larger inclination angles due to stronger absorption effects and also kinematic effects caused by lower Doppler factors ($D=[\gamma_{\rm b}(1-\beta_{\rm b}\cos\xi)]^{-1}$).  
For fast blobs, the $\gamma$-ray emission is strongly limited to the small inclination angles ($\xi<10^\circ$). The cut-offs in the $\gamma$-ray spectra also clearly depend on the location of the injection region of electrons within the jet.
However, the spectral index in the lower energy part of the $\gamma$-ray spectra (the GeV energy range) does not depend strongly on the observation angle.

\subsubsection{Spectra after propagation in the EBL}

\begin{figure}
\includegraphics[scale=0.53, trim= 0 15 0 0, clip]   {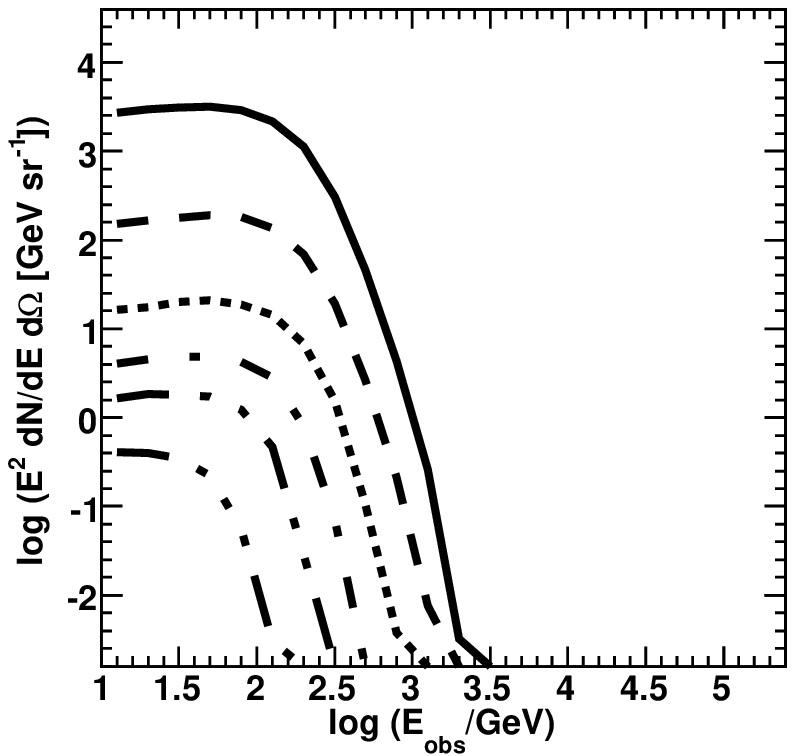}\figlab{$H_0=1-30r_{in}$}{3.4cm}\figlab{$v=0.99c$}{3.1cm}
\includegraphics[scale=0.53, trim= 20 15 0 0, clip] {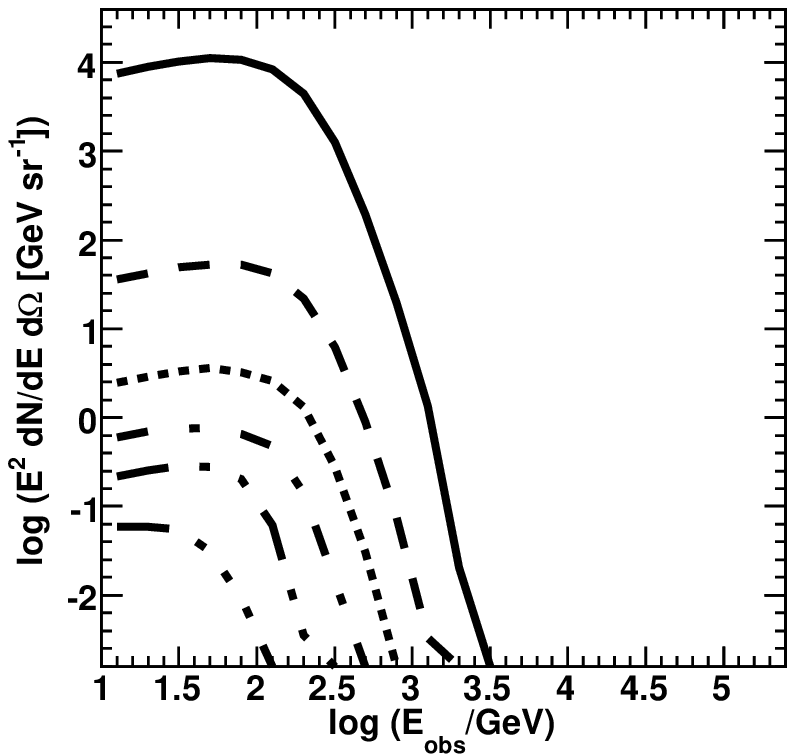}\figlab{$H_0=1-30r_{in}$}{3.4cm}\figlab{$v=0.998c$}{3.1cm}\\   
\includegraphics[scale=0.53, trim= 0 15 0 0, clip]  {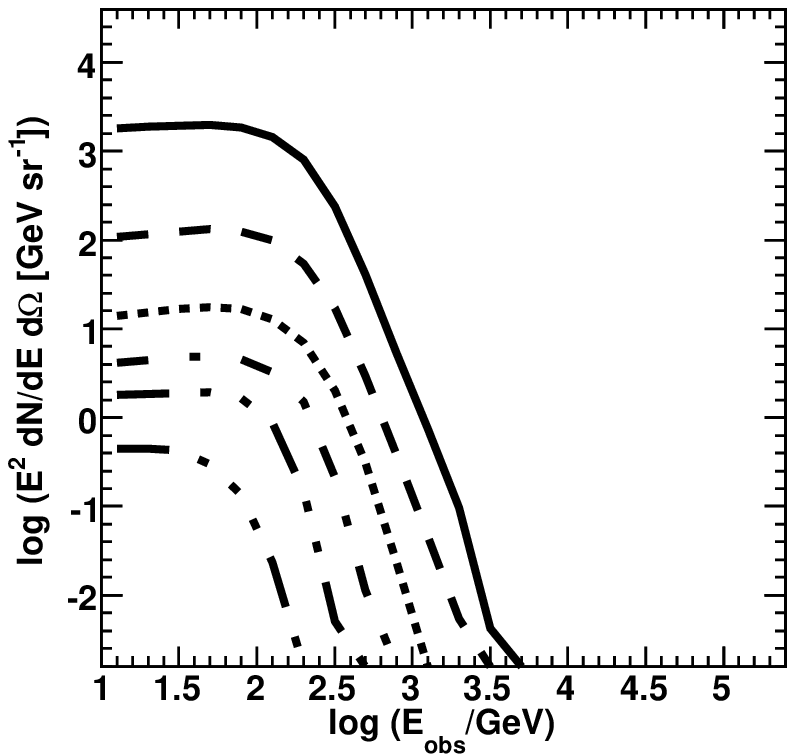}\figlab{$H_0=1-100r_{in}$}{3.4cm}\figlab{$v=0.99c$}{3.1cm}     	 
\includegraphics[scale=0.53, trim= 20 15 0 0, clip]{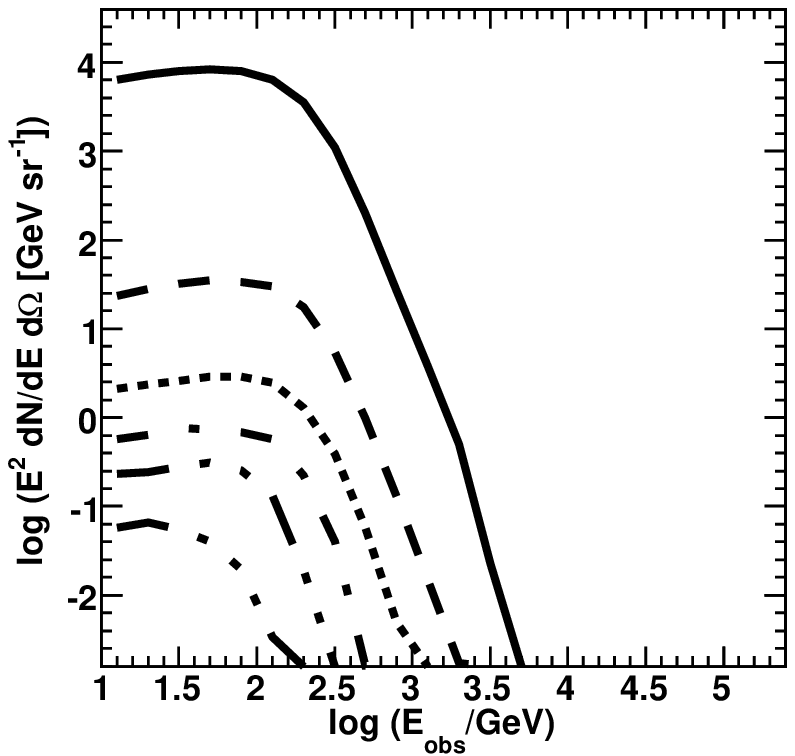}\figlab{$H_0=1-100r_{in}$}{3.4cm}\figlab{$v=0.998c$}{3.1cm}\\  
\includegraphics[scale=0.53, trim= 0 0 0 0, clip]   {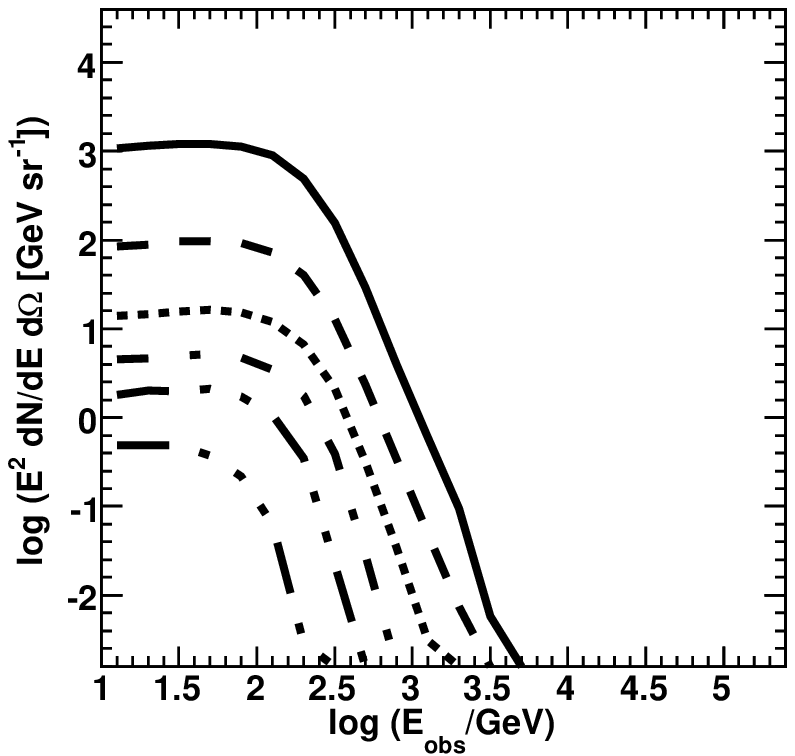}\figlab{$H_0=1-300r_{in}$}{3.65cm}\figlab{$v=0.99c$}{3.35cm}   	 
\includegraphics[scale=0.53, trim= 20 0 0 0, clip] {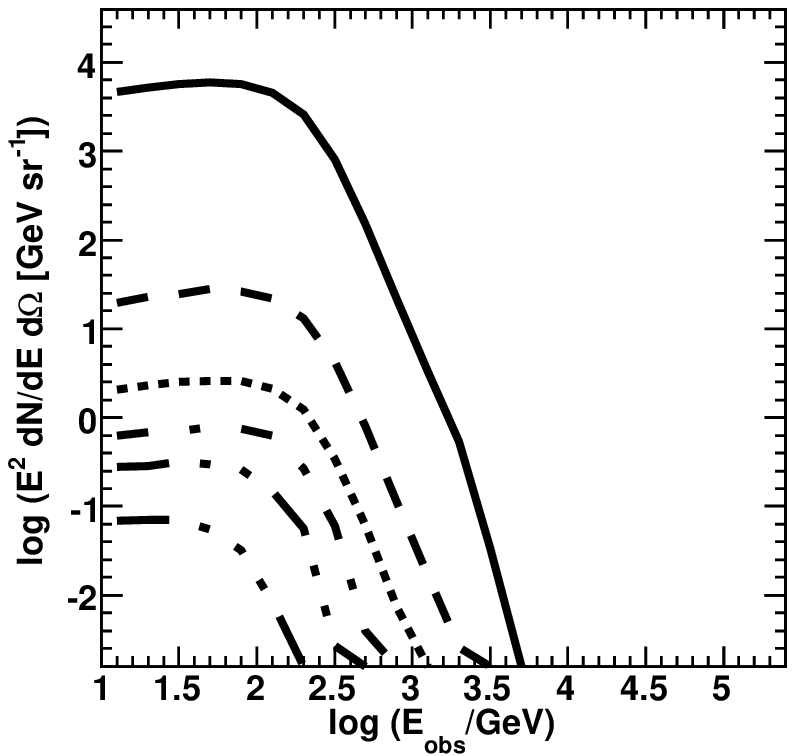}\figlab{$H_0=1-300r_{in}$}{3.65cm}\figlab{$v=0.998c$}{3.35cm}  
\caption{Gamma-ray spectra from 3C 279 for the parameters as in Fig.~\ref{spectra_3c279_cont}, 
but after the propagation in extragalactic background light described by the model \protect\citet{fr08}.}
\label{spectra_3c279_ebl}
\end{figure}

Since 3C~279 is a distant source, the observed $\gamma$-ray spectra differ from the spectra at the source presented in Fig.~\ref{spectra_3c279_cont}. 
Due to the cosmological effects, energies of $\gamma$-ray photons are reduced. 
Moreover, the part of the $\gamma$-ray spectrum at a few hundred GeV is strongly absorbed in the EBL. 
In order to obtain spectra and light curves as seen by a distant observer one has to correct for these propagation effects on the way from the source to the observer. 
The corrections depend on the redshift $z$ of the source. The observed energy $E_{obs}$ of the $\gamma$-ray photon is reduced according to $E_{\rm obs}=E_{\rm em}/(1+z)$,
where $E_{\rm em}$ is the energy of the emitted $\gamma$-ray in the disk frame.  
The observed $\gamma$-ray spectra and the light curves are attenuated by a survival probability $e^{-\tau_{EBL}}$, where $\tau_{EBL}$ is the energy dependent optical depth in the EBL radiation. 
The cosmological effects also influence the arrival time of $\gamma$-ray flare
according to $\Delta t_{\rm obs}=\Delta t_{\rm em}(1+z)$, where
$\Delta_{\rm em}$ is the time interval measured in the disk frame and $\Delta t_{\rm obs}$ is the time interval measured in the observer frame. 
We took all those effects into account when obtaining the $\gamma$-ray spectra in 
Fig.~\ref{spectra_3c279_ebl}.

The $\gamma$-ray optical depths in the EBL radiation field, $\tau_{EBL}$, are calculated according to \cite{fr08} model. Due to a large redshift of 3C~279, the $\gamma$-ray spectra above $\sim 200$ GeV are strongly absorbed resulting in a strong steepening of the spectrum. Therefore, the shape of the $\gamma$-ray spectra above $\sim$100 GeV are quite independent on the blob Lorentz factor and the extend of the injection region
within the jet. This creates problems for extracting physical parameters of the $\gamma$-ray emission region from the comparison of the observed and calculated spectra.

\begin{figure}
\includegraphics[scale=0.53, trim=  0 0 0 0, clip] {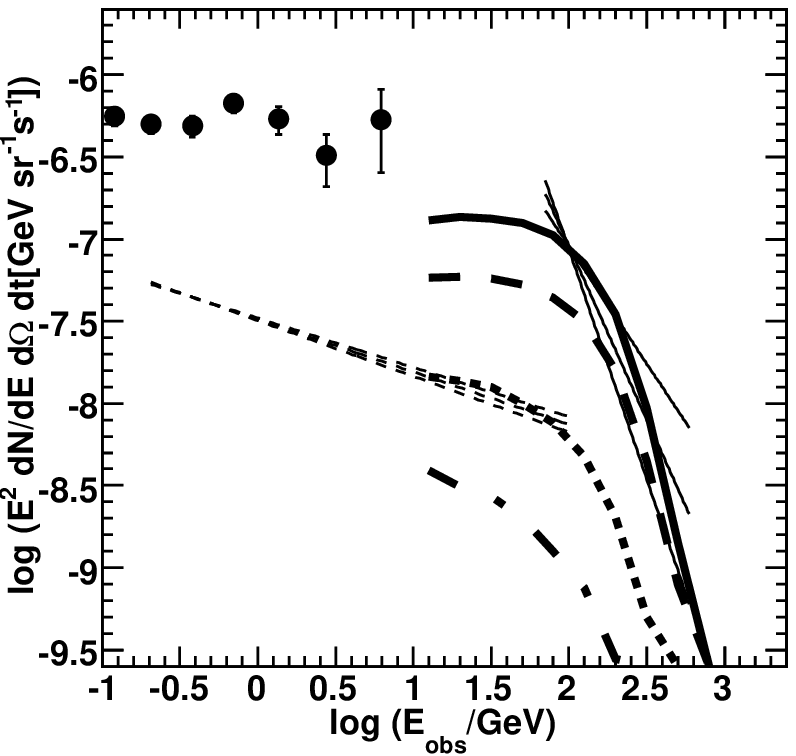}\figlabl{$v=0.998c$}{1.05cm}\figlabl{$H_0=1-300r_{in}$}{0.65cm}\figlab{$\frac{dN_e}{dE_e}\!\!\sim\!\!E_e^{-2.5}$}{3.6cm}
\includegraphics[scale=0.53, trim= 12 0 0 0, clip] {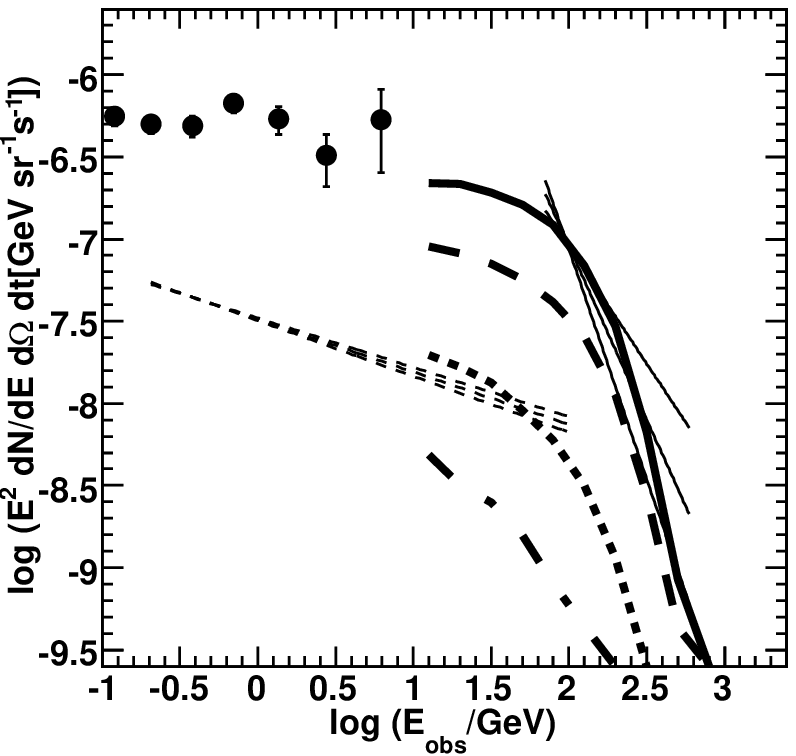}\figlabl{$v=0.998c$}{1.05cm}\figlabl{$H_0=1-300r_{in}$}{0.65cm}\figlab{$\frac{dN_e}{dE_e}\!\!\sim\!\!E_e^{-3}$}{3.6cm}
\caption{
The IC $e^\pm$ pair cascade $\gamma$-ray spectra at specific time intervals calculated for 3C~279 after the propagation in the intergalactic radiation, compared with various observations. 
Primary electrons are injected with the power law spectrum and spectral index $-2.5$ (left figure) or $-3$ (right figure) between 10~GeV and 10~TeV in the blob frame of reference.
The injection occurs in the range of distances from the base of the jet: $H_0=1$--300$r_{in}$. $\gamma$-ray  spectra are shown for the time intervals:
$t=0$--23 days (solid), 23--45 days (dashed), 45--68 days (dotted), 68--90 days (dotted).
Blob propagates with a velocity $v=0.998c$ ($\gamma=15.8$).
Thin lines with error bow-ties show observational results in a high state of 3C~279 collected by 
the MAGIC \citep[solid, ][]{al08b}, the EGRET \citep[black points, ][]{h01a}, and the mean spectrum from 11 month observations by the {\it Fermi}-LAT \citep[dashed, ][]{ab10}.
The solid curve is normalized to the $\gamma$-ray spectrum observed by the MAGIC telescope.
}
\label{spectra_time_3c279_obs_comp}
\end{figure}

The IC $e^\pm$ pair cascade $\gamma$-ray spectra, after absorption effects in the EBL, are compared with the observations of the two high states of the source seen by the EGRET and the MAGIC, and with mean flux level spectrum seen by the {\it Fermi}-LAT (see Fig.~\ref{spectra_time_3c279_obs_comp}). 
The spectrum calculated for the beginning of the flare ($t=0-23$~days) and  for the range of injection distances $H_0=1$--300$r_{in}$ can  fit the EGRET and the MAGIC measurements provided that electron are injected with a power law spectrum and spectral index in the range from $-2.5$ to $-3$. 
Note that the model predicts how the level of emission should evaluate in time.
After $\sim$(60-90) days, it should drop to the mean emission level determined by the {\it Fermi}-LAT observations.
Thus a single large flare every several months can explain the mean emission level.

\section{Time structure of the gamma-ray emission}

As shown in the previous section the disk radiation field seen by electrons in the blob 
moving along the jet changes significantly. As a result, $\gamma$-rays are produced at different places above the disk and at different moments. Therefore, the $\gamma$-ray spectra arriving to the distant observer show significant dependence on the observation time. 
Here we calculate the $\gamma$-ray light curves expected from such IC $e^\pm$ pair cascade for specific parameters of the model and as a function of the location of the observer in respect to the disk axis.
$\gamma$-rays are produced in a cascade initiated by electrons which are injected with a power law spectrum continuously along the jet and cool in the IC process (see for details Section~\ref{sec_3}).
As before, we consider separately the cases of slowly moving jets, but observed at a large inclination angle, and fast moving jets, observed nearly along the jet direction.
For each case we calculate the expected $\gamma$-ray light curves and spectra 
arriving to the observer at different moments for specific propagation distances of the blob with fixed velocity. The $\gamma$-ray light curves calculated for different parameters are normalized to 1 erg of injected energy of electrons (in the blob frame).

\subsection{Blazars at large inclination angles}

In order to investigate the time dependence of $\gamma$-ray emission let's first consider a simple scenario in which electrons are injected at a fixed distance from the accretion disk.
Electrons with a differential power law spectrum are instantaneously injected inside a compact blob at a given height $H_0$ above the accretion disk. The blob is either at rest, or it is moving along the jet with a velocity $v$. The electrons are cooled down in the accretion disk radiation field producing $\gamma$-rays, which initiate the IC $e^\pm$ pair cascade in the whole volume above the accretion disk.
These $\gamma$-rays arrive to the observer at different times from two reasons.
At first, primary $\gamma$-rays are produced at different distances from the base of the jet. 
Afterwards, secondary cascade $\gamma$-rays are produced at different locations above the accretion disk passing different distances on their way to the observer.

\begin{figure}
\centering
\includegraphics[scale=0.50, trim=  0 12 0 0, clip]{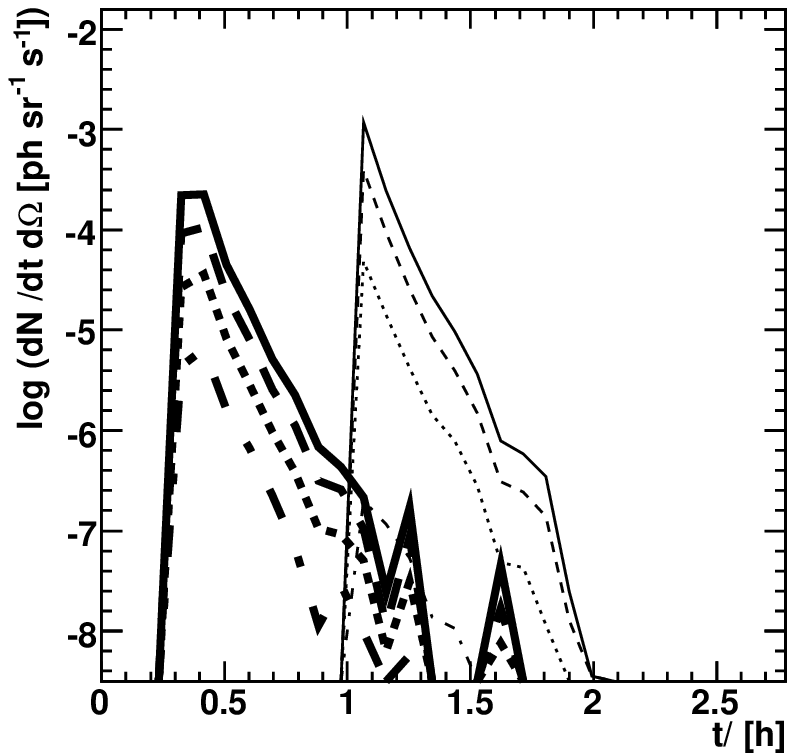}\figlab{$v=0, H_0=10r_{in}$}{3.2cm}     
\includegraphics[scale=0.50, trim= 20 12 0 0, clip]{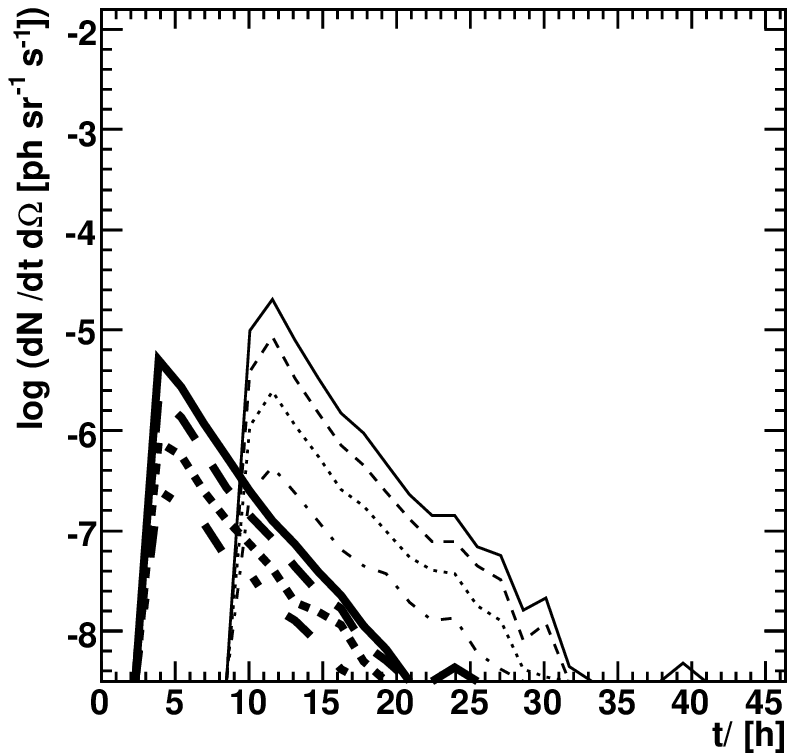}\figlab{$v=0, H_0=100r_{in}$}{3.2cm}\\
\includegraphics[scale=0.50, trim=  0  0 0 0, clip]{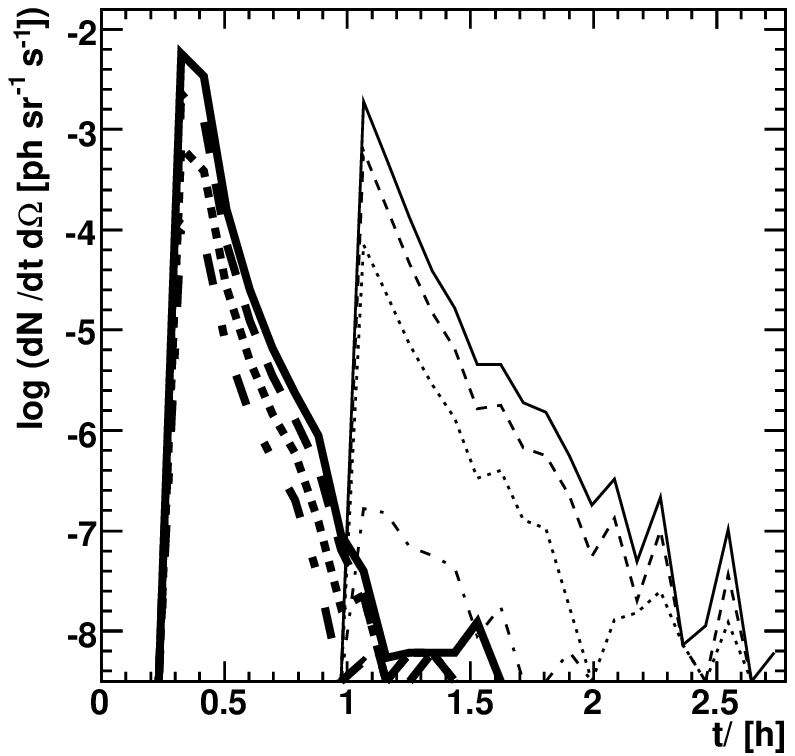}\figlab{$v=0.9c, H_0=10r_{in}$}{3.4cm}
\includegraphics[scale=0.50, trim= 20  0 0 0, clip]{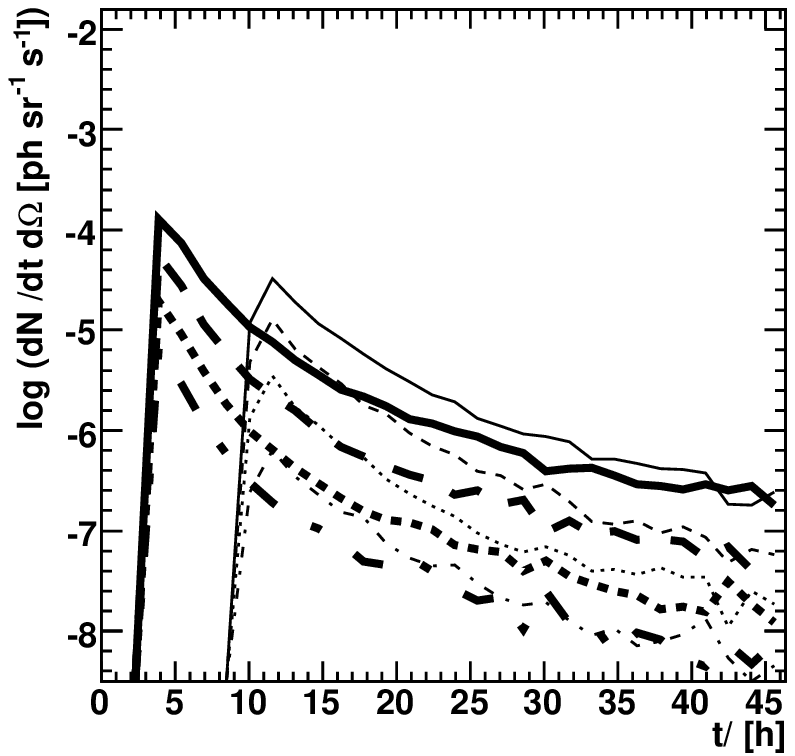}\figlab{$v=0.9c, H_0=100r_{in}$}{3.4cm} 
\caption{The $\gamma$-ray light curves from the IC $e^\pm$ pair cascade initiated by electrons, injected 
isotropically in a blob, with spectral index $-2$ between 10 GeV and 10 TeV (in the blob frame) for Cen~A at different energy ranges $E>10$ (solid), $>30$ (dashed), $>100$ (dotted), and $>300$ GeV (dot-dashed). Injection of electrons occurs at $H=10$ $r_{in}$ (left figures), and 100 $r_{\rm in}$ (right figures).
Stationary blob ($v=0$) (top figures) and  blob propagating with $v=0.9c$ (bottom figures). 
Two observation angles are considered $32^\circ$ (thick lines) and $57^\circ$ (thin lines).}
\label{fig:CenA_lc_point}
\end{figure}

In Fig.~\ref{fig:CenA_lc_point} we show the $\gamma$-ray light curves obtained in such a scenario for the blob at rest or moving with the velocity $v=0.9c$, and for two injection heights along the jet $H_0=10$ and $100r_{\rm in}$.
In most of the cases, the $\gamma$-ray light curves show a strong initial peak with an exponential decay. The characteristic decay time strongly depends on the injection height. If the injection place is located far away from the disk, electrons cool down in a relatively weak radiation field. 
Therefore, cooling process of electrons takes longer producing broader peak (see right figures in Fig.~\ref{fig:CenA_lc_point}).

It is interesting to evaluate the effect of the blob velocity on the $\gamma$-ray light curves. 
For an injection place of electrons close to the disk, the emission peak is sharper in the case of a faster blob and low inclination angles, i.e. the flare is characterized by shorter time scales. This effect is connected with the relativistic beaming of the primary $\gamma$-rays along the jet direction.
During typical cooling time scale of electrons, the blob moves in the direction of the observer, effectively reducing the time scale in the reference frame of the observer.
In the case of the electron injection place located farther away from the disk, the $\gamma$-ray light curves show a long tail for a fast blob. 
This is caused by the comparable time scales for the electron cooling in the disk radiation with the travel time scale of the blob along the jet. 
Therefore, the $\gamma$-rays are produced on much longer distances and the corresponding $\gamma$-ray emission time scale in the observer's reference frame increases.
It results in a long tail with a slower than exponential decay. 

In the next step, we consider the injection of the electrons in the blob at a fixed range of distances from the disk. 
We are considering both the cases of a slow (velocity $v=0.5c$) and a fast ($v=0.9c$) moving blob. It is assumed that the injection process of electrons occurs in the range of distances between $1r_{\rm in}$ and 30, 100, and 300 $r_{\rm in}$. 

\begin{figure}
\centering
\includegraphics[scale=0.53, trim=  0 10 0 0, clip] {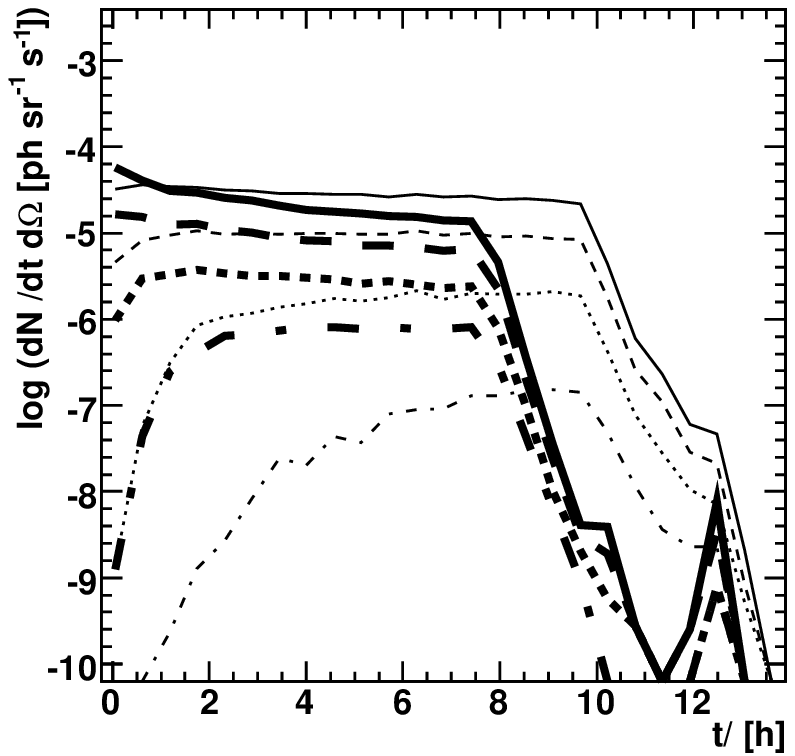}\figlab{$v=0.5c, H_0=1-30r_{in}$}{3.4cm}     
\includegraphics[scale=0.53, trim= 20 10 0 0, clip]  {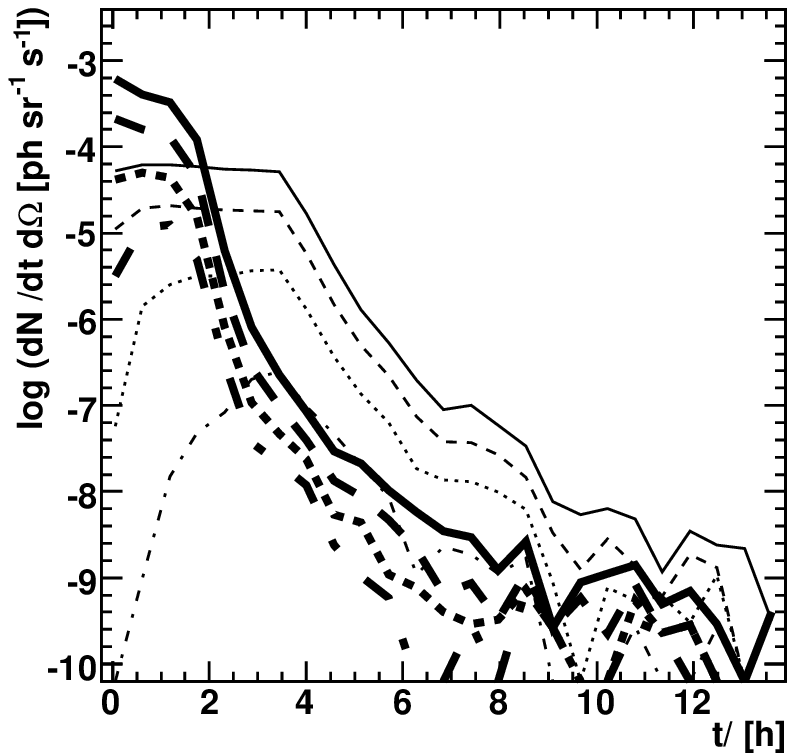}\figlab{$v=0.9c, H_0=1-30r_{in}$}{3.4cm}\\	
\includegraphics[scale=0.53, trim=  0 10 0 0, clip]{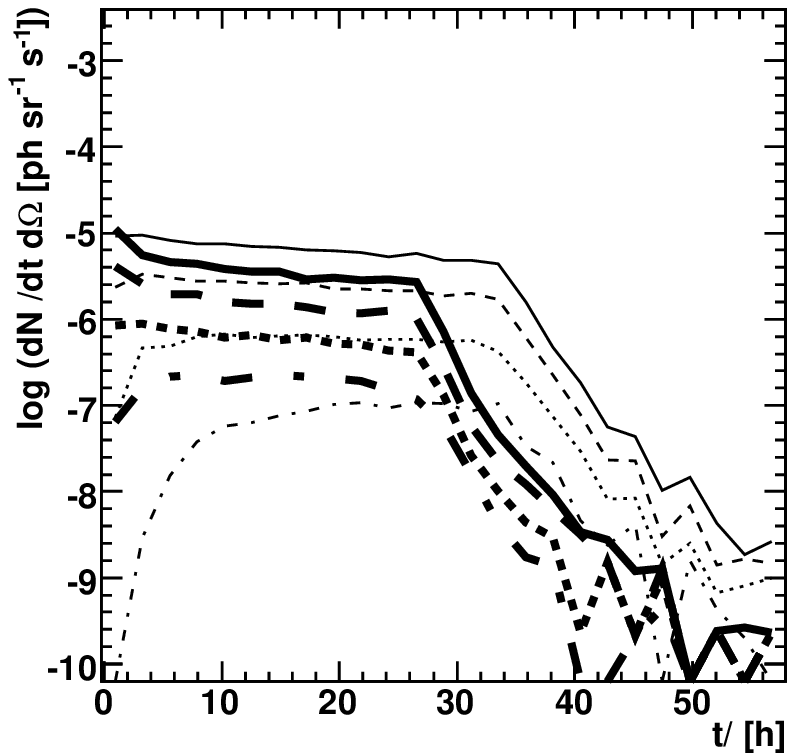}\figlab{$v=0.5c, H_0=1-100r_{in}$}{3.4cm}	
\includegraphics[scale=0.53, trim= 20 10 0 0, clip] {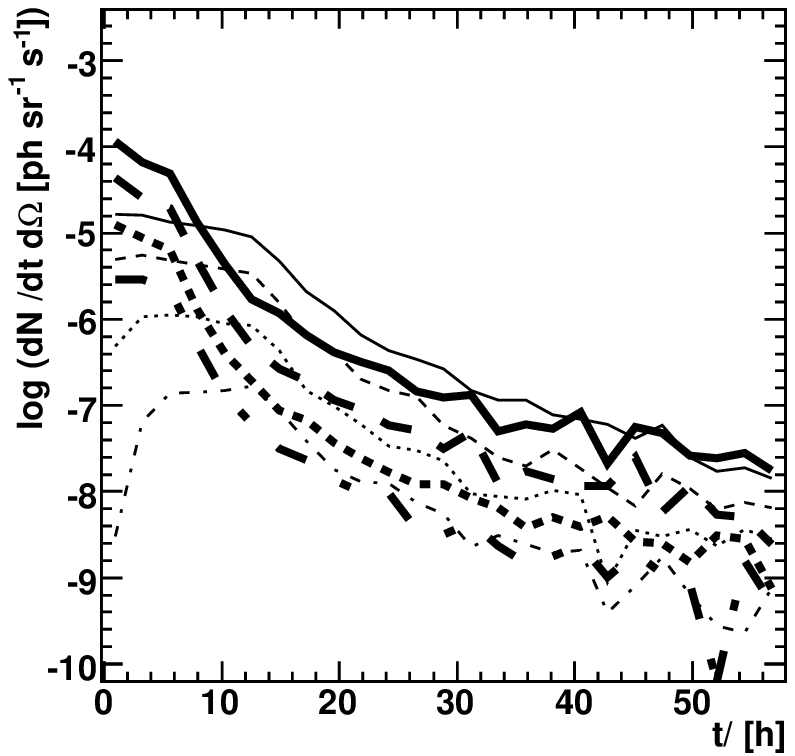}\figlab{$v=0.9c, H_0=1-100r_{in}$}{3.4cm}\\
\includegraphics[scale=0.53, trim=  0 0 0 0, clip] {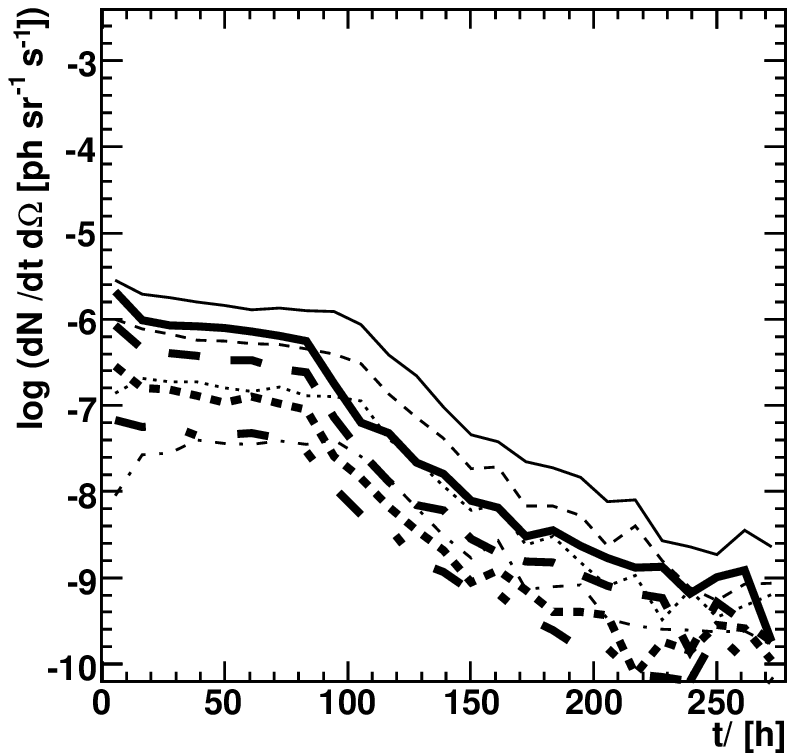}\figlab{$v=0.5c, H_0=1-300r_{in}$}{3.6cm}    
\includegraphics[scale=0.53, trim= 20 0 0 0, clip] {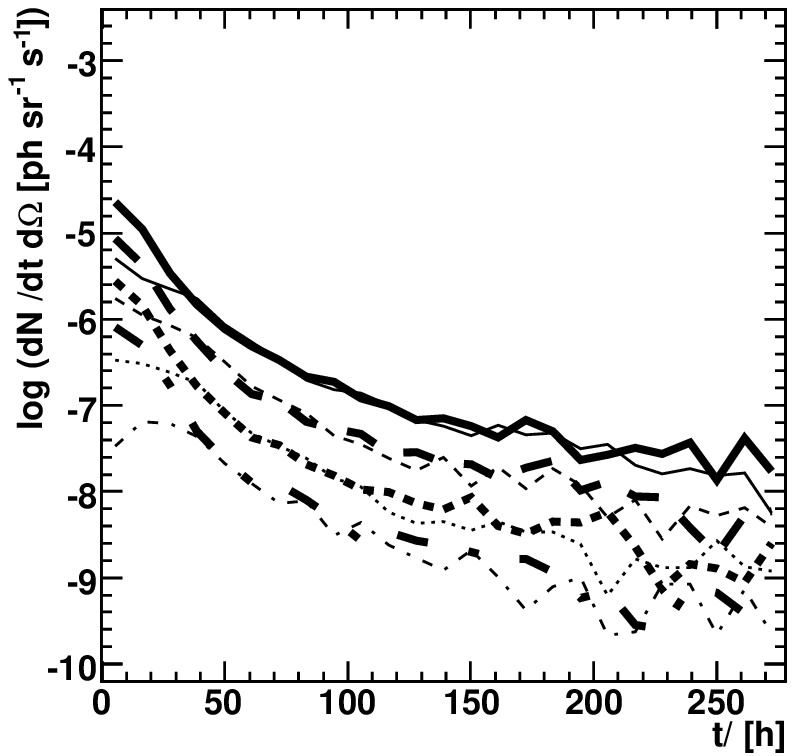}\figlab{$v=0.9c, H_0=1-300r_{in}$}{3.6cm}   
\caption{
The $\gamma$-ray light curves produced in IC $e^\pm$ pair cascade initiated by electrons injected isotropically in a blob. Electrons are injected with a spectral index $-2$ between 10 GeV and 10 TeV (in the blob frame). The parameters of Cen~A are applied.
The injection process of electrons occurs between 1--30 $r_{\rm in}$ (top figures), 1--100 $r_{in}$ (middle figures) and 1--300 $r_{in}$ (bottom figures).
The blob propagates with the velocity: $0.5c$ (left figures) and  $0.9c$ (right figures).  
The light curves are shown for energy thresholds and angles as in Fig.~\ref{fig:CenA_lc_point}.}
\label{fig:CenA_lc_along}
\end{figure}

The results of calculations are shown in Fig.~\ref{fig:CenA_lc_along}.
In general the $\gamma$-ray light curves show at first a plateau with a strong exponential decay.
Depending on the injection parameters, it can be followed by a long tail of slowly decaying emission. 
The flat part of the light curve is produced by the fast cooling electrons close to their injection place.
Depending on the range of injection heights, the emission can last at a fairly constant level for from a few up to a few tens of hours. 
The tails, as in the case of a fixed injection, are caused by electrons escaping into region of a constantly decreasing radiation field.
The duration of the flat part of the light curve is determined mostly by the range of the distances on which the injection of electrons takes place. 
Moreover, for a fast moving blob observed at low inclination angles, it is significantly reduced due to the relativistic effects. 
If a source is observed at large inclination angles, then the relativistic effects are less pronounced, resulting in a weaker but more extended plateau in the $\gamma$-ray light curve (see thin curves in right figures in Fig.~\ref{fig:CenA_lc_along}).

The properties of the light curve depend on the energies of $\gamma$-ray photons. 
The beginning of the flare, corresponds to the time in which the blob is close to the disk. In such a strong radiation field, the very high energy (VHE) $\gamma$-rays are strongly attenuated.
Therefore, the VHE emission ($> 300$ GeV) is weaker at the beginning of the flare (see dot-dashed curves in top Fig.~\ref{fig:CenA_lc_along}). 
The absorbed $\gamma$-rays are reprocessed in the IC $e^\pm$ pair cascade into a lower energy $\gamma$-rays.
Those secondary $\gamma$-rays create an additional peak in the light curve above $10$ GeV at the beginning of the flare (see top left Fig.~\ref{fig:CenA_lc_along}). 
This leads to an interesting consequence. 
VHE $\gamma$-ray emission (in the Cherenkov telescopes energy range) should appear after the GeV $\gamma$-ray emission (detected by $\gamma$-ray satellites). 
Therefore, the model predicts a soft to hard $\gamma$-ray evolution in the case of $\gamma$-ray flares observed from blazars.
This effect is even more pronounced for large observation angles (see dot-dashed thin lines in Fig.~\ref{fig:CenA_lc_along}). In this case, due to a more favorable angle between the direction of the $\gamma$-ray photon and the hot inner region on the accretion disk, the absorption of high energy $\gamma$-rays is even stronger. 
Thus, primary TeV and sub-TeV $\gamma$-ray photons are reprocessed more efficiently into the multi-GeV range. 

\begin{figure}
\centering
\includegraphics[scale=0.53, trim= 0 11 0 0, clip]{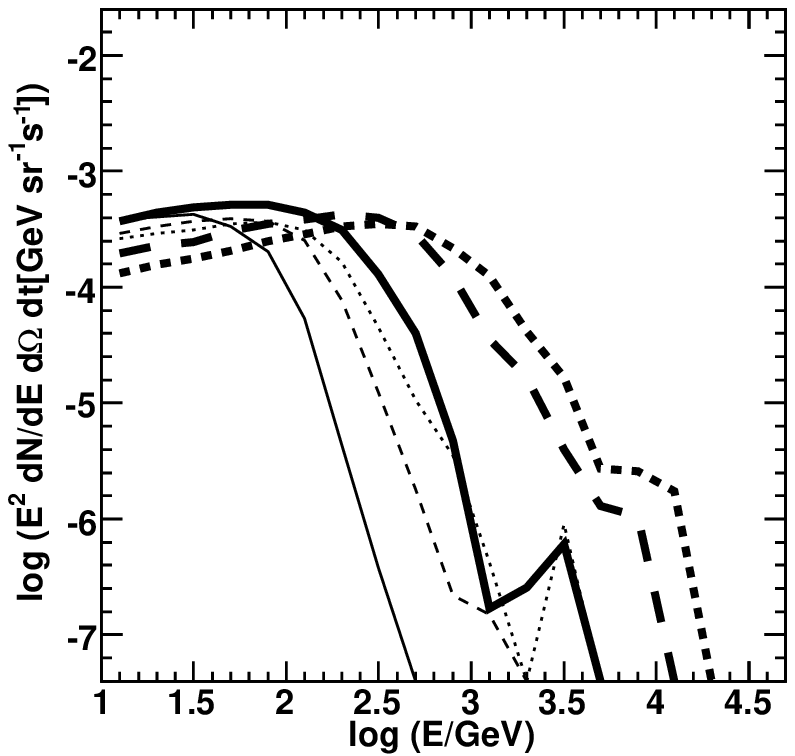}\figlab{$v=0.5c, H_0=1-30r_{in}$}{3.5cm}    
\includegraphics[scale=0.53, trim= 20 11 0 0, clip]{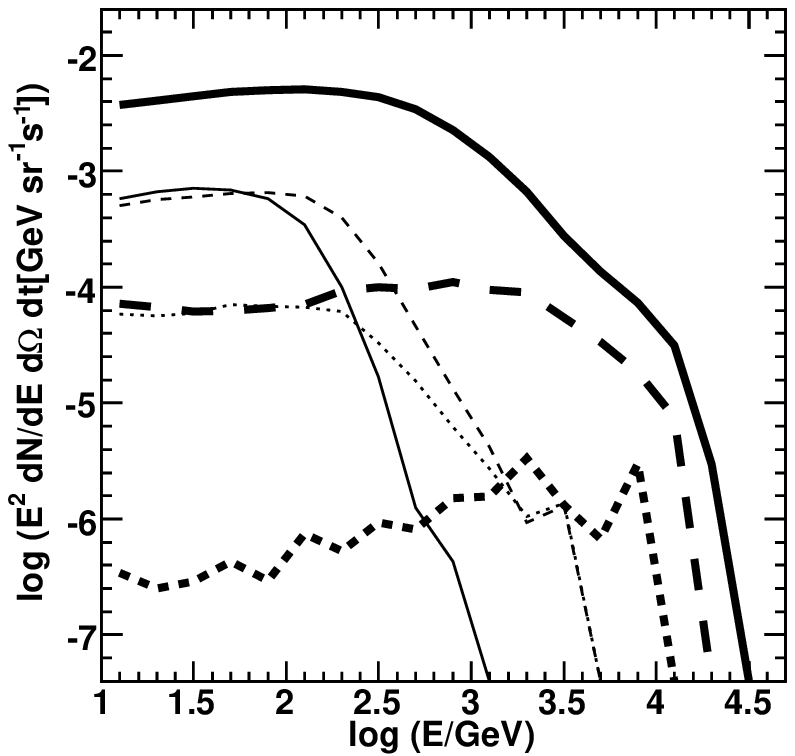}\figlab{$v=0.9c, H_0=1-30r_{in}$}{3.5cm}\\ 
\includegraphics[scale=0.53, trim= 0 11 0 0, clip]{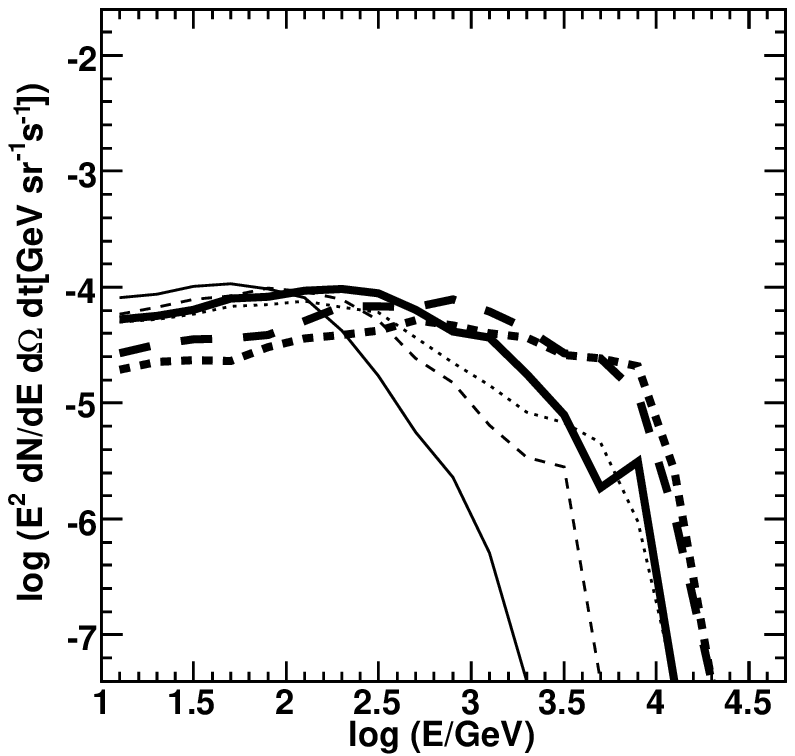}\figlab{$v=0.5c, H_0=1-100r_{in}$}{3.5cm}    
\includegraphics[scale=0.53, trim= 20 11 0 0, clip]{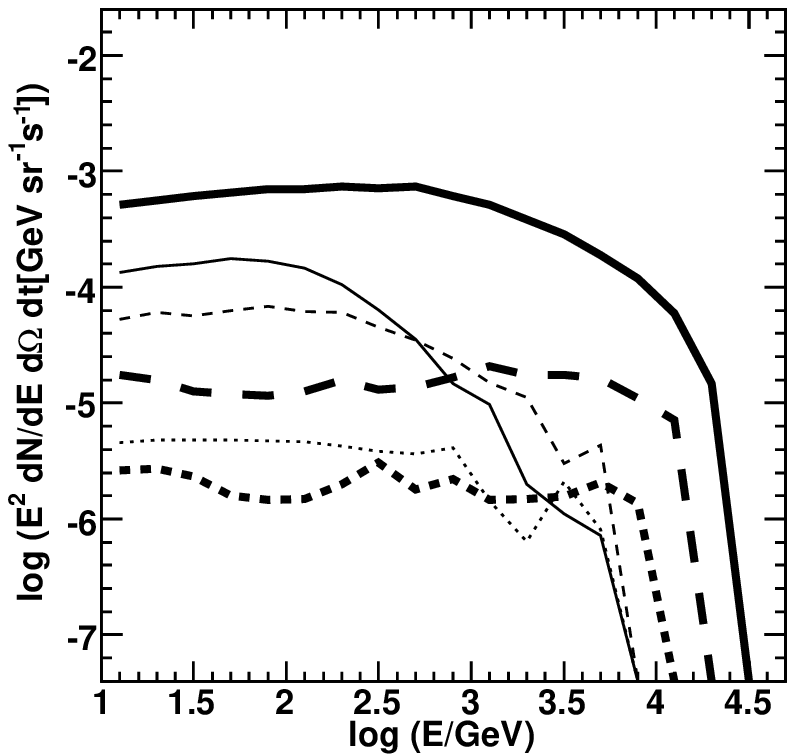}\figlab{$v=0.9c, H_0=1-100r_{in}$}{3.5cm}\\
\includegraphics[scale=0.53, trim= 0 0 0 0, clip]{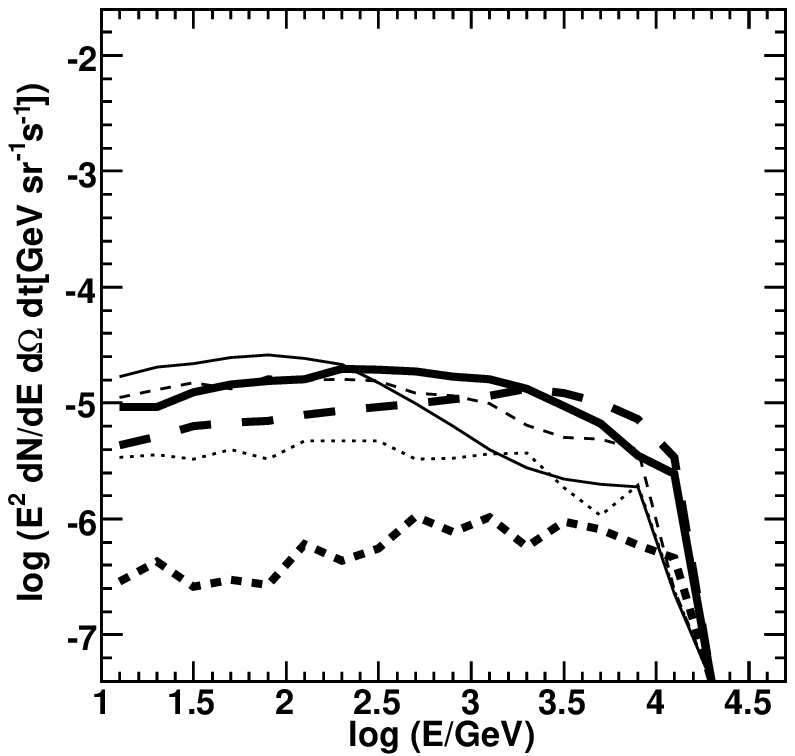}\figlab{$v=0.5c, H_0=1-300r_{in}$}{3.6cm}   
\includegraphics[scale=0.53, trim= 20 0 0 0, clip]{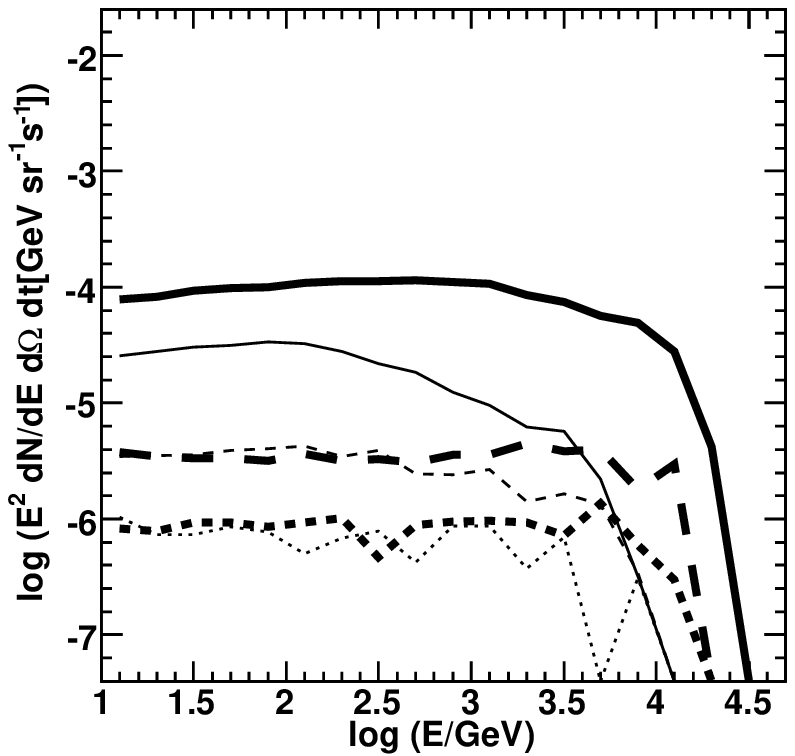}\figlab{$v=0.9c, H_0=1-300r_{in}$}{3.6cm}  
\caption{The $\gamma$-ray
spectra at a specific time intervals $t$ from the IC $e^\pm$ pair cascade initiated by electrons, injected isotropically in a blob, with a differential spectrum with a spectral index $-2$ between 10 GeV and 10 TeV (in the blob frame) for the parameters of Cen~A.
Blob propagates with the velocity $0.5c$ (left figures) and  $0.9c$ (right figures). 
Observation angle $32^\circ$ (thick lines) and $57^\circ$ (thin lines).
The $\gamma$-ray spectra are accumulated in the time intervals: 
$t=0$--2h (solid curves), 2--4h (dashed), and 4--6h (dotted) for the injection heights in the 
range $H_0=1$--30$r_{in}$ (upper panel); 
$t=0$--10h (solid), 10--20h (dashed), 20--30h (dotted) for $H_0=1$--100$r_{in}$ (middle panel); 
and $t=0$--50h (solid), 50--100h (dashed), 100--150h (dotted) for $H_0=1$--300$r_{in}$ 
(bottom panel).}
\label{spectra_time_cena}
\end{figure}

In Fig.~\ref{spectra_time_cena} we show the $\gamma$-ray spectra for different time intervals in the case of the model I for the maximum energy of electrons.
In the case of electron injection close to the disk, the multi-GeV $\gamma$-ray flux is slowly dropping with time in the plateau phase. 
However, the spectrum extends to higher energies with the development of the flare reaching TeV energies after a few hours (see the upper left Fig.~\ref{spectra_time_cena}). 
On the other hand, for the injection region of electrons extending far away from the disk, the $\gamma$-ray spectrum keeps a constant shape in time.
The level of emission decreases significantly while entering the tail phase in the time structure of the emission (see dotted curves in bottom panels in Fig.~\ref{spectra_time_cena}). 
This is a result of a slower cooling of electrons in the constantly weakening radiation field without strong absorption of $\gamma$-rays produced in the IC process. 
We note that the higher value of blob velocity result in larger variations of the spectrum. 

\begin{figure}
\includegraphics[scale=0.53, trim=  0 11 0 0, clip]{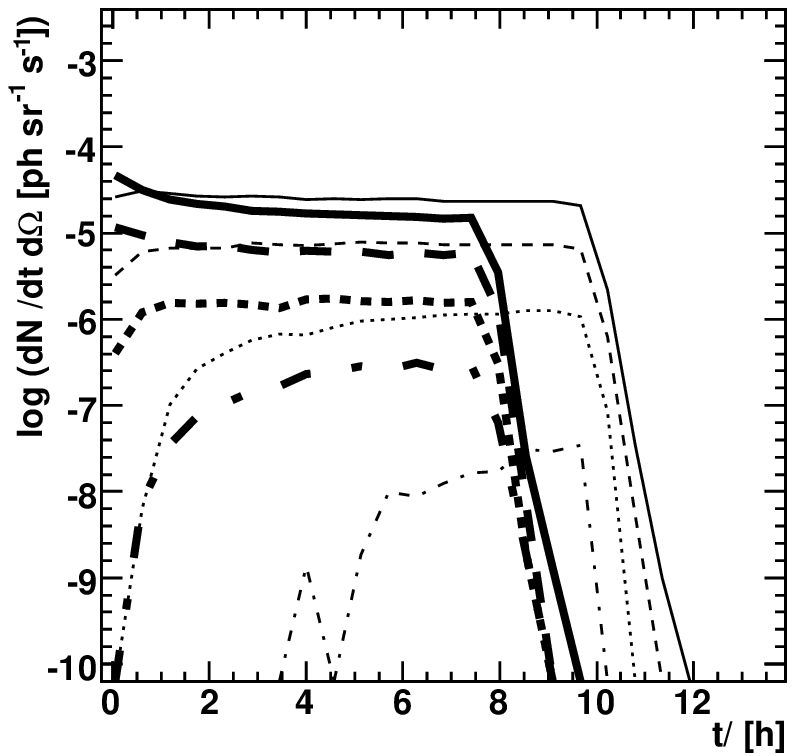}\figlab{$v=0.5c, H_0=1-30r_{in}$}{3.4cm}   
\includegraphics[scale=0.53, trim= 20 11 0 0, clip]{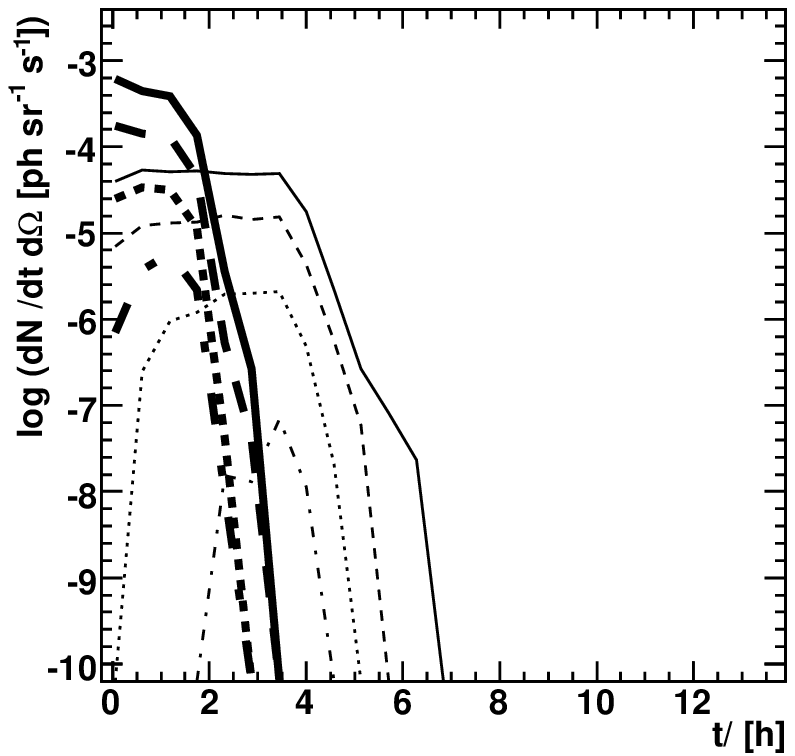}\figlab{$v=0.9c, H_0=1-30r_{in}$}{3.4cm}\\
\includegraphics[scale=0.53, trim=  0 11 0 0, clip]{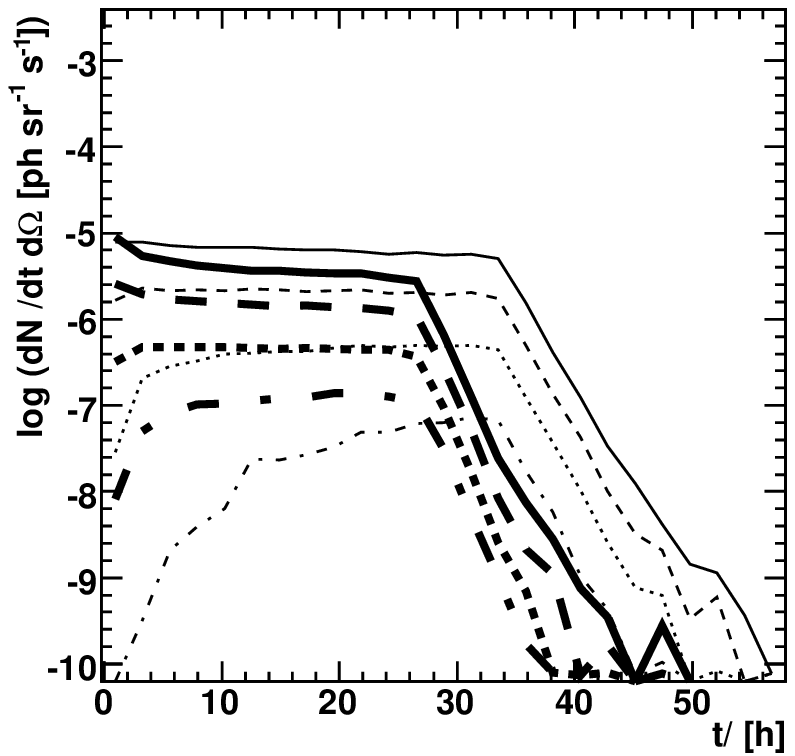}\figlab{$v=0.5c, H_0=1-100r_{in}$}{3.4cm}	 
\includegraphics[scale=0.53, trim= 20 11 0 0, clip]{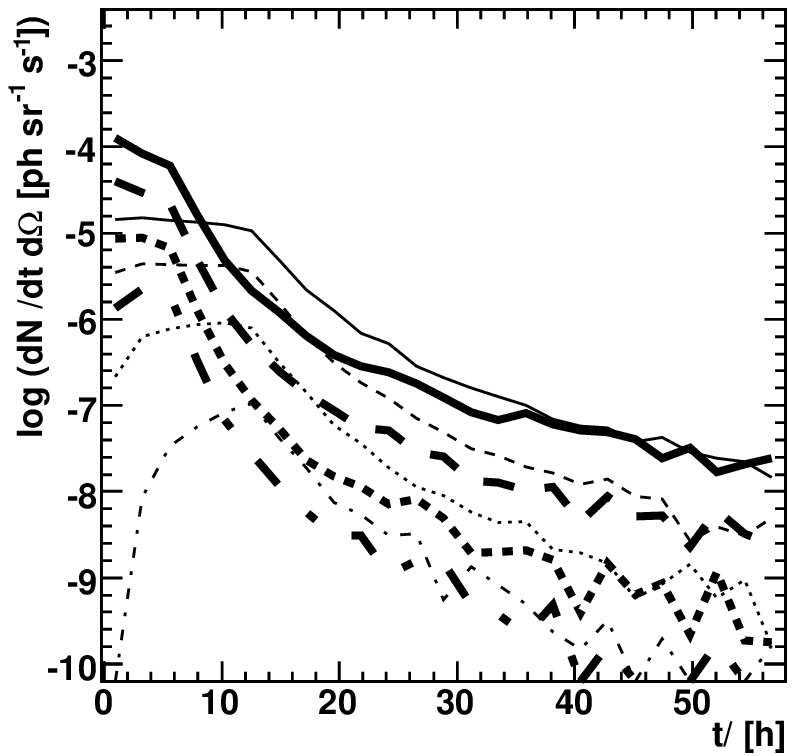}\figlab{$v=0.9c, H_0=1-100r_{in}$}{3.4cm}\\
\includegraphics[scale=0.53, trim=  0 0 0 0, clip]{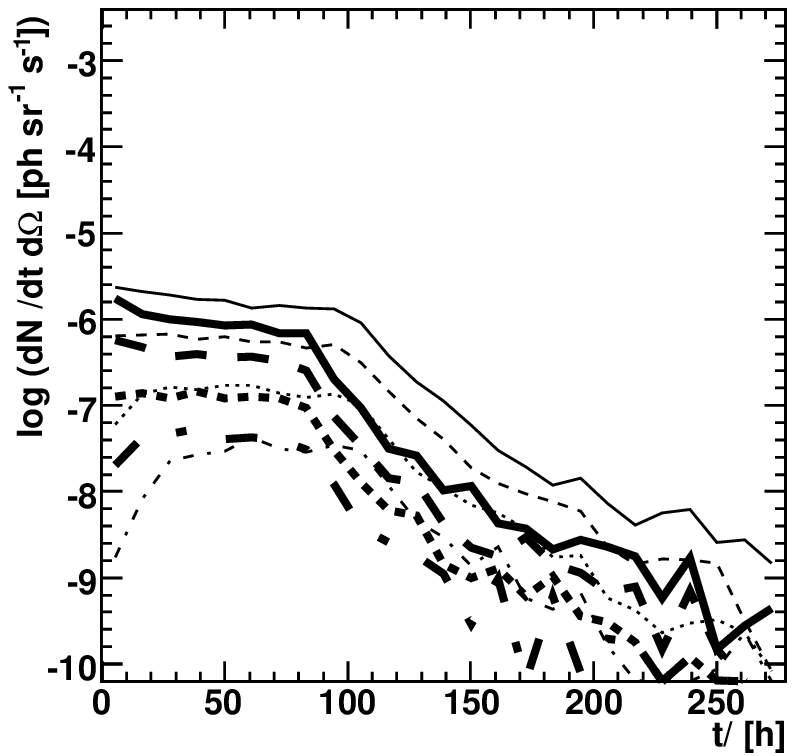}\figlab{$v=0.5c, H_0=1-300r_{in}$}{3.6cm}  
\includegraphics[scale=0.53, trim= 20 0 0 0, clip]{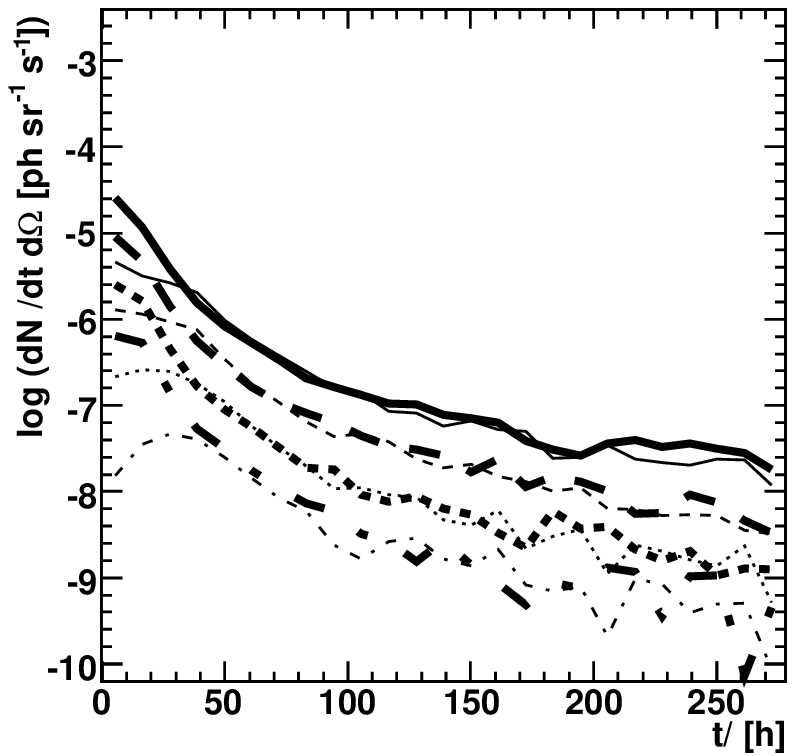}\figlab{$v=0.9c, H_0=1-300r_{in}$}{3.6cm}  
\caption{As in Fig.~\ref{fig:CenA_lc_along}, but for the maximum energies of electrons obtained according to model II (see text).
}\label{fig:CenA_lc_model2_along}
\end{figure}

\begin{figure}
\includegraphics[scale=0.53, trim=  0 11 0 0, clip]{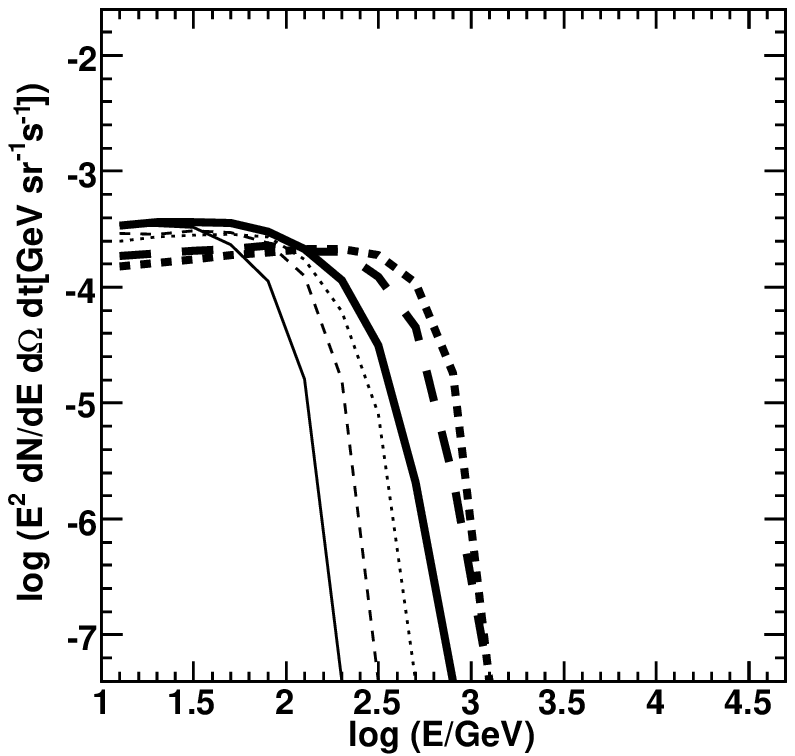}\figlab{$v=0.5c, H_0=1-30r_{in}$}{3.4cm}   
\includegraphics[scale=0.53, trim= 20 11 0 0, clip]{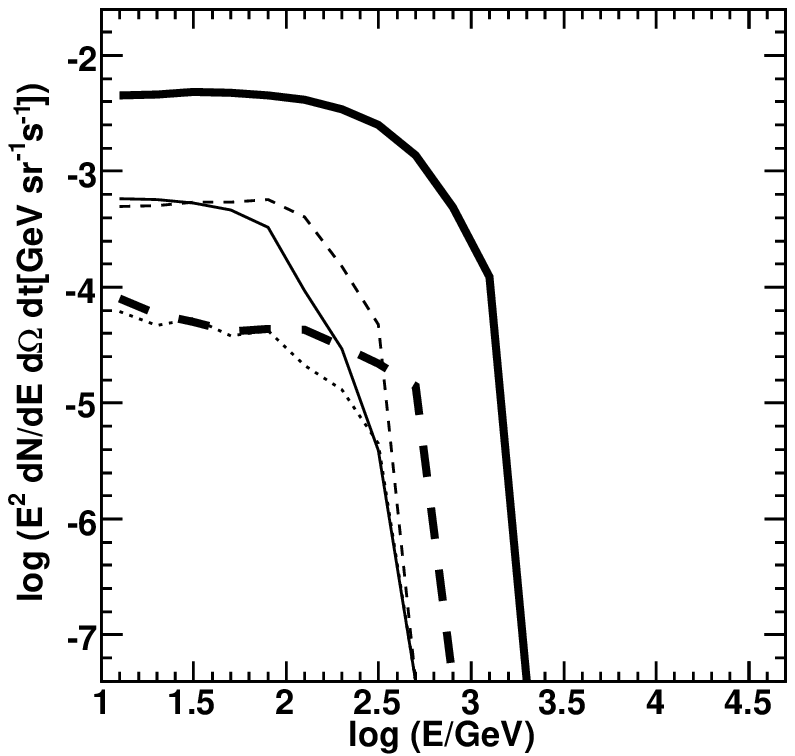}\figlab{$v=0.9c, H_0=1-30r_{in}$}{3.4cm}\\
\includegraphics[scale=0.53, trim=  0 11 0 0, clip]{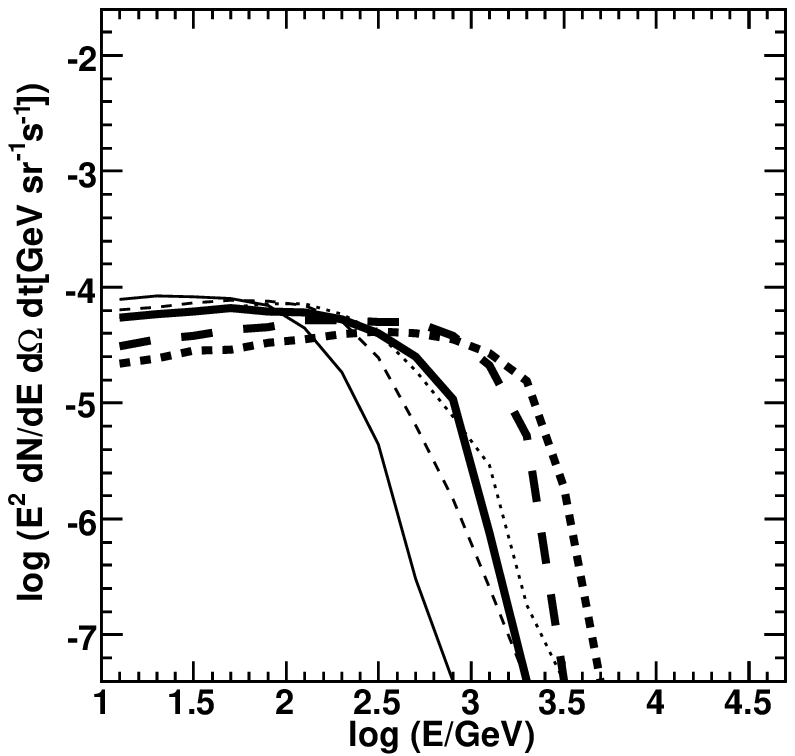}\figlab{$v=0.5c, H_0=1-100r_{in}$}{3.4cm}	 
\includegraphics[scale=0.53, trim= 20 11 0 0, clip]{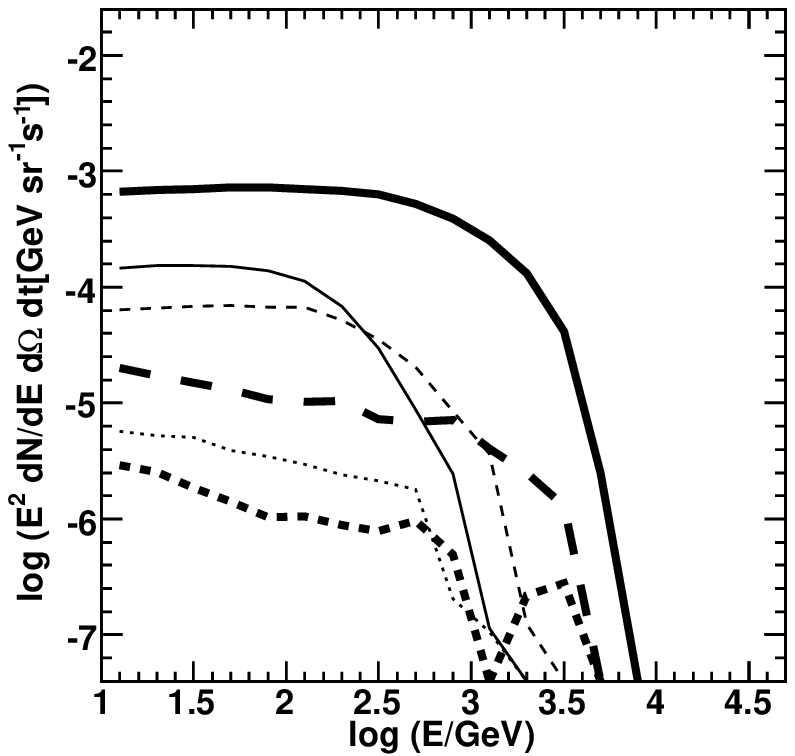}\figlab{$v=0.9c, H_0=1-100r_{in}$}{3.4cm}\\
\includegraphics[scale=0.53, trim=  0 0 0 0, clip]{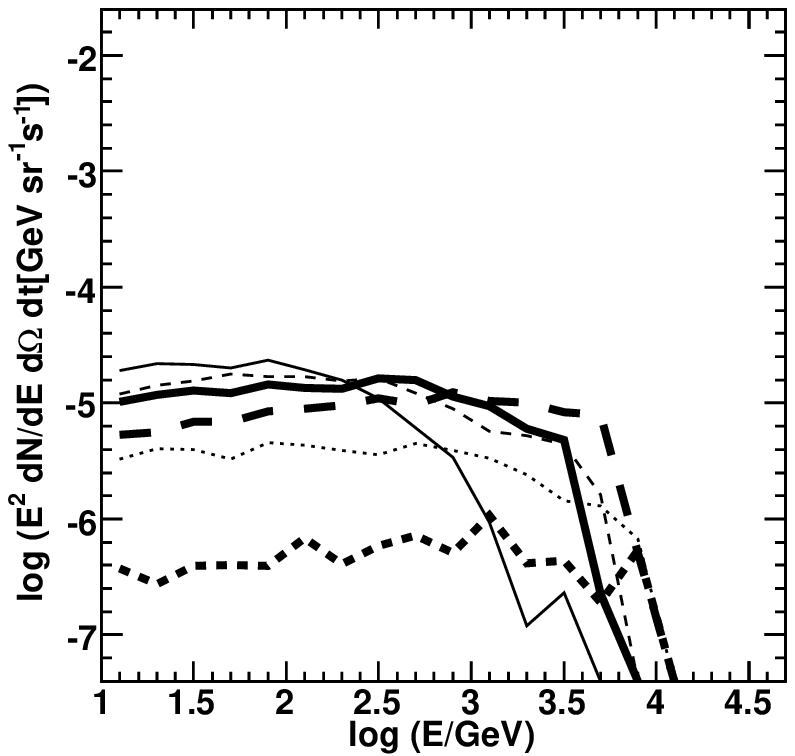}\figlab{$v=0.5c, H_0=1-300r_{in}$}{3.6cm}  
\includegraphics[scale=0.53, trim= 20 0 0 0, clip]{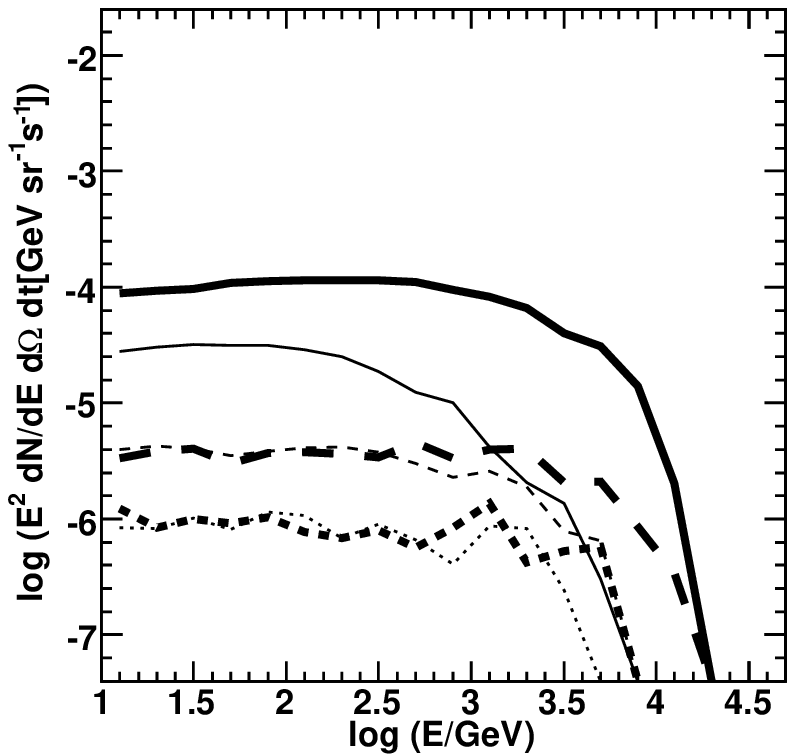}\figlab{$v=0.9c, H_0=1-300r_{in}$}{3.6cm}  
\caption{As in Fig.~\ref{spectra_time_cena}, but for model II of maximum energy of injected electron.}
\label{spectra_time_model2_cena}
\end{figure}

In Fig.~\ref{fig:CenA_lc_model2_along} and Fig.~\ref{spectra_time_model2_cena}, we present the 
$\gamma$-ray light curves and the time dependent gamma-ray spectra for the maximum energies of 
electrons obtained in terms of the model II. 
The basic properties of the $\gamma$-ray light curves and spectra are similar to those obtained 
in term of the model I. In general, electrons are accelerated to lower energies in the model II.
Therefore the spectra also extent to lower energies.
The differences are most pronounced in the case of electron injection close to the disk.
Then, the maximum energies of electron determined by the synchrotron energy losses are 
clearly below 10 TeV. This results in the $\gamma$-ray spectra limited to lower energies 
than calculated in model I (see upper plots in Fig.~\ref{spectra_time_model2_cena}). 

\subsection{Blazars at small inclination angles}

\begin{figure}
\centering
\includegraphics[scale=0.53, trim=  0 10 0 0, clip]{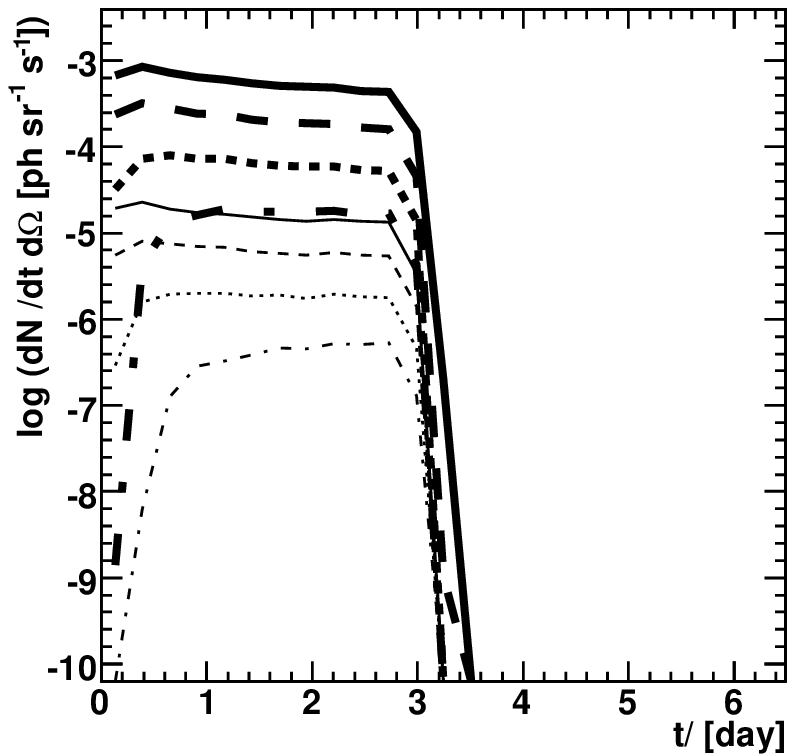}\figlab{$H_0=1-30r_{in}$}{3.4cm}\figlab{$v=0.99c$}{3.1cm}
\includegraphics[scale=0.53, trim= 20 10 0 0, clip]{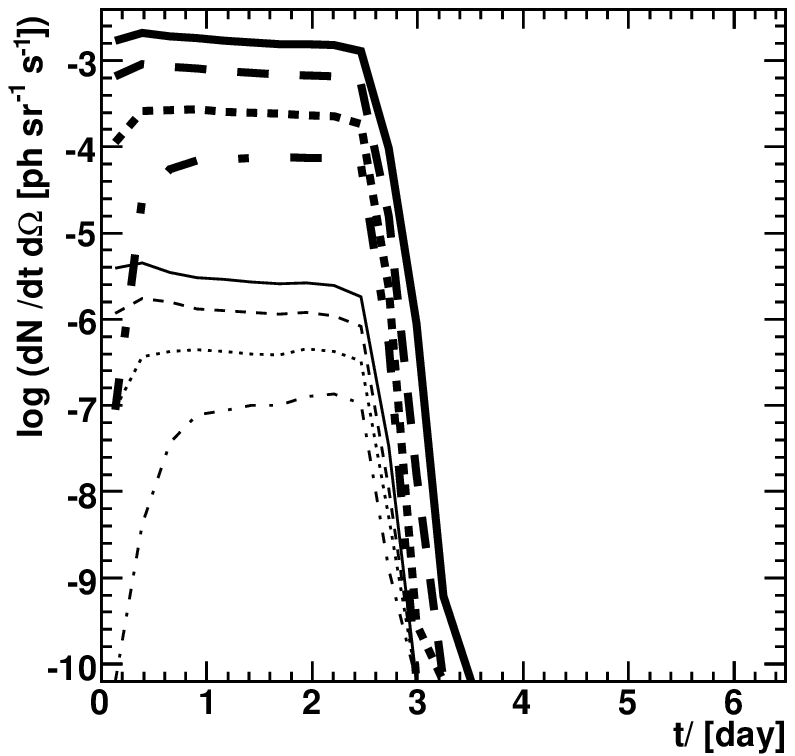}\figlab{$H_0=1-30r_{in}$}{3.4cm}\figlab{$v=0.998c$}{3.1cm}\\
\includegraphics[scale=0.53, trim=  0 10 0 0, clip]{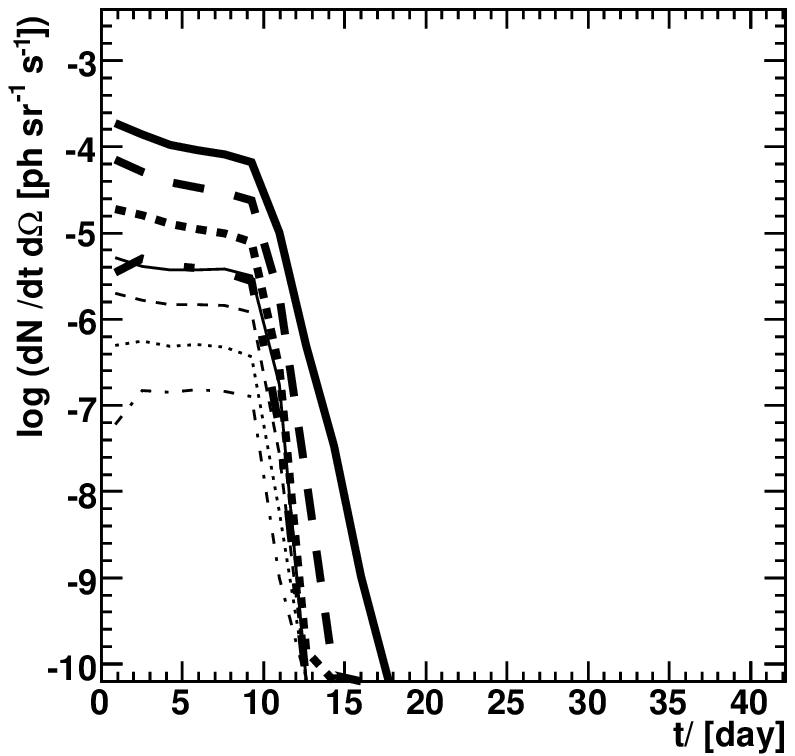}\figlab{$H_0=1-100r_{in}$}{3.4cm}\figlab{$v=0.99c$}{3.1cm}
\includegraphics[scale=0.53, trim= 20 10 0 0, clip]{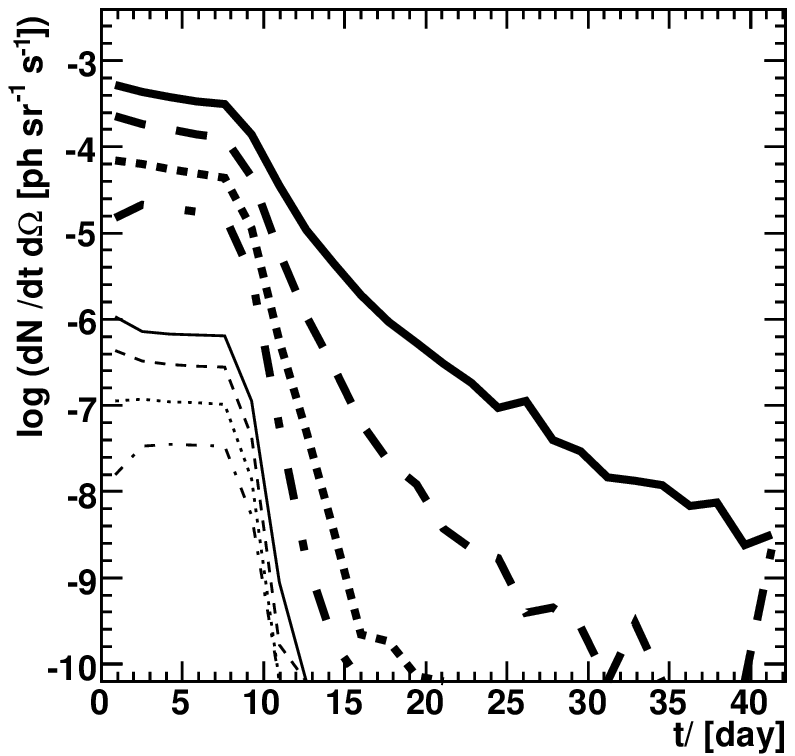}\figlab{$H_0=1-100r_{in}$}{3.4cm}\figlab{$v=0.998c$}{3.1cm}\\
\includegraphics[scale=0.53, trim=  0 0 0 0, clip] {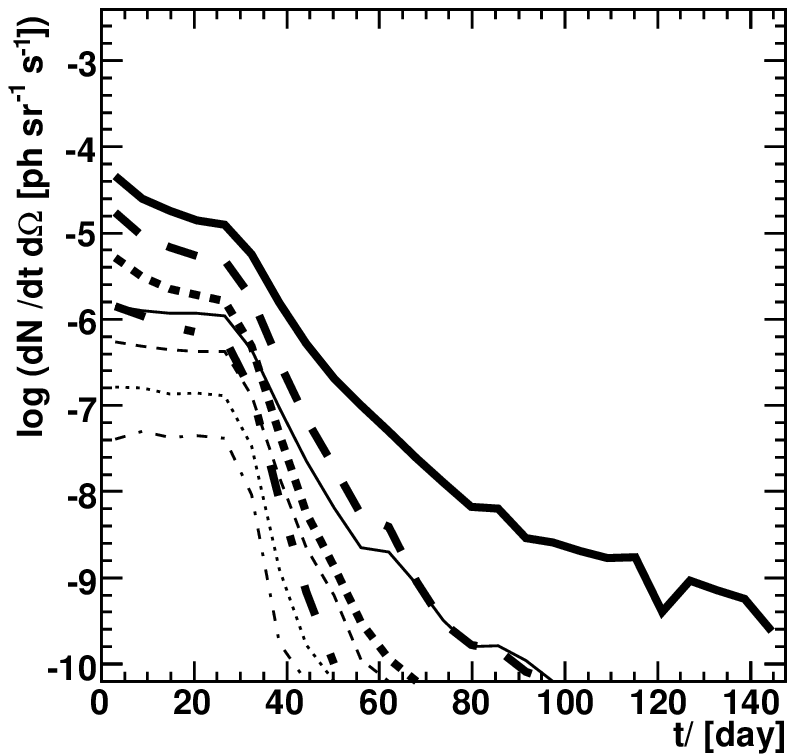}\figlab{$H_0=1-300r_{in}$}{3.6cm}\figlab{$v=0.99c$}{3.3cm}
\includegraphics[scale=0.53, trim= 20 0 0 0, clip] {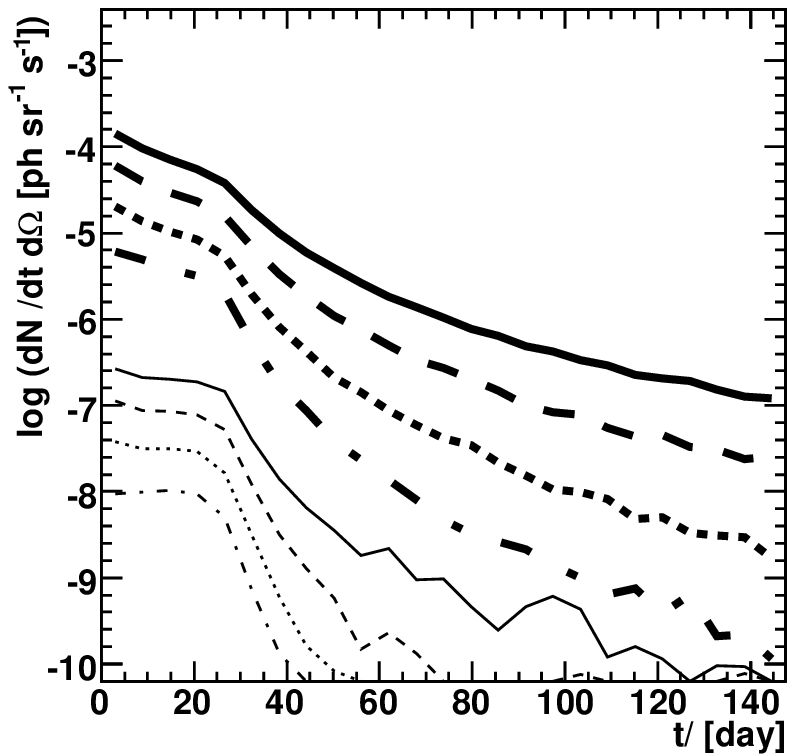}\figlab{$H_0=1-300r_{in}$}{3.6cm}\figlab{$v=0.998c$}{3.3cm}
\caption{The $\gamma$-ray light curves calculated for the electron spectrum and their injection 
distances as in Fig.~\ref{fig:CenA_lc_along} but for the parameters of 3C~279. 
The blob is moving with the velocity $v=0.99c$ ($\gamma=7.1$, left figures), or  0.998c ($\gamma=15.8$, right). The line styles correspond to these same energy ranges as in Fig.~\ref{fig:CenA_lc_along}. 
The values of the observation angles have been fixed on $12^\circ$ (thick lines) and 
$22^\circ$ (thin lines).
}\label{fig:lc_3c279}
\end{figure}

The $\gamma$-ray light curves from blazars observed at low inclination angles are analyzed taking 
as an example the parameters of 3C~279 (see Fig.~\ref{fig:lc_3c279}).
In general, the light curves show at the beginning slow dependence on time (a plateau) followed by a 
strong exponential cut-off.
In the case of faster blobs the cooling process of electrons occurs in a more extended region along the jet. 
Then, the $\gamma$-ray light curves show longer and slowly decaying tails.  
If the injection region is limited to the inner part of the jet, then the light curves show only a flat plateau extending for a few days. 

The duration of the $\gamma$-ray flare depends on the observation angle. 
For small angles, the tail emission can even last a few times longer than the duration of the plateau
(see e.g. the $\gamma$-ray light curves in the middle and bottom panel in Fig.~\ref{fig:lc_3c279}). 
However, since the tail emission is dropping fast, depending on the total flare luminosity and the sensitivity of the instrument, only it's beginning may be observable.
On the other hand, for larger inclination angles the $\gamma$-ray emission is much more confined in 
time. The $\gamma$-ray flares observed at lower energies lasts significantly longer than at higher 
energies. Therefore, we should expect shorter flares at TeV energies than at GeV energies.   

\begin{figure}
\includegraphics[scale=0.53, trim=  0 10 0 0, clip]{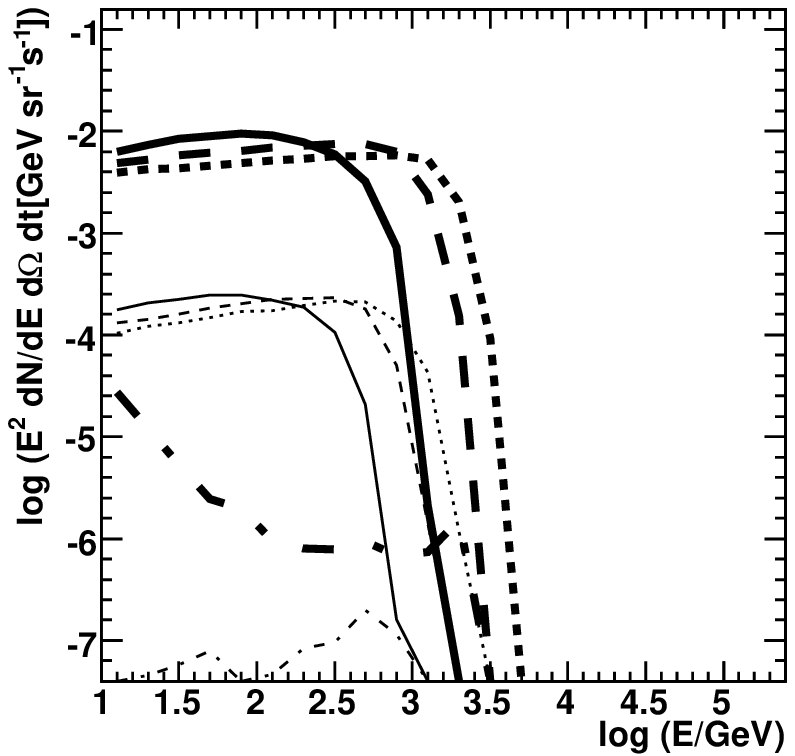}\figlab{$H_0=1-30r_{in}$}{3.4cm}\figlab{$v=0.99c$}{3.1cm}	
\includegraphics[scale=0.53, trim= 20 10 0 0, clip]{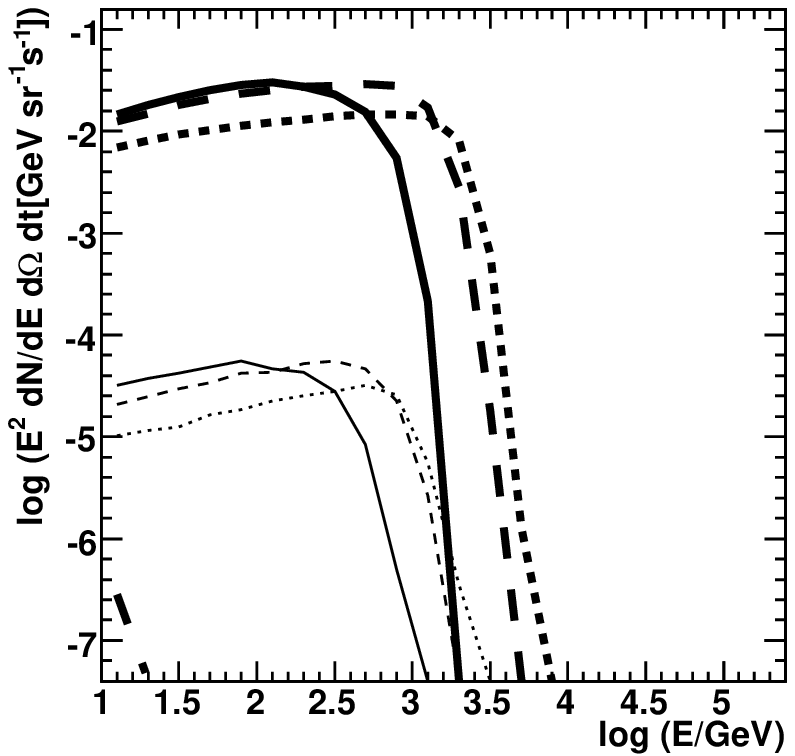}\figlab{$H_0=1-30r_{in}$}{3.4cm}\figlab{$v=0.998c$}{3.1cm}\\	
\includegraphics[scale=0.53, trim=  0 10 0 0, clip]{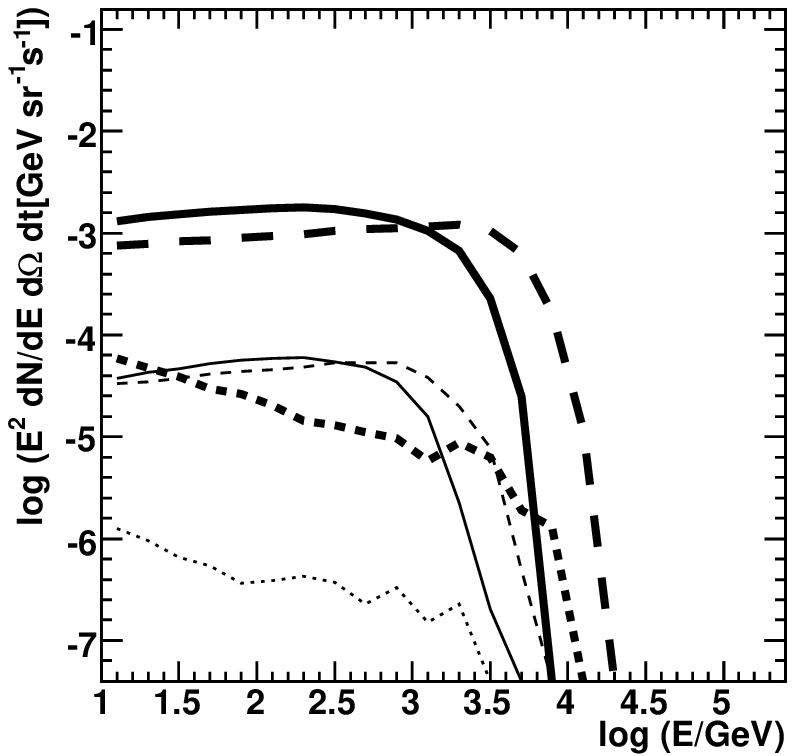}\figlab{$H_0=1-100r_{in}$}{3.4cm}\figlab{$v=0.99c$}{3.1cm}	
\includegraphics[scale=0.53, trim= 20 10 0 0, clip]{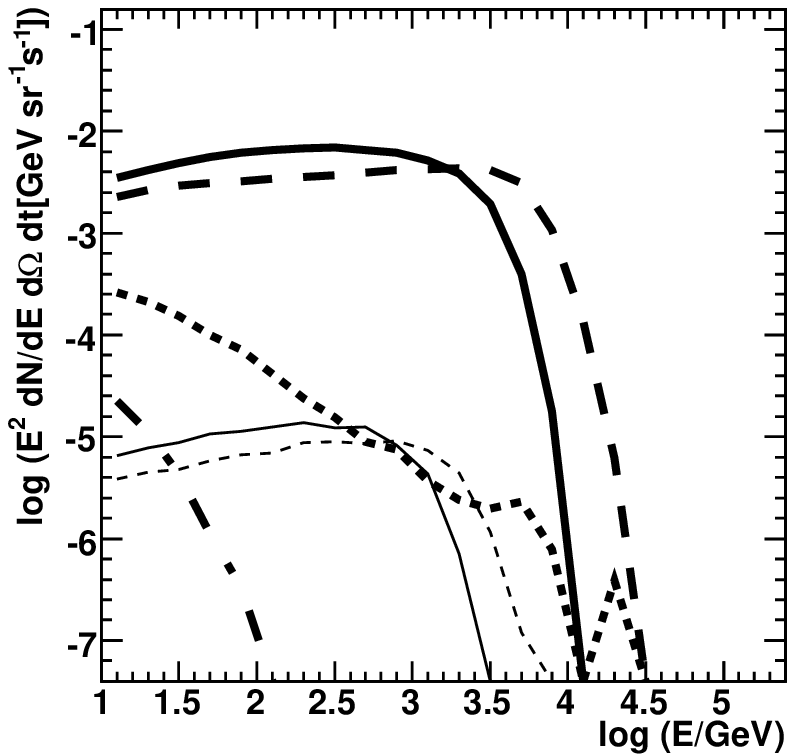}\figlab{$H_0=1-100r_{in}$}{3.4cm}\figlab{$v=0.998c$}{3.1cm}\\
\includegraphics[scale=0.53, trim=  0 0 0 0, clip] {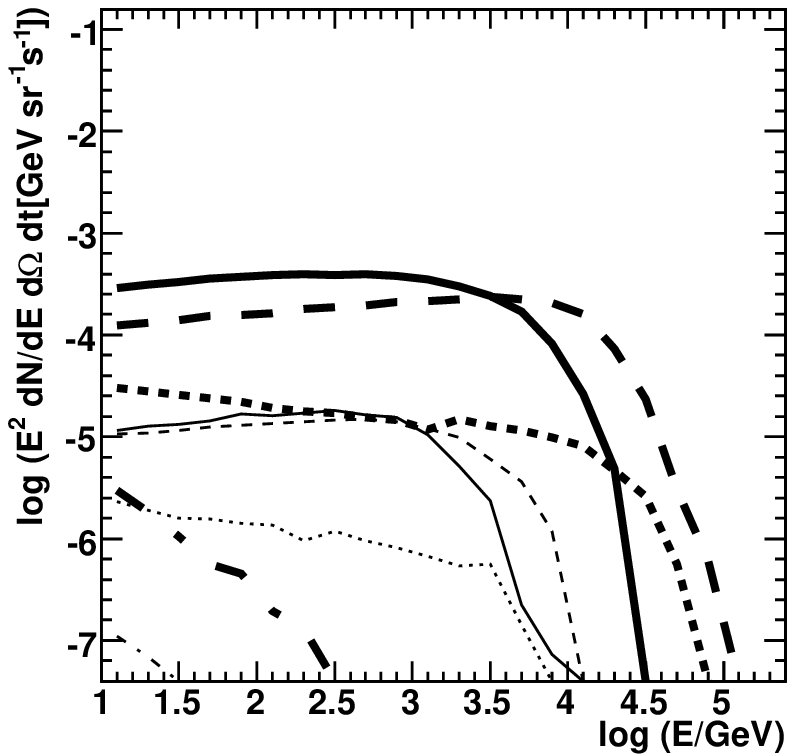}\figlab{$H_0=1-300r_{in}$}{3.6cm}\figlab{$v=0.99c$}{3.3cm}	
\includegraphics[scale=0.53, trim= 20 0 0 0, clip] {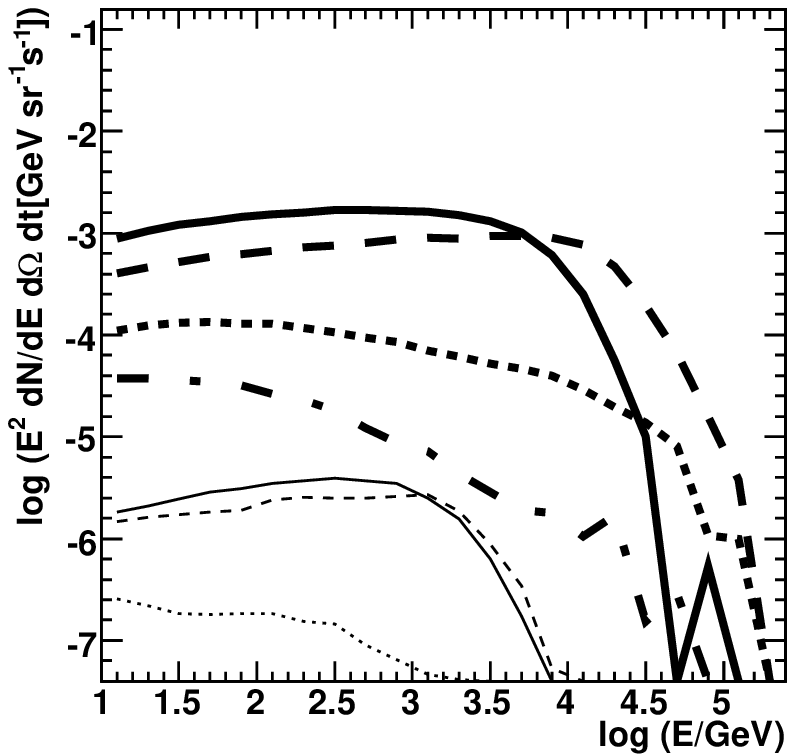}\figlab{$H_0=1-300r_{in}$}{3.6cm}\figlab{$v=0.998c$}{3.3cm}   
\caption{Time dependent spectra for 3C~279. Blob propagates with the velocity $0.99c$ ($\gamma=7.1$, left figures) and  $0.998c$ ($\gamma=15.8$, right).  
The upper figures are for the electrons injected within the range of distances from the base of the jet $H_0=1$--30$r_{in}$ (spectra for times 
(in the source frame of reference) $t=0$--1 day (solid curves), 1--2 days (dashed), 2--3 days 
(dotted), 3--4 days (dot-dashed)).
The middle figures are for $H_0=1$--100$r_{in}$ (spectra for times $t=0$--5 days (solid), 5--10 days 
(dashed), 10--15 days (dotted), 15--20 days (dot-dashed)).
The bottom figures are for $H_0=1$--300$r_{in}$ (spectra for times $t=0$--15 days (solid), 15--30 days (dashed), 30--45 days (dotted), 45--60 days (dot-dashed)). The observation angle is fixed on $12^\circ$ (thick lines) and $22^\circ$ (thin lines).}\label{spectra_time_3c279}
\end{figure}

The $\gamma$-ray spectra at the source, expected at specific time intervals, are shown in 
Fig.~\ref{spectra_time_3c279}. 
In the case of electron injection close to the accretion disk (injection range  1--30 $r_{in}$) 
the $\gamma$-ray spectra have similar shape during the plateau phase (see upper panel in 
Fig.~\ref{spectra_time_3c279}). 
For electrons injected along larger range of distances from the base of the jet, the $\gamma$-ray spectra during the plateau and exponential decay phase differ significantly.
During the plateau phase, the spectra has similar shape and slightly extend to higher energies with 
development of the flare. However, during the exponential decay phase, the $\gamma$-ray
spectra become significantly steeper with a cut-off at lower energies. 
We note that the $\gamma$-ray emission during the tail phase is produced by electrons that
are in the pure cooling phase above the injection region limited by the distance $H_2$.
Therefore, spectrum of these electrons is clearly steeper than the assumed one, $\propto E^{-2}$,
in the acceleration region.

\begin{figure}

\includegraphics[scale=0.53, trim=  0 10 0 0, clip]{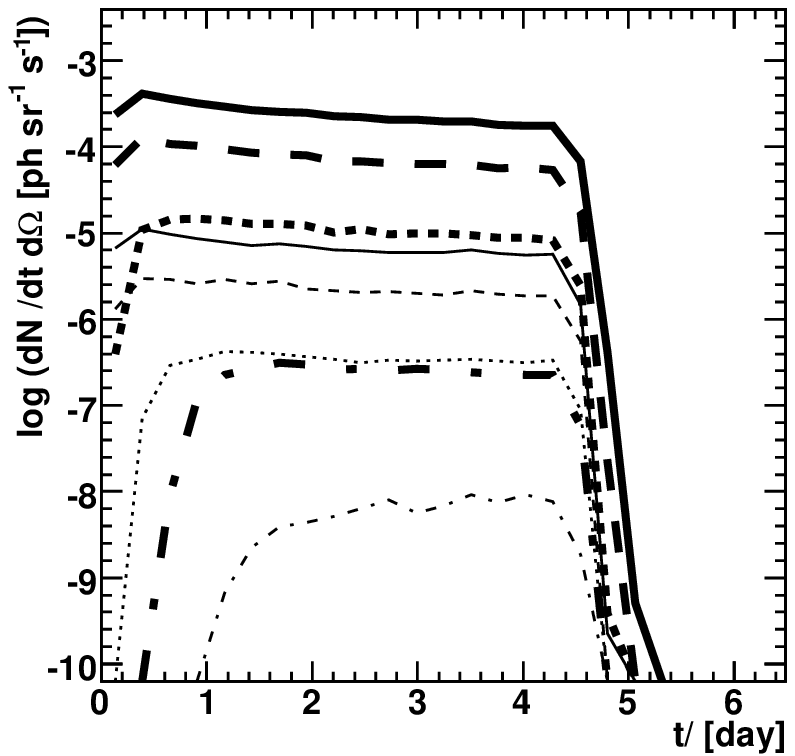}\figlab{$H_0=1-30r_{in}$}{3.5cm}\figlab{$v=0.99c$}{3.2cm}    
\includegraphics[scale=0.53, trim= 20 10 0 0, clip]{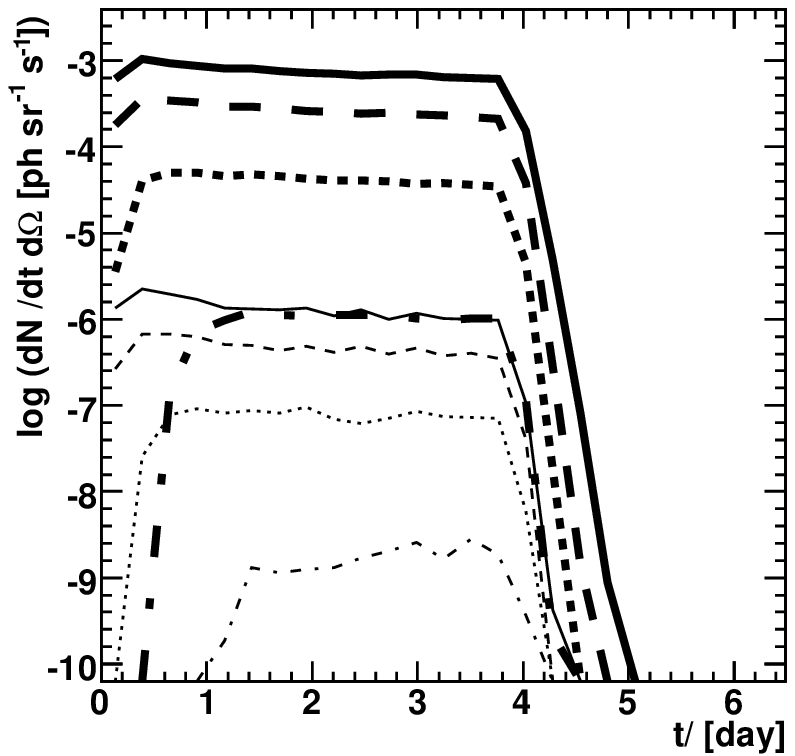}\figlab{$H_0=1-30r_{in}$}{3.5cm}\figlab{$v=0.998c$}{3.2cm}\\ 
\includegraphics[scale=0.53, trim=  0 10 0 0, clip]{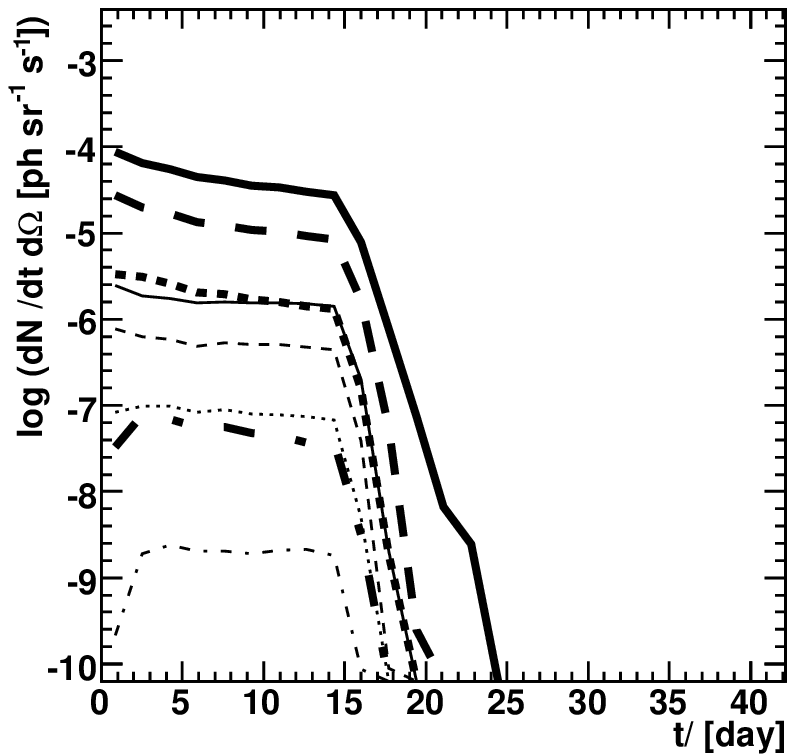}\figlab{$H_0=1-100r_{in}$}{3.5cm}\figlab{$v=0.99c$}{3.2cm}   
\includegraphics[scale=0.53, trim= 20 10 0 0, clip]{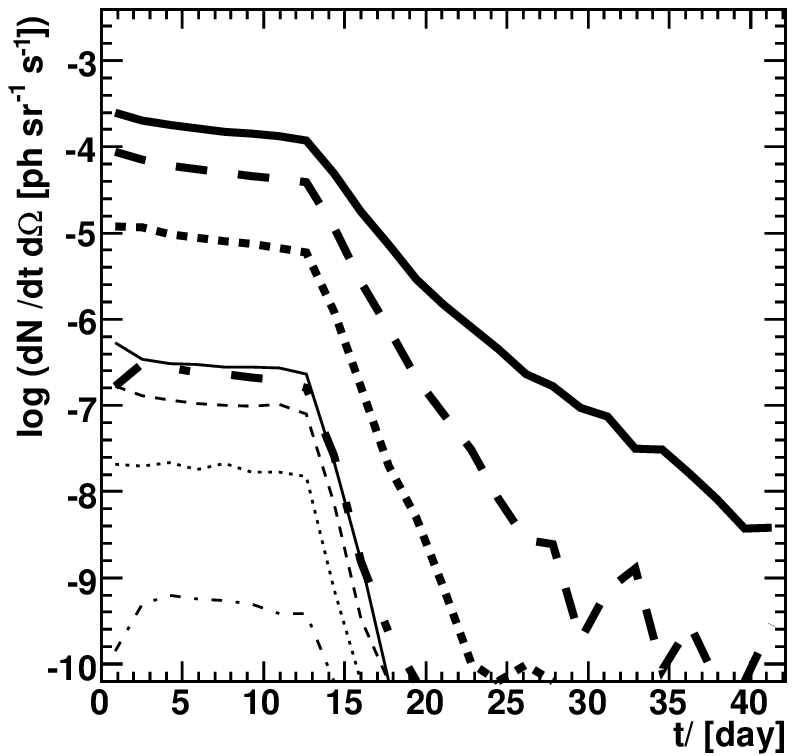}\figlab{$H_0=1-100r_{in}$}{3.5cm}\figlab{$v=0.998c$}{3.2cm}\\
\includegraphics[scale=0.53, trim=  0 0 0 0, clip] {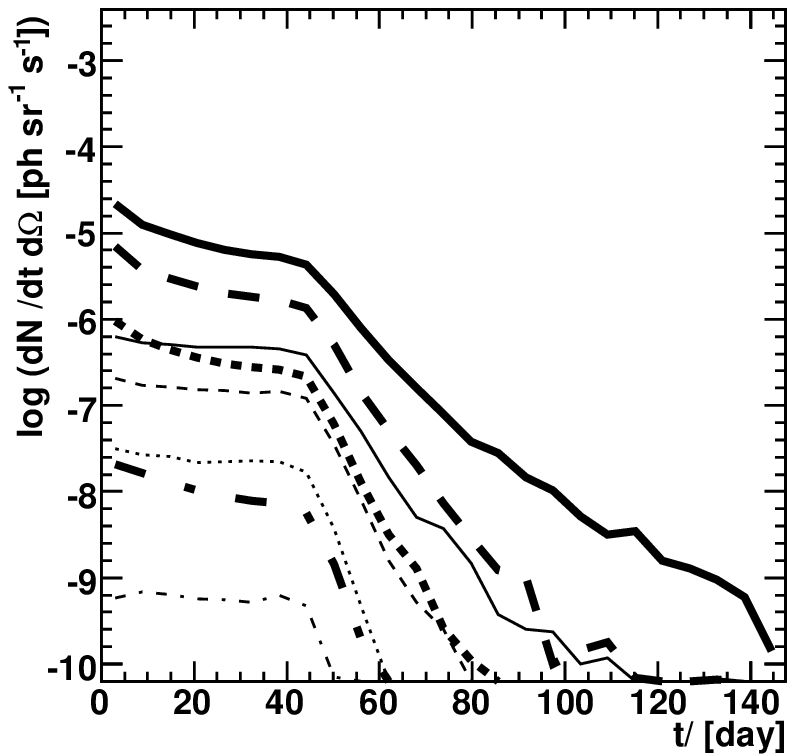}\figlab{$H_0=1-300r_{in}$}{3.6cm}\figlab{$v=0.99c$}{3.3cm}   
\includegraphics[scale=0.53, trim= 20 0 0 0, clip] {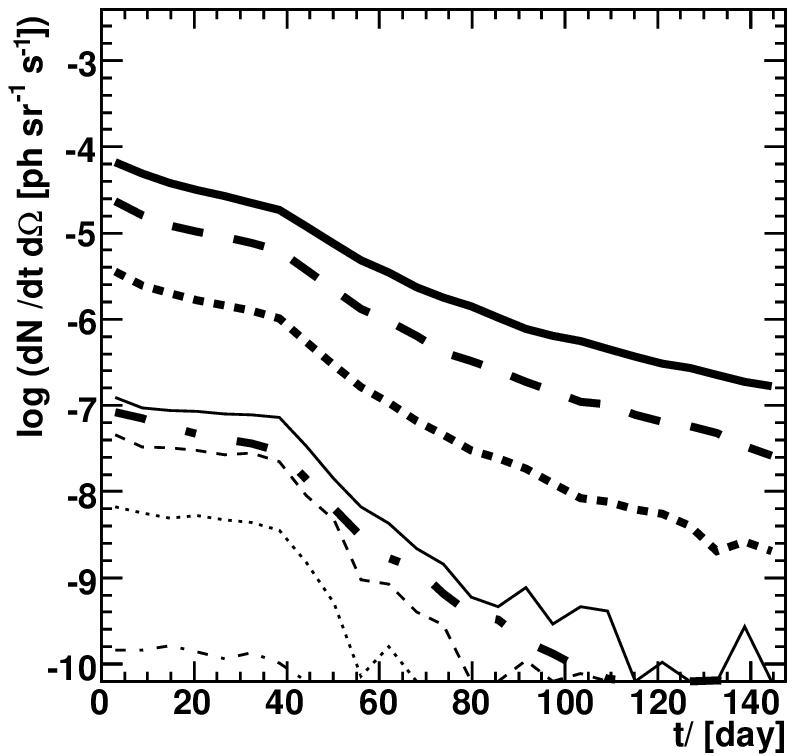}\figlab{$H_0=1-300r_{in}$}{3.6cm}\figlab{$v=0.998c$}{3.3cm}  
\caption{The $\gamma$-ray light curves calculated for 3C~279 (for the parameters as shown in Fig.~\ref{fig:lc_3c279}) but after the propagation through the intergalactic radiation field which is taken from the  model by \protect\citet{fr08}. The light curves are presented in the observer's frame on the Earth.}
\label{lc_3c279_obs}
\end{figure}

\begin{figure}
\includegraphics[scale=0.53, trim=  0 14 0 0, clip]{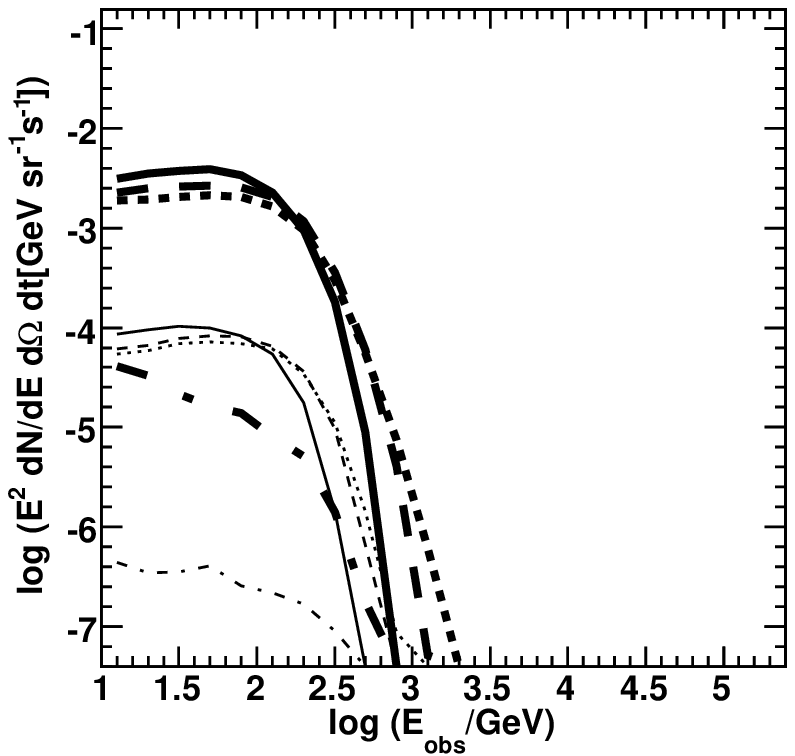}\figlab{$H_0=1-30r_{in}$}{3.4cm}\figlab{$v=0.99c$}{3.1cm}    
\includegraphics[scale=0.53, trim= 20 14 0 0, clip]{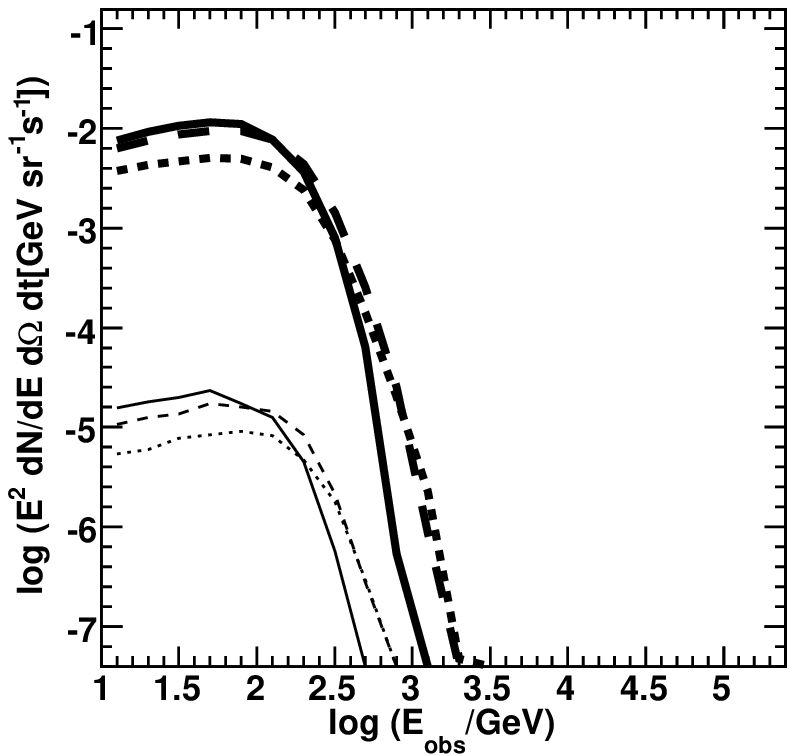}\figlab{$H_0=1-30r_{in}$}{3.4cm}\figlab{$v=0.998c$}{3.1cm}\\ 
\includegraphics[scale=0.53, trim=  0 14 0 0, clip]{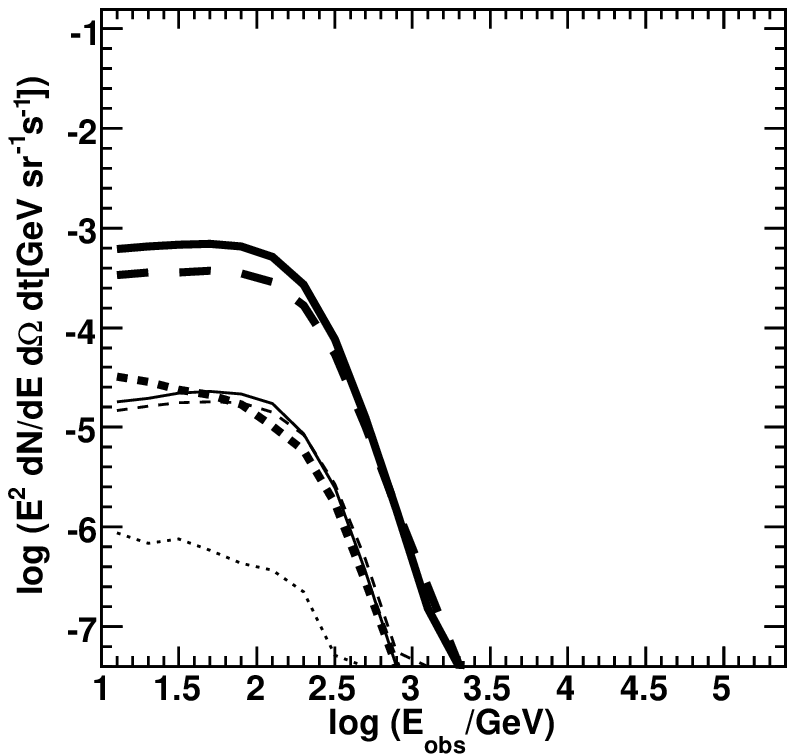}\figlab{$H_0=1-100r_{in}$}{3.4cm}\figlab{$v=0.99c$}{3.1cm}   
\includegraphics[scale=0.53, trim= 20 14 0 0, clip]{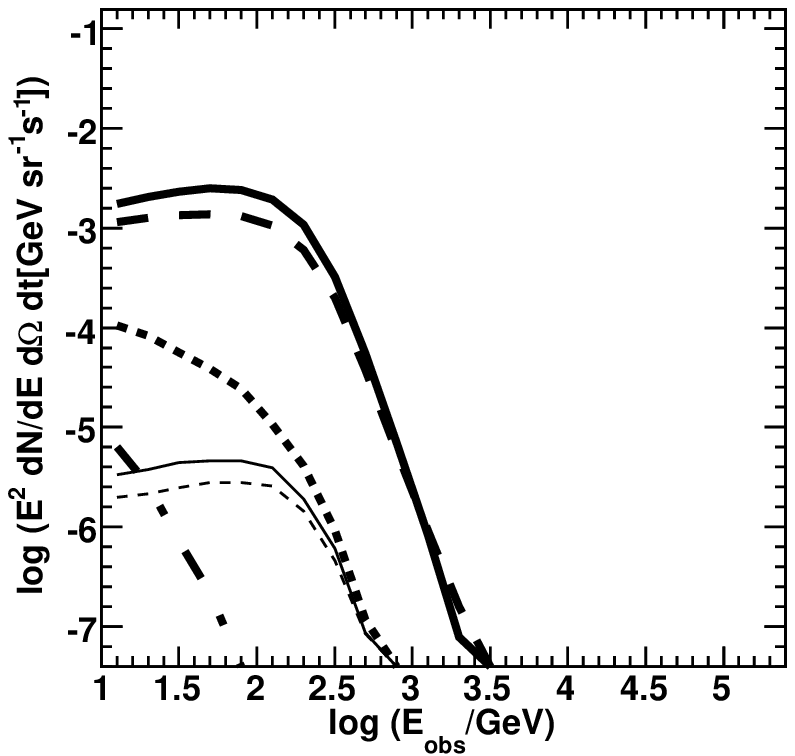}\figlab{$H_0=1-100r_{in}$}{3.4cm}\figlab{$v=0.998c$}{3.1cm}\\
\includegraphics[scale=0.53, trim=  0 0 0 0, clip] {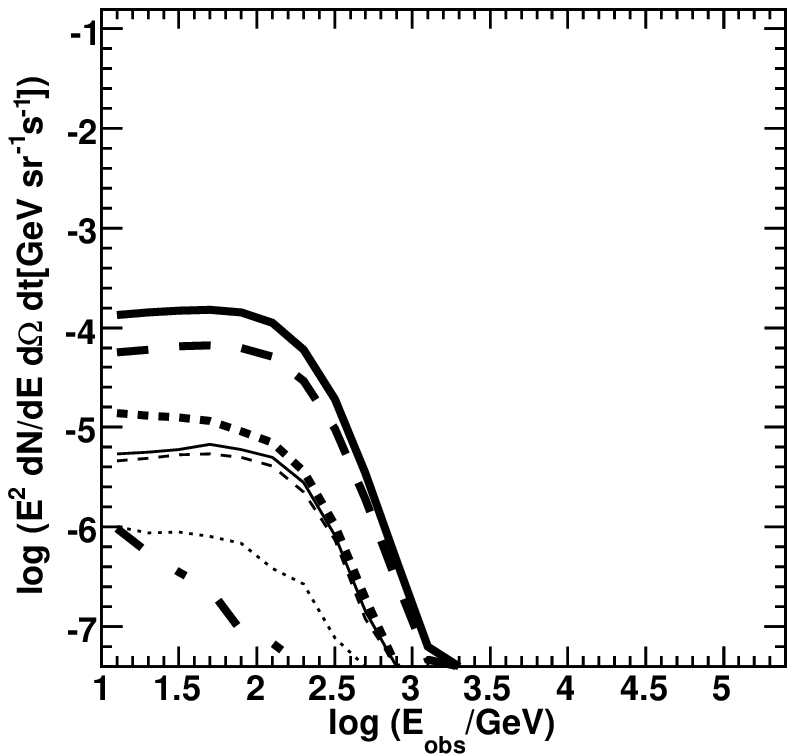}\figlab{$H_0=1-300r_{in}$}{3.6cm}\figlab{$v=0.99c$}{3.3cm}   
\includegraphics[scale=0.53, trim= 20 0 0 0, clip] {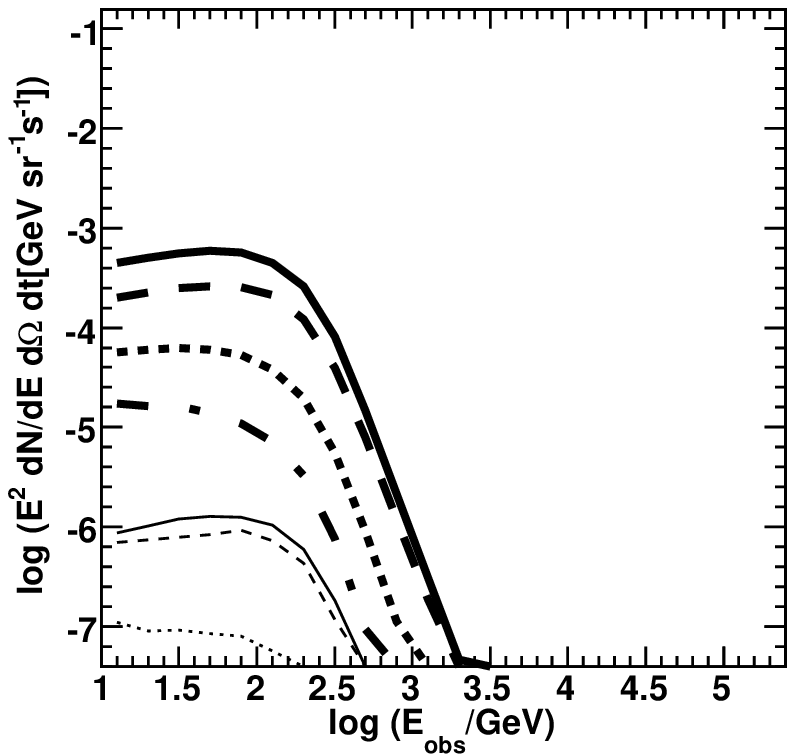}\figlab{$H_0=1-300r_{in}$}{3.6cm}\figlab{$v=0.998c$}{3.3cm}   
\caption{The $\gamma$-ray spectra at specific observed time intervals calculated for 3C~279
(for the parameters as shown in Fig.~\ref{spectra_time_3c279}) but after propagation in the 
intergalactic radiation.
The observed times are $1+z$ larger than the local times i.e.:
top figures (the injection height  range $H_0=1$--30$r_{in}$) (spectra for times $t=0$--1.5 days 
(solid curves), 1.5--3 days (dashed), 3--4.5 days (dotted), 4.5--6 days (dot-dashed)).
Middle figures $H_0=1$--100$r_{in}$ ($t=0$--7.5 days (solid), 7.5--15 days (dashed), 15--23 days 
(dotted), 23--30 days (dot-dashed)).
Bottom figures $H_0=1$--300$r_{in}$ ($t=0$--23 days (solid), 23--45 days (dashed), 45--68 days 
(dotted), 68--90 days (dot-dashed)).
}\label{spectra_time_3c279_obs}
\end{figure}

The $\gamma$-ray light curves and spectra at the specific time intervals, after correction for the propagation effects in the intergalactic radiation field, are shown in Fig.~\ref{lc_3c279_obs} 
and Fig.~\ref{spectra_time_3c279_obs}. 
The time intervals in the Fig.~\ref{spectra_time_3c279_obs} correspond to those shown in  Fig.~\ref{spectra_time_3c279}, calculated for 3C~279 in its reference frame.
The clear differences between these spectra are caused by the cosmological propagation effects. Because of these effects, the observed time scales of the flares are also diluted by a factor of $1+z$ with respect to the time measured at the source. 

The propagation in the EBL over the distance scale of 3C~279 results in an almost complete absorption of the $\gamma$-ray spectrum above $\sim 1$ TeV and a very steep spectrum above $\sim 100$ GeV. 
The shapes of the $\gamma$-ray spectra at specific time intervals become more similar to each other showing weaker dependence on time with the development of the flare.
Still, the spectral index at 100 GeV of the tail emission is significantly steeper than in the plateau emission phase. 

\begin{figure}
\centering
\includegraphics[scale=0.53, trim=  0 0 0 16, clip] {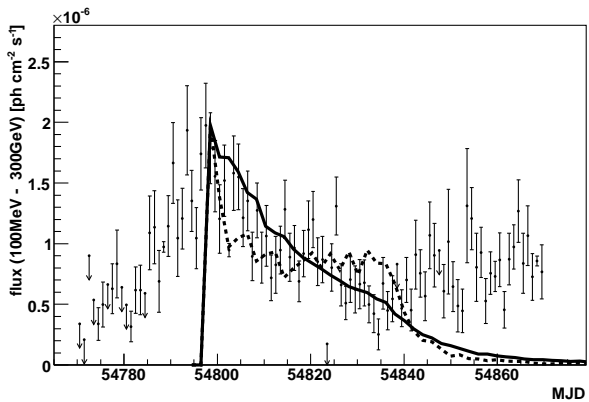}
\caption{
The cascade $\gamma$-ray light curve (above 10~GeV) in the observer's reference frame (as in Fig.~\ref{lc_3c279_obs}) is confronted with the shape of the $\gamma$-ray flare observed from 3C~279 by the {\it Fermi}-LAT at energies above $0.1 GeV$ (the LAT monitored source list, \citet{cc08}). 
Primary electrons are injected with the power law spectrum and spectral index $-2.5$ at a constant rate inside the blob moving along the jet with the velocity $0.998c$ through the range of distances from the base of the jet $H_0=1$--300$r_{in}$. The observation angle is assumed to be $9^\circ$ (solid curve) or $15^\circ$ (dashed).}\label{lc_3c279_obs_comp}
\end{figure}

For the illustration, we confront the calculated $\gamma$-ray light curve for specific parameters with the recent observations of 3C~279.
The {\it Fermi}-LAT instrument revealed a large flare lasting for several weeks.
The LAT $\gamma$-ray light curve above the energy threshold of  $0.1$~GeV resulting from the LAT bright source monitoring \citep[see also ][]{cc08} has been used. 
This light curve has a similar shape as the energy light curve observed above $1$~GeV, but due to larger statistics, it has a more precise sampling. 
It is compared with the $\gamma$-ray light curve at the energy threshold $>10$~GeV calculated in the framework of our model  (see Fig.~\ref{lc_3c279_obs_comp}). 
The behavior of the dropping part of the $\gamma$-ray light curve from 3C~279 is consistent with our calculations for the observation angle equal to $9^{\rm o}$. 
However, the rising part of the calculated light curve is very steep which is due to our assumption on the injection of electrons into a point-like blob at fixed moment of time. 
In the case of an extended blob or the injection mechanism operating at the base of the jet for a specific period of time with rising efficiency, the finite rise time of the $\gamma$-ray flare from 3C~279 should be naturally explained in terms of our model. 
Therefore, we conclude that the several weeks $\gamma$-ray flares, observed in the GeV range from the blazars of the 3C~279 type, can be explained by our IC $e^\pm$ pair cascade scenario initiated by electrons in the radiation field of the accretion disk.
\section{Discussion and Conclusions}

We discuss the most complete version of the external IC model for the $\gamma$-ray production in blazars. 
It is assumed that electrons are injected in the inner part of the jet and comptonize thermal radiation coming from the accretion disk surface. 
As a result a first generation of $\gamma$-rays is produced. 
These $\gamma$-rays initiate IC $e^\pm$ pair cascade in the anisotropic disk radiation. 
We calculate the $\gamma$-ray spectra escaping to the observer located at an arbitrary angle in respect to the direction of the jet. 
In the framework of this model we calculate also the $\gamma$-ray light curves expected from blazars in which blobs containing relativistic electrons propagate with arbitrary velocities.
The angular dependent $\gamma$-ray light curves and spectra are calculated for blazars
observed at large and small inclination angles. 
As an example we have applied the parameters of the two well known AGNs: Cen~A and 3C~279, which have been recently observed in the TeV $\gamma$-rays.
The first source is located relatively nearby, and has a jet which propagates at a relatively large angle in respect to the observer with a mild velocity.
In contrast, the second source is more distant. 
It's inclined at a small angle and the jet propagates with a large Doppler factor.

In order to perform the cascade calculations with a limited number of free parameters, we made some simplified assumptions.
At first, we assume that the disk radiation is the only source of soft photons with which relativistic electrons can interact. 
The calculations with other radiation fields (e.g. infrared radiation from the molecular torus or the disk and jet radiation scattered by the matter around the accretion disk) are straightforward. 
But the interpretation of the final results will be more difficult due to the large number of free parameters describing such more realistic model. 
Therefore, the analysis of such an anisotropic cascade model with a more complicated soft radiation field is left for a future paper.
We also assume the simplest possible (i.e. constant along the jet) dependence of the injection rate of relativistic electrons along the jet. 
This allows us to study the basic features of this complicated cascade model.
These features may change significantly for the more complicated forms of the injection rates along the jet. 
In fact such more complicated dependence for the injection rate is suggested by the observed $\gamma$-ray light curves of flares observed from blazars at different energy ranges.
The simultaneous future observations of $\gamma$-ray light curves at different energy ranges, e.g. GeV and TeV, will put constraints on the detailed models of $\gamma$-ray emission from specific blazars. 

We considered two different models for the maximum energies of injected electrons.
In the model I, the maximum energies of electrons are fixed on 10 TeV independent on the distance from the base of the jet. In the second model, electron maximum energies are determined by their synchrotron energy losses in the local magnetic field of the jet. The electrons have clearly lower energies in the second model.

Based on our calculations, we predict a few characteristic features of the IC $e^\pm$ pair cascade model which can be tested by the future multiwavelength GeV and  TeV $\gamma$-ray observations.
In the case of blazars inclined at relatively large angles such as Cen A, we predict the $\gamma$-ray spectra extending through the TeV energy range even at large inclination angles provided that the blob velocities are mildly relativistic (as observed in the case of Cen A). For slowly moving blobs, the GeV $\gamma$-ray emission at large inclination angles can be even larger than for small angles. However, the TeV $\gamma$-ray emission behaves in the opposite way. Therefore, the model predicts even higher chances of detection of nearby blazars with blobs moving with mild velocities and inclined at large angles in respect to those once inclined at small angles.

Assuming that the injection rate of primary electrons within the blob is constant during the propagation within the jet, the $\gamma$-ray flares last for a few hours up to a few days depending on the range of
injection distances from the accretion disk. However, the $\gamma$-ray spectrum does not change
significantly with the development of the flare. 
It is expected that at large inclination angles the $\gamma$-ray flares observed at the GeV energies should precede the $\gamma$-ray flares at TeV energies. 
This is caused by the strong absorption of TeV $\gamma$-rays propagating at large
inclination angles at the beginning of the flare (when the relativistic electrons in the blob are
still relatively  close to the accretion disk).

In the case of blazars observed at relatively small inclination angles and with fast moving blobs
(e.g 3C 279), the $\gamma$-ray emission features differ significantly. Such blazars are typically located  at cosmological distances. So then, additional effects are caused by the propagation of $\gamma$-rays through
the Extragalactic Background Light. Let us at first concentrate on the $\gamma$-ray emission features
in the source frame (before propagating in the EBL). The $\gamma$-ray emission during the initial
stage has similar intensity (the plateau phase) which is followed by an exponential decay whose 
time scale depends on the range of injection distances along the jet and the velocity of the blob.
The $\gamma$-ray spectrum has similar shape during the plateau phase extending to larger
energies with the development of the phase. However, during the exponential decay of the flare,
the $\gamma$-ray spectrum becomes steeper and cut-offs at lower energies. Therefore we predict
significantly longer $\gamma$-ray flares at the GeV energies than at the TeV energies.
The $\gamma$-ray spectra in the observer's reference frame (after propagation in the EBL) are 
drastically absorbed at energies above a few hundred GeV. 
Therefore, the shapes of the $\gamma$-ray spectra at different stages of the flare become more similar to each other.

In order to envisage whether the considered model parameters are reasonable, keeping in mind the limitations of the simplicity of the scenario, we compare our results with the spectral features of two blazars visible at a small and large inclination angles, Cen A and 3C~279. 
The GeV-TeV $\gamma$-ray spectrum observed from Cen A  can be well described by our model in the case of the observer located at the angle $<40^{\rm o}$ for the electron injection spectrum with the spectral index $-3$ at the range of distances between $1-300$ r$_{\rm in}$ from the base of the jet (see Fig.~3). Note that the estimates of the inclination angle of the jet in this source lay in the range $10^{\rm o}-80^{\rm o}$. So then, our calculations are consistent with the lower values of the inclination angle. The spectra calculated for larger angles are clearly too steep. 
Unfortunately, Cen A is a relatively weak source. Therefore, the $\gamma$-ray spectral information at a flare state and the light curves are at present unavailable. 
We have to wait for the next generation Cherenkov instrument (e.g. CTA) and/or a strong flare from this object in order to test some predictions of our model mentioned above (e.g. hard to soft evolution of the high energy $\gamma$-ray spectra with developing of the flare).

We also show satisfactory fit to the GeV-TeV $\gamma$-ray spectrum reported by the EGRET and the MAGIC telescopes from 3C~279 during the flare state and predict how these spectra should evolve in time (see Fig.~8). The spectral information from distant blazars (such as 3C~279) is partially hidden by the propagation effects in the EBL. Therefore, it will be more difficult to confront predictions for the spectra (e.g. its evolution in time) and light curves (e.g. the plateau and tail emission) in the $\gamma$-ray range expected in terms of our model with the observations. 

In the case of Cen A, we modeled the GeV-TeV $\gamma$-ray spectrum in terms of the cascade model initiated by primary electrons (pure external Compton $e^\pm$ pair cascade model, this work) and the cascade model initiated by primary $\gamma$-rays (produced in the jet, e.g. in the SSC model, see \citet{sb10}).
There aren't any definitive differences between these two scenarios. 
Both can describe the spectral features equally well (compare Fig.~3 in this paper with Fig.~4 in \citet{sb10}). However, the injection spectrum of primary particles have to differ significantly. Moreover, it is expected that the application of the SSC model as a source of primary $\gamma$-rays for considered here cascade scenario may have additional consequences. For example, SSC scenario may produce additional soft radiation field which may compete
with the accretion disk radiation. As a result, other component in the broad band high energy spectrum might appear and some effects on the IC $e^\pm$ cascade spectrum produced only in the disk radiation is 
expected due to an additional cooling mechanism of secondary leptons.   
They have not been included in the cascade calculations by \citet{sb10}.

We have discussed the $\gamma$-ray emission features in terms of the anisotropic, time dependent, IC $e^\pm$ pair cascade model in the case of a simple, constant lepton injection for a range of parameters.
The $\gamma$-ray light curves can change significantly in the case of a more complicated model which includes, e.g. the change of the injection rate of electrons along the jet, the extended injection of electrons at the base of the jet, or the escape of electrons from the blob.
Moreover, blobs can propagate along the jet in a more complicated manner than considered in this simple scenario. 
For example, the blob may decelerate during its propagation along the jet and its viewing angle may also change as postulated in some works. 
Therefore, detailed modeling of the development of $\gamma$-ray flares from specific blazars in such a more complicated version of the external Compton model have to wait for a more complete knowledge on the processes within the jets of AGNs.

We would like to stress an importance of the simultaneous multiwavelength observations of blazars observable at different inclination angles. 
They would allow a more detailed modeling of the processes responsible for the production of $\gamma$-rays in blazars.  
Simultaneous GeV and TeV observations of flares from blazars can efficiently limit the acceleration region and the dynamics of the injection rate.

\section*{Acknowledgments}
This work is supported by the Polish MNiSzW grants: \\
N N203 390834 and N N203 510138.


\appendix

\section[]{IC \electron$^\pm$ pair cascade in the disk radiation}\label{dodatek}

\begin{figure}
\centering
\includegraphics{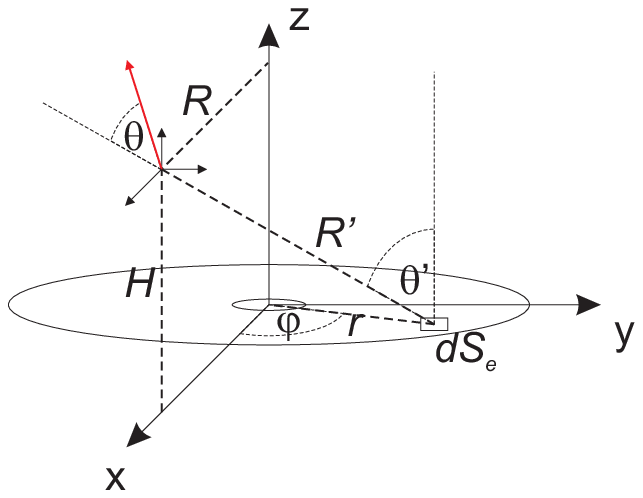}
\caption{
Schematic presentation of the geometry of interaction of the $\gamma$-ray photon with the soft photon with energy $\epsilon$ coming from the accretion disk.
Soft thermal photon is emitted from the unit surface $dS_{\rm e}$ at the angle,~$\theta'$, to the disk surface. 
The emission place is defined by the distance,~$r$, from the center of the accretion disk and the angle, $\varphi$.
The location of the $\gamma$-ray photon is defined by the distance from the axis of the accretion disk $R$ and the height above the disk surface $H$. The distance between $dS_{\rm e}$
and the location of $\gamma$-ray photon is marked by $R'$.
Photons interact at the angle $\theta$.
}
\label{figa1}
\end{figure}

We consider the process of interaction of electrons injected close to the jet base with the soft thermal photons emitted from the surface of the accretion disk.
Due to the large optical depths, electrons interact efficiently with soft photons in the inverse Compton process producing $\gamma$-rays. These $\gamma$-rays again interact with the same disk radiation producing  next generation of leptons. As a result, the IC $e^\pm$ pair cascade develops in the anisotropic radiation field created by an accretion disk in the whole volume above the disk, i.e. outside the jet volume. Here we show the method which is used to track the development of particles in such a cascade.
The method allows us to produce $\gamma$-ray spectra seen by external observer at arbitrary direction with respect to the accretion disk.

\setlength\arraycolsep{2pt}
\subsection{$e^\pm$ pairs from $\gamma-\gamma$ interaction}

Let's consider a $\gamma$-ray photon with energy, $E_\gamma$, propagating at an arbitrary direction at a given distance from the accretion disk (see Fig~\ref{figa1}). Differential energy density of soft photons from the disk, seen by the $\gamma$-ray photon, is 
\begin{equation}
n=\frac{dn}{dS_e d\epsilon dV}=
\frac{2}{h^3 c^3}\frac{\epsilon^2}{\exp(\epsilon/kT)-1}
\frac{\cos\theta'}{{R'}^2},
\label{gestosc fotonow}
\end{equation}
\noindent 
where $h$ is the Planck constant, 
$k$ is the Boltzmann constant 
$c$ is the velocity of light, 
$T = T(r)$ is the temperature profile on the disk surface at the distance, $r$, from its center (see Sect.~\ref{sec:model}), and other parameters are defined in Fig.~\ref{figa1}.

At first, we determine the propagation distance, $L_{\rm int}$, after which, a $\gamma$-ray photon with energy $E_\gamma$ interacts with the soft photon from the disk,  
\begin{equation}
\int_{\rm 0}^{L_{\rm int}}\lambda(x)^{-1}dx = -\ln(1 - P_{\rm r}),
\label{dist}
\end{equation}
\noindent
where $P_{\rm r}$ is the random number, and $\lambda(x)$ is the mean free path for 
$e^\pm$ pair production in $\gamma$-$\gamma$ collision at the propagation distance,~$x$. 
$\lambda(x)^{-1}$ is calculated following by general formula,
\begin{equation}
\lambda^{-1}(x) = \iiint\sigma_{\gamma\gamma}(1-\cos\theta)n \,d\epsilon \,d\varphi \,rdr,
\label{lambda1}
\end{equation}
\noindent
where $\sigma_{\gamma\gamma}$ is the cross section for $\gamma-\gamma\rightarrow e^+e^-$ pair production \citep{jr76}, and $r_{\rm in}$ and $r_{\rm out}$ are the inner and outer radius of the accretion disk.
$\lambda(x)^{-1}$ can be re-written into the form suitable for numerical calculations,
\begin{equation}
\lambda^{-1}={{3\sigma_T}\over{8(hc)^3}}
\int_{r_{\mathrm{in}}}^{r_{\mathrm{max}}} dr\, r (kT)^3 
\int_0^{2\pi} d\varphi I(b)(1-\cos\theta)\frac{\cos\theta'}{{R'}^2},
\label{lambda2}
\end{equation}
\noindent
where $\sigma_{\rm T}$ is the Thomson cross section.
In the above formula the internal integration over the energy $\epsilon$ was substituted by integration over resulting electron velocity $\beta$ in the center of mass system. 
The value of internal integral $I(b)$ is parametrized with a single parameter $b=2m_{\rm e}^2/[E_\gamma(1-\cos\theta)kT]$, $m_{\rm e}$ is the electron rest energy:
\begin{eqnarray}
I(b)&=&2 b^3 \int_0^1 \frac{\beta}{(1-\beta^2)^4}
\frac{1}{\exp(b/(1-\beta^2))-1} g(\beta) d\beta, \label{parametryzacja} \\
g(\beta)&=&\left(1-\beta^2\right)
\left(2\beta\left(\beta^2-2\right)+\left(3-\beta^4\right)
\ln\left(\frac{1+\beta}{1-\beta}\right)\right).
\end{eqnarray}
\noindent
If $L_{\rm int}$ is within the simulation area (defined by the maximum distance $10^4 r_{in}$), the $\gamma$-ray is absorbed in the $e^\pm$ pair production process, otherwise it escapes the accretion disk radiation field.

We determine the energies of secondary electron and positron in the case of absorption of $\gamma$-ray photon.
At first, we draw the place on the disk from which interacting soft photon originates and afterwards its energy. 
The probability of an interaction of $\gamma$-ray photon with soft photons coming from a given location on the disk is proportional to the function under integral in Eq.~(\ref{lambda2}),
\begin{equation}
dP\propto r T^3I(b)(1-\cos\theta)\frac{\cos\theta'}{{R'}^2}drd\varphi 
= f_{\gamma\gamma}(r, \varphi)drd\varphi.
\label{place}
\end{equation}
We use the rejection method in order to determine the parameters of the soft photon emission  place on the disk i.e., $r$ and $\varphi$. 
To do that, we estimate the maximum value of $f_{\gamma\gamma}$.
Using the known temperature profile of the disk surface and numerically determined maximum possible value for $I(b) < 2.4$, we get the condition
$f_{\gamma\gamma}(r,\varphi)<f_{\max} = 2\cdot 2.4\cdot T_{in}^3 r_{in}^2/H^2$.
Afterwards we draw random values of $r$, $\varphi$ and $P_{\rm r}$ until the condition  $f_{\gamma\gamma}(r,\varphi)<P_{\rm r}\cdot f_{\max}$ is fulfilled.

In the next step, we determine the velocity, $\beta$, of leptons produced in the center of mass system.
It is directly linked to the energy of interacting soft photon emitted from the specific place on the disk by the formula $\epsilon = bkT/(1-\beta^2)$.  
Distribution of $\beta$ is proportional to the function under integral in Eq.~\ref{parametryzacja}. 
We determine the maximum value of, $f_{2, \max}$, for fixed value of parameter $b$. 
Now, we randomly sample the velocity of lepton produced in the center of mass system.

In order to determine energies of secondary leptons, at first the scattering angle (in the center of mass system), $\theta_{\rm sc}$ ($\cos\theta_{\rm sc} = \mu_{\rm sc}$), of the produced lepton has to be found. 
$\mu_{\rm sc}$ is randomly sampled from the angular distribution (see \citet{jr76}),
\begin{equation}
\frac{d\sigma}{d\mu_{sc}} \propto
\frac{\beta( 1-\beta^4\mu_{sc}^4+2\beta^2(1-\beta^2)(1-\mu_{sc}^2))} 
     {(1-\beta^2\mu_{sc}^2)^2},
\label{dsigma dmisc}
\end{equation}
\noindent
by reversal method of the above probability distribution. For known $\beta$, the energy $E'$ and momentum $p'$ of lepton in the center of mass system is determined. 
The velocity of the center of mass system, $\beta_{\rm cms}$, and its Lorentz factor $\gamma_{\rm cms}$, is calculated from $\beta_{cms} = \sqrt{E_\gamma^2+\epsilon^2+2E_\gamma\epsilon\cos\theta}/(
E_\gamma+\epsilon)$. This allows us to determine energies of leptons in the accretion disk reference frame by applying obvious transformation,
\begin{eqnarray}
E_{\rm e^\pm}= \gamma_{cms}(E'\pm \beta_{cms}\mu_{sc}p').
\label{energyep}
\end{eqnarray}
\noindent
It is assumed that directions of secondary $e^\pm$ pairs produced in $\gamma-\gamma$ interaction
are the same as the original direction of the primary $\gamma$-ray photon.

\subsection{$\gamma$-rays from IC scattering}
The method of simulation of the parameters of the secondary $\gamma$-rays produced by leptons in the inverse Compton Scattering process is essentially the same as described above for the production of $e^\pm$ pair in $\gamma - \gamma$ collision. 
Let's consider a lepton with a Lorentz factor, $\gamma$ (and the corresponding velocity $\beta$) interacting with disk thermal photon.
The energy of photon in the lepton's frame before and after scattering is equal to $\epsilon'=\omega' m_e$ and $\epsilon'_{sc}=\omega'_{sc} m_e$ respectively, $m_e=511$~keV. 
$\omega'$ is related to the photon energy $\epsilon$ in the disk frame by a transformation:
$\omega'= \epsilon'/m_{\rm e} = \gamma\epsilon(1-\beta\mu)/m_{\rm e}=a\epsilon$,
$\mu$ is a cosine of the angle between direction of the soft photon and the lepton. 

We use general Eq.~(\ref{lambda1}) for calculating the mean free path, but with the cross section replaced by the total cross section for the IC process (see \citet{jr76}). For numerical calculations, we use the following formula for the IC mean free path,
\begin{equation}
\lambda_{\mathrm{IC}}^{-1} =\frac{3\sigma_T}{2(hc)^3}
\int d(\ln r)\,r^2
\int d\varphi (1-\cos\theta)\frac{\cos\theta'}{{R'}^2}I_\mathrm{IC}, \label{eq:lambda_ICS}
\end{equation}
\noindent
where
\begin{equation}
  I_\mathrm{IC}=\frac{1}{a^3}\int^{\infty}_{0}
  s(\omega')\frac{{\omega'}^2}{\exp(\omega'/b)-1}d\omega', \label{eq:sigmaICS_razy_planck}
\end{equation} 
\noindent
$b = akT$.
The total cross section for IC is $ \sigma_{\mathrm{IC}}=2\pi r_0^2 s(\omega')$.
The interaction place of lepton with the soft photon is determined according to Eq.~(\ref{dist}). In the next step, we simulate the place ($r, \varphi$) from which the scattered soft photon has been emitted. Again the rejection method is applied to the probability function proportional to the function under integrals in 
Eq.~(\ref{eq:lambda_ICS}), 
\begin{equation}
dP \propto r^2 (1-\cos\theta)\frac{\cos\theta'}{{R'}^2}I_\mathrm{IC} d\ln r d\varphi.
\end{equation}
After the interaction place is fixed, we simulate the energy of interacting soft photon in the lepton's frame by drawing $\omega'$ from the distribution (applying the rejection method),
\begin{equation}
dP \propto s(\omega')\frac{{\omega'}^2}{\exp(\omega'/b)-1}d\omega'.
\end{equation}
\noindent
Afterwards the scattering angle of the soft photon in the electron reference frame is evaluated from the distribution,
\begin{equation}
\frac{d\sigma}{d\mu_{sc}} \propto\frac{
\mu_{sc}^2+\omega'-\mu_{sc}\omega'+\frac{1}{1+\omega'-\mu_{sc}\omega'}}
{(1+\omega'-\mu_{sc}\omega')^2},
\label{scatang}
\end{equation}
\noindent
where $\mu_{\rm sc} = \cos\theta_{\rm sc}$. 
We rewrite Eq.~(\ref{scatang}) in a variable $x = 1 + \omega' -\mu_{\rm sc}\omega'$ and apply the reversal of the cumulative probability distribution method,
{
\begin{eqnarray}
D[x]&=&\int_1^x \frac{d\sigma}{d\mu_{sc}} \frac{d\mu_{sc}}{dx}dx=
\frac{1}{2{\omega'}^3x^2}
\Big( 2({\omega'}^2-2\omega'-2)x^2\ln(x) \nonumber\\
&+&\big((x-1)(4\omega' x+{\omega'}^2(1+x)+2x(1+x)\big)
\Big). 
\end{eqnarray}
The value of $x$ (and related to it scattering angle $\mu_{\rm sc}$) is obtained by solving the following equation for random number $P_{\rm r}$,
\begin{equation}
D[x]=P_{\rm r}\cdot D[1+2\omega'].
\end{equation}

\begin{figure}
\centering
\includegraphics[trim=150 27 0 0, clip]{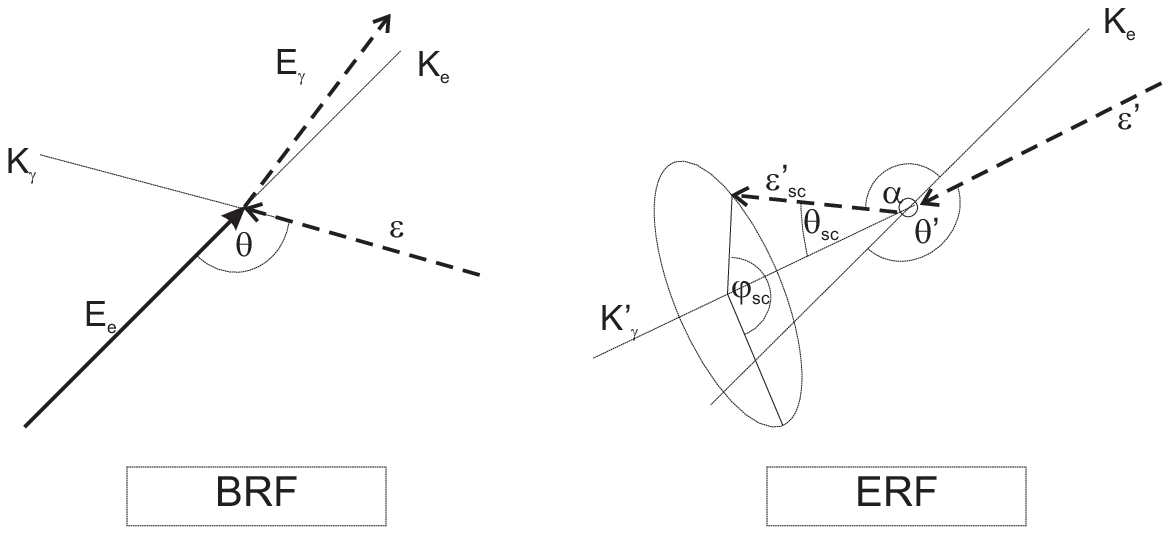}
\caption{IC scattering as seen in the electron rest frame. 
The direction of the relativistic boost (the electron movement direction in the disk frame) is shown with $K_e$, $\gamma$-ray direction is $K'_\gamma$.
The disk photon is interacting at the angle $\theta'$ to the boost direction. 
After scattering its direction is determined by the angles $\theta_{sc}$ and $\varphi_{sc}$. 
The scattered photon is inclined at the angle $\alpha$ to the boost direction. 
}
\label{rys:ics_uklady}
\end{figure}
Now, we are able to calculate the energy of $\gamma$-ray photon produced in the IC process in the disk frame from,
\begin{equation}
E_\gamma=\gamma\epsilon'_{sc}(1+\beta\cos\alpha),
\end{equation}
\noindent
where the energy of the scattered soft photon in the electron rest frame is $\epsilon'_{sc}=\epsilon'/[1+\epsilon'(1-\mu_{sc})/m_{\rm e}]$.
The angle between directions of relativistic boost and scattered photon is calculated as $\cos\alpha=\cos\theta_{sc}\cos\theta'+\sin\theta_{sc}\sin\theta'\cos\varphi_{sc}$ (see Fig.~\ref{rys:ics_uklady}).
The interaction angle in the electron rest frame is related to the corresponding angle in the disk frame $\theta$ with the transformation:
$\cos\theta'=c p_x'/\epsilon'=\gamma\epsilon(\cos\theta-\beta)/\epsilon'$.

\section[]{Time development of the IC $e^\pm$ pair cascade}

In order to obtain the gamma-ray light curves and the time dependent spectra we have to track the time development of the cascade above the accretion disk.
Each particle has an assigned clock, $t$, running in the reference frame of the accretion disk. 
The injection process occurs continuously in a blob moving with the velocity $v$ along the jet in the range of distances between $H_1$ and $H_2$. 
The injection time of electrons at the height, $H_0$, is equal to
\begin{equation}
t_0=(H_0-H_1)/v.
\end{equation}
If a new particle is created in the cascade, its clock is set to the time of its parent particle during the interaction. 

In order to switch from particle clock time, to a time of external observer we have to apply a correction for the emission distance and direction. 
In reality the source is observed under a specified inclination angle $\xi_0$.
In order to obtain enough statistics, the $\gamma$-ray spectra and the light curves are obtained by summation over a small range of cosine of observation angles (e.g. $\Delta(\cos\xi)=0.1$). 
Within such a small $\Delta(\cos\xi)$ interval, the time and spectral characteristics of the $\gamma$-ray emission does not change significantly.
Below we describe the method which has been used for the correct arrival time evaluation in order not to introduce any artificial delays due to integration over a range of $\xi$ angles. 
\begin{figure}
\centering
\includegraphics{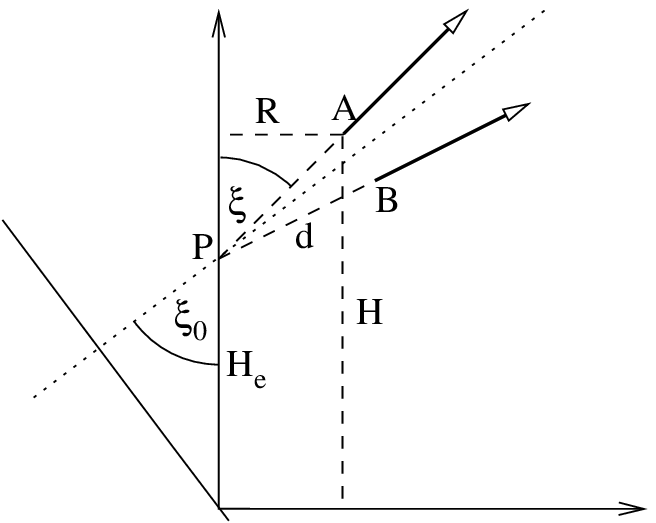}
\caption{
Position dependent time correction applied in order to obtain the  $\gamma$-ray
light curves at the observer reference frame. 
Two $\gamma$-ray photons, A and B (thick arrows), are produced in two sub-cascades at similar inclination angles and at a distance, $d$, from the location of the parent electron ($H_e$ above the disk surface).
The $\gamma$-ray marked by, A, is characterized by the height over the accretion disk $H$, the distance from the jet axis $R$, and the inclination angle $\xi$.
The inclination angle $\xi_0$ describes the center of the cosine angle interval 
in which $\gamma$-rays escaping to the observer are collected.
}
\label{rys:time_corr}
\end{figure}

In case of the lack of the  magnetic field in the region above the disk, the secondary leptons propagate
exactly as $\gamma$-rays i.e., along straight lines.
Therefore, each cascade initiated by a primary electron in the jet propagates along a straight line starting from a point on a jet.
The extrapolated height above the accretion disk $H_e$ can be calculated from (see Fig.~\ref{rys:time_corr}):
\begin{eqnarray}
H_e&=&H-d\cos\xi, \\
d&=&R\sin^{-1}\xi.
\end{eqnarray}
The gamma-ray A, produced at the time $t$, is equivalent to a gamma-ray sent from the point P (at the height $H_e$ above the disk) at the time $t'=t-d/c$.

After extrapolation of directions of escaping $\gamma$-rays to the locations in the jet, we calculate a second correction for the height of $H_e$. 
Instead of using the inclination angle of the individual $\gamma$-rays, we use the central value,~$\xi_0$, of a range of the inclination angles $\Delta(\cos\xi)$.
In this way, both photons A and B will be corrected by the same time difference of $H_e\cos\xi_0$.
Therefore the final correction which is applied to all $\gamma$-ray photons is
\begin{equation}
t_{corr}=t-\frac{1}{c}\left(d+H_e\cos\xi_0\right).
\label{eq:tcorr}
\end{equation}
\label{lastpage}
\end{document}